# Charting the velocity of brain growth and development

Johanna M. M. Bayer[1,2*], Augustijn A. A. de Boer[1,2] Barbora Rehák-Bučková[1,2], Charlotte J. Fraza[1,2], Tobias Banaschewski[3]; Gareth J. Barker[4]; Arun L.W. Bokde[5], Rüdiger Brühl[6], Sylvane Desrivières[7], Herta Flor[8], Hugh Garavan[9], Penny Gowland[10], Antoine Grigis[11], Andreas Heinz[13], Herve Lemaitre,[14], Jean-Luc Martinot[15,17], Marie-Laure Paillère Martinot[15,16], Eric Artigues 15,17, Frauke Nees[18], Dimitri Papadopoulos Orfanos[12], Tomáš Paus[19,20], Luise Poustka[21], Michael N. Smolka[22], Nathalie Holz[1], Nilakshi Vaidya[23], Henrik Walter[24], Robert Whelan[25], Paul Wirsching[24], Gunter Schumann[23,26,27], Alzheimer's Disease Neuroimaging Initiative[+], Nina Kraguljac[28], Christian F. Beckmann[1,2], Andre F. Marquand[1,2]

[1*]Donders Institute for Brain, Cognition and Behaviour, Kapittelweg 29, Nijmegen, 6511SK, Gelderland, The Netherlands.
[2]Radboud University Medical Center, Kapittelweg 29, Nijmegen, 6511SK, Gelderland, The Netherlands.
[3] Department of Child and Adolescent Psychiatry and Psychotherapy, Central Institute of Mental Health, Medical Faculty Mannheim, Heidelberg University, Square J5, 68159 Mannheim, Germany;
[4]Department of Neuroimaging, Institute of Psychiatry, Psychology & Neuroscience, King's College London, United Kingdom;
[5]Discipline of Psychiatry, School of Medicine and Trinity College Institute of Neuroscience, Trinity College Dublin, Dublin, Ireland;
[6]Physikalisch-Technische Bundesanstalt (PTB), Braunschweig and Berlin, Germany;
[7]Social, Genetic and Developmental Psychiatry Centre, Institute of Psychiatry, Psychology & Neuroscience, King's College London, United Kingdom;
[8]Institute of Cognitive and Clinical Neuroscience, Central Institute of Mental Health, Medical Faculty Mannheim, Heidelberg University, Square J5, Mannheim, Germany;
[9]Department of Psychology, School of Social Sciences, University of Mannheim, 68131 Mannheim, Germany;
[10]Departments of Psychiatry and Psychology, University of Vermont, 05405 Burlington, Vermont, USA;
[11]Sir Peter Mansfield Imaging Centre School of Physics and Astronomy, University of Nottingham, University Park, Nottingham, United Kingdom;
[12]NeuroSpin, CEA, Université Paris-Saclay, F-91191 Gif-sur-Yvette, France;
[13]Department of Psychiatry and Psychotherapy, University of Tübingen, Germany; German Center for Mental Health (DZPG), Site Tübingen, Germany;
[14]Institut des Maladies Neurodégénératives, UMR 5293, CNRS, CEA, Université de Bordeaux, 33076 Bordeaux, France;
[15]Institut National de la Santé et de la Recherche Médicale, INSERM U 1299 "Developmental Trajectories, Psychiatry & Addictology", University Paris-Saclay, Ecole Normale Supérieure Paris-Saclay, Dept. mathematics, CNRS UMR9010; Centre Borelli, Gif-sur-Yvette, France;
[16]AP-HP. Sorbonne University, Department of Child and Adolescent Psychiatry, Pitié-Salpêtrière Hospital, Paris, France;
[17]Psychiatry research Department, Barthélémy Durand Hospital, Etampes, France;
[18]Institute of Medical Psychology, Ludwig-Maximilians-Universität (LMU) in Munich, Munich, Germany;
[19]Department of Psychiatry, Faculty of Medicine and Centre Hospitalier Universitaire Sainte-Justine, University of Montreal, Montreal, Quebec, Canada;
[20]Departments of Psychiatry and Psychology, University of Toronto, Toronto, Ontario, Canada;
[21]Department of Child and Adolescent Psychiatry, Center for Psychosocial Medicine, University Hospital Heidelberg, Heidelberg, Germany;
[22]Department of Psychiatry and Psychotherapy, Technische Universität Dresden, Dresden, Germany;
[23]Centre for Population Neuroscience and Stratified Medicine (PONS), Department of Psychiatry and Psychotherapy, Charité Universitätsmedizin Berlin, Germany;
[24]Department of Psychiatry and Psychotherapy CCM, Charité – Universitätsmedizin Berlin, corporate member of Freie Universität Berlin, Humboldt-Universität zu Berlin, and Berlin Institute of Health, Berlin, Germany;
[25]School of Psychology and Global Brain Health Institute, Trinity College Dublin, Ireland;
[26]Centre for Population Neuroscience and Precision Medicine (PONS), Institute for Science and Technology of Brain-inspired Intelligence (ISTBI), Fudan University, Shanghai, China.
[27]German Center for Mental Health (DZPG), Site Berlin-Potsdam, Germany;
[28]Ohio State College of Medicine, Columbus, Ohio, USA

†Data used in preparation of this article were obtained from the Alzheimer's Disease Neuroimaging Initiative(ADNI) database (adni.loni.usc.edu). As such, the investigators within the ADNI contributed to the designand implementation of ADNI and/or provided data but did not participate in analysis or writing of this report.A complete listing of ADNI investigators can be found at:http://adni.loni.usc.edu/wp-content/uploads/how_to_apply/ADNI_Acknowledgement_List.pdf

*Corresponding author(s). E-mail(s): johanna.bayer@donders.ru.nl; andre.marquand@donders.ru.nl

Abstract

Brain charts have emerged as a highly useful approach for understanding brain development and aging on the basis of brain imaging and have shown substantial utility in describing typical and atypical brain development with respect to a given reference model. However, all existing models are fundamentally cross-sectional and cannot capture change over time at the individual level. We address this using velocity centiles, which directly map change over time and can be overlaid onto cross-sectionally derived population centiles. We demonstrate this by modelling rates of change for 24,062 scans from 10,795 healthy individuals with up to 8 longitudinal measurements across the lifespan. We provide a method to detect individual deviations from a stable trajectory, generalising the notion of 'thrive lines', which are used in pediatric medicine to declare 'failure to thrive'. Using this approach, we predict transition from mild cognitive impairment to dementia more accurately than by using either time point alone, replicated across two datasets. Last, by taking into account multiple time points, we improve the sensitivity of velocity models for predicting the future trajectory of brain change. This highlights the value of predicting change over time and makes a fundamental step towards precision medicine.

## 1. Main

Modeling change over time is a central goal in biology, both for understanding the mechanisms underlying growth and development and for monitoring disease progression. Indeed, in many clinical contexts, it has been suggested that time is the best diagnostician (Irving and Holden 2013). For example, brain disorders are increasingly recognised in terms of alterations to an expected trajectory of brain growth or ageing; psychiatric disorders are frequently associated with atypical trajectories of growth and development (Insel 2014) and neurological disorders can be understood as deviations from a healthy trajectory of ageing (Hou et al. 2019; Jack et al. 2013; Lorenzi et al. 2015; Liu et al. 2024).

Decades of development in the field of brain imaging has produced many methods to study age-related differences on the basis of cross-sectional and longitudinal data (Franke et al. 2010; Reuter et al. 2012; Ashburner and Ridgway 2012; Fjell and Walhovd 2010), and has revealed intricate patterns of brain changes across development and into old age (Paus et al. 2008; Tau and Peterson 2010; Tamnes et al. 2010). However, the majority of this work has focussed on the estimation of group level trajectories, which essentially describe the population average curve for development and ageing across the lifespan. Whilst these approaches are useful for understanding general trends, they do not provide the ability to understand individual differences, nor to predict deviations from an expected trajectory at an individual level, nor integrate a person's historical trajectory, which is often essential for clinical decision-making.

More recently, inspired by pediatric growth charts that have long been used to chart child development (Cole 2012), brain charting (normative modeling) approaches have emerged to understand individual differences in brain development and ageing in clinical and non-clinical populations (Marquand, Wolfers, et al. 2016; Marquand, Rezek, et al. 2016; Marquand et al. 2019). These techniques have been used to chart changes across the human lifespan with greater precision than has previously been possible (Rutherford et al. 2022; Bethlehem et al. 2022; Ge et al. 2024) and across many different clinical populations (Marquand 2019; Fraza 2024). One of the key features of normative models is that they enable individual level characterisation by indicating the population centile at which a given study participant lies at a given timepoint, or in other words whether a given individual is more extreme than would be expected for their age and sex. However, it is crucial to recognise that this does not provide any information about individual change over time, even if the normative models themselves are estimated on the basis of longitudinal data. Rather, what is needed is the ability to detect centile *crossings* at the level of the individual participant and a method to include past measurements into the forecast. Centile crossings indicate significant movements up, down and across centiles across multiple measurements (Fig 1A). Finally, classical models do not take into account a person's individual developmental history. This is important because integrating multiple time points, that is, conditioning on a previous history, can potentially improve the predictions. This also requires longitudinal data.

In more detail, the centiles in existing normative models essentially describe the group-level distribution at each timepoint and do not capture movements of individuals through the growth chart as they transition between time points, as illustrated in Fig 1B. This is crucial because individuals often do not exactly track cross-sectional centiles through time. Rather, they frequently cross centiles, even in the absence of a brain disorder. This is illustrated in Fig 1C, which shows two different pairs of trajectories that are indistinguishable under conventional cross-sectional normative models.

There are other reasons why cross-sectional normative modeling techniques are unsuitable for measuring longitudinal change. First, the significance of individual changes depends on both the position in the lifespan period and the position within the population distribution. For example, in Fig 1D all the red lines have the same slope, but not all provide evidence for a trajectory that deviates from what would be expected by the population level centiles. Second, they cannot differentiate asynchronous developmental trajectories, which may temporarily push an individual to a more extreme centile, without indicating a truly meaningful change, as shown in Fig1E. Third, it is important to account for regression to the mean, such that extreme values at the first measurement timepoint are statistically more likely to be followed by more moderate ones at subsequent timepoints (Fig 1E,F). Finally, current cross-sectional normative models do not incorporate an individual's longitudinal trajectory and therefore cannot use prior measurements to inform predictions of future observations. This is shown in Fig 1G, where the forecast changes depending on whether only the last timepoint or the whole trajectory is accounted for. Taken together, these factors show that additional approaches are necessary to properly evaluate longitudinal change including a person's history at the level of the individual person.

We refer to conventional cross sectional normative models as *population centiles* ('*distance centiles*' in the growth charting literature (van Buuren 2023)). To identify centile crossings, a different approach is needed, that accounts for movements within the population through time. In this paper, we modernise a general framework for estimating change over time at the level of the individual person using *velocity centiles* (van Buuren 2023) (Berkey et al. 1993) and involve mapping the distribution of the normative *rate of change* between consecutive time points rather than the cross-sectional value at each time point (van Buuren 2023) and demonstrate the utility of this for structural neuroimaging data. Concretely, his framework allows us to infer the distribution of rates of change between two consecutive time points, and thereby to derive *thrive lines* (Cole 2012), which are statistical bounds on the expected rate of change and allow inferences about the expected longitudinal change at the individual level (including atypical development or disease progression). For example, these allow us to identify when the rate exceeds a given threshold and hence benchmark a significant change between two measurements. These are used in pediatric medicine to declare 'failure to thrive', which would correspond to a rate of change more extreme than, say, 95% of the population between those two measurements (van Buuren 2023; Cole 2012). This same rationale can be applied to many clinical problems in brain disorders. For example, detecting a structural deviation in the left lateral ventricle from a stable trajectory could signal transitioning from an at-risk state (e.g. mild cognitive impairment) to a disease state (e.g. dementia). We provide an example of these thrive lines for the brain in Fig 1H.

Estimating velocity centiles is considerably more challenging for brain imaging data than for simple measures such as height or weight because: (i) the range and complexity of changes that occur across different brain regions and lifespan is substantial; (ii) no single cohort exists that spans the lifespan with sufficient sampling density; (iii) longitudinal neuroimaging data that do exist are very sparsely sampled (typically two or a few data points per participant), (iv) strong site effects arise when aggregating cohorts (Bayer, Dinga, et al. 2022; Bayer, Thompson, et al. 2022) and (v) velocity centiles need to be able to quantify a meaningful change beyond the noisiness of brain scans. While two previous reports have started to address some of those issues (Di Biase et al. 2023; Rehák Bučková et al. 2025), no general-purpose approaches currently exist to quantify change over time in brain measures at the level of the individual.

In this manuscript, we provide a comprehensive framework for assessing change over time in brain imaging data. To achieve this, we (i) estimate purpose built pre-trained normative models (Rutherford et al. 2022) on the basis of large samples, carefully processed using harmonised longitudinal pipelines; (ii) provide means to estimate velocity centiles and thrive lines for across heterogeneous clinical cohorts, which (iii) enables statistical quantification of the unexpectedness of change in brain imaging derived features at the level of the individual person given velocity centiles. We validate these approaches by predicting transition from mild cognitive impairment to dementia on the basis of longitudinal imaging data.

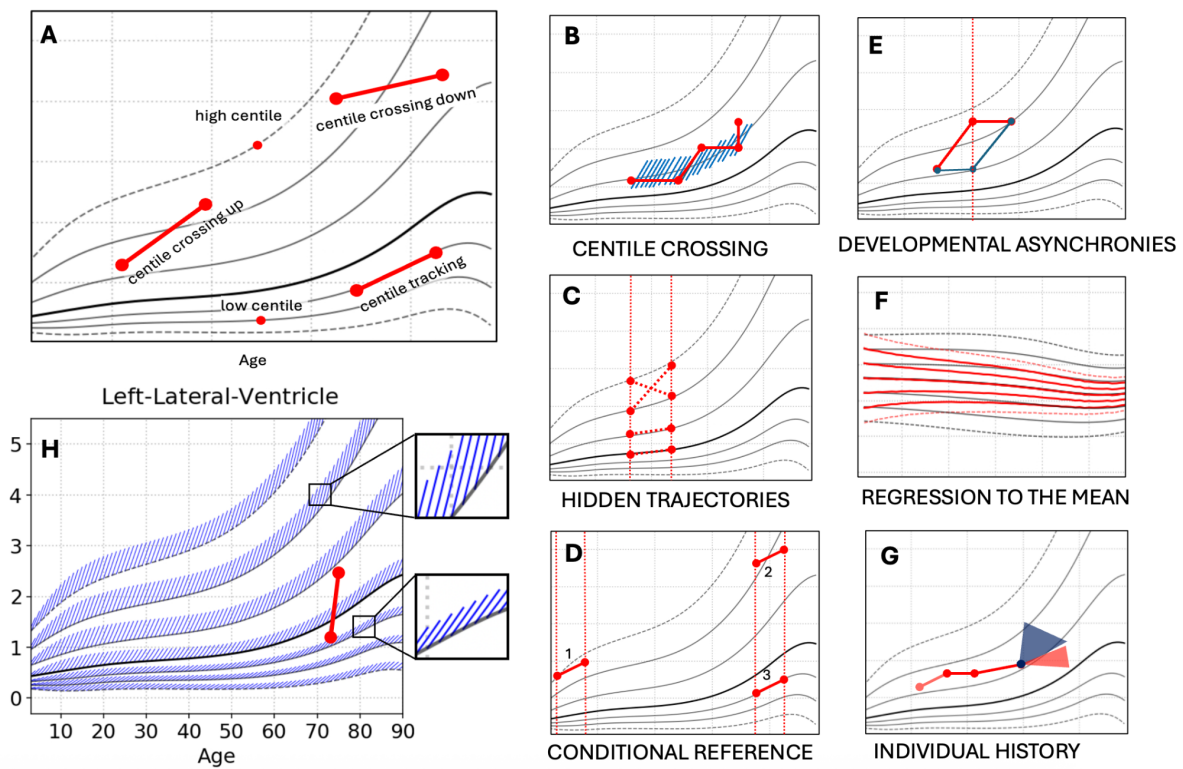

Figure. 1: Velocity centiles vs. population centiles: A: Example of centile crossing up, down and centile tracking on top of cross-sectional population centiles. B-F: Conditions under which population centiles

fail. B: *Centile Crossing*: illustration of centile crossings during development, that are not statistically larger than would be expected due to chance. C: *Hidden trajectories:* since classical models only provide a distribution at each timepoint (i.e. centiles that vary smoothly through time), they cannot differentiate different trajectories that cross between timepoints (top two lines) from trajectories that stick to their centiles (bottom two lines). D: *Conditional Reference*: an equal magnitude longitudinal change can have varying significance, depending on the developmental period (1 vs 2 & 3) and on the starting point (2 vs 3). E: *Developmental Asynchronies*: during periods of developmental change, asynchronous development spurts can lead to (temporary) shuffling of z-scores of individuals. F: *Regression to the Mean*: the probability of the next measurement to regress to the mean increases with the extremeness of the previous measurement. G Individual History: the forecast window for the next time point changes whether a person's personal history over multiple time points is taken into account (red shadow) or just the last available measurement (blue shadow) H: Examples of thrive lines, overlaid onto population centiles of the Left-Lateral Ventricle. The blue lines indicate the 95% thrive lines that indicate both the slope and expected magnitude of change after two years (see text for details). The red subject is an individual from the ADNI dataset that transitions from mild cognitive impairment to Alzheimer's disease and shows a clear centile crossing event. Zoomed in sections: illustration of the conditional nature of thrive.

2. Results

First, we generated a set of cross-sectional normative reference models for cortical thickness, surface area and subcortical brain volume. To achieve this, we compiled an aggregate dataset containing n = 58,793 subjects from 100 scanning sites (72,758 scans, out of which n = 28,550 subjects, 29,181 scans, were used as training set) of which a substantial proportion of subjects had longitudinal data (n = 10,795 Subjects, 24,062 scans) (Fig 2A, B, see supplementary table S3.1, S3.2). Some pipelines failed for some subjects, which is why the total number of subjects differs slightly between pipelines. All data were processed using harmonised longitudinal pipelines (see methods), which we have shown substantially improves normative model estimation (Rehák Bučková et al. 2025). To ensure accurate modeling of both nonlinear effects and complex, non-Gaussian distributional structure, we employed a recently proposed flexible modeling approach based on the 'sin-arcsinh' (SHASH) distribution (de Boer et al. 2024; Kia et al. 2022). Overall, the fit of these models was excellent, both in terms of the mean and the shape of the distribution (supplementary fig. S5.1-3). These models therefore complement prior normative reference models we have provided (Rutherford et al. 2022) through more flexible modeling of the shape of distribution, the possibility to learn models from distributed data (federated learning), advanced modeling of site effects and nonlinear, non-Gaussian age effects, and broader coverage across the lifespan.

Next, we sought to demonstrate the feasibility of estimating velocity centiles for tracking temporal change using longitudinal data. To achieve this, we developed a method to accommodate natural variation in the degree to which individuals cross population-level centiles throughout development. This is necessary because individuals do not exactly track a given cross-sectional centile as they progress through development, due to a variety of factors such as inter-individual variability in growth rate and random measurement noise (Fig 1E). To achieve this, we use the longitudinal data to determine stability (i.e. correlation) of individual subject centile estimates (Z scores) across time. We then developed a banded matrix completion algorithm (see methods) to enable us to infer the distribution of transitions across the lifespan on the basis of these inferred correlations (supplementary Figure S1)(van Buuren 2023). From these estimates, we can then derive thrive lines, that is cut-off points for normal longitudinal development) for a given velocity centile (e.g. 5% velocity centile), which can be overlaid on the population centiles for a given follow-up period. Fig 2C shows two examples: for the left lateral ventricles and left hippocampus. These are shown for an example follow up period of two years, although other follow-up periods can be computed as needed. Because neurodegeneration implies reduction of volume in the hippocampus, we plot the lower (i.e. downward 2.5% velocity centile). In contrast, for the ventricle, we plot the upper 97.5% velocity centile, because in this case neurodegeneration results in an increase in volume for the ventricles. The length of these lines reflects the expected change within the follow up period (in the aforementioned examples, 2 years) and the slope of each of the blue or red lines indicates the statistical limit on the slope at a given point). Thrive lines therefore permit statistical statements at the level of the individual person. For example, 97.5% of individuals within the normative model do not have a change that exceeds the red or blue shaded areas in figure 2C, or a slope greater than one of the individual blue or red lines. Notice these have the desirable properties that they are adaptive according to the lifespan stage and position within the population (c.f. Fig 1D) and therefore account for the base rate of centile crossings that occur naturally during development. Thrive lines for other brain regions are shown in the supplement (S2.1-3).

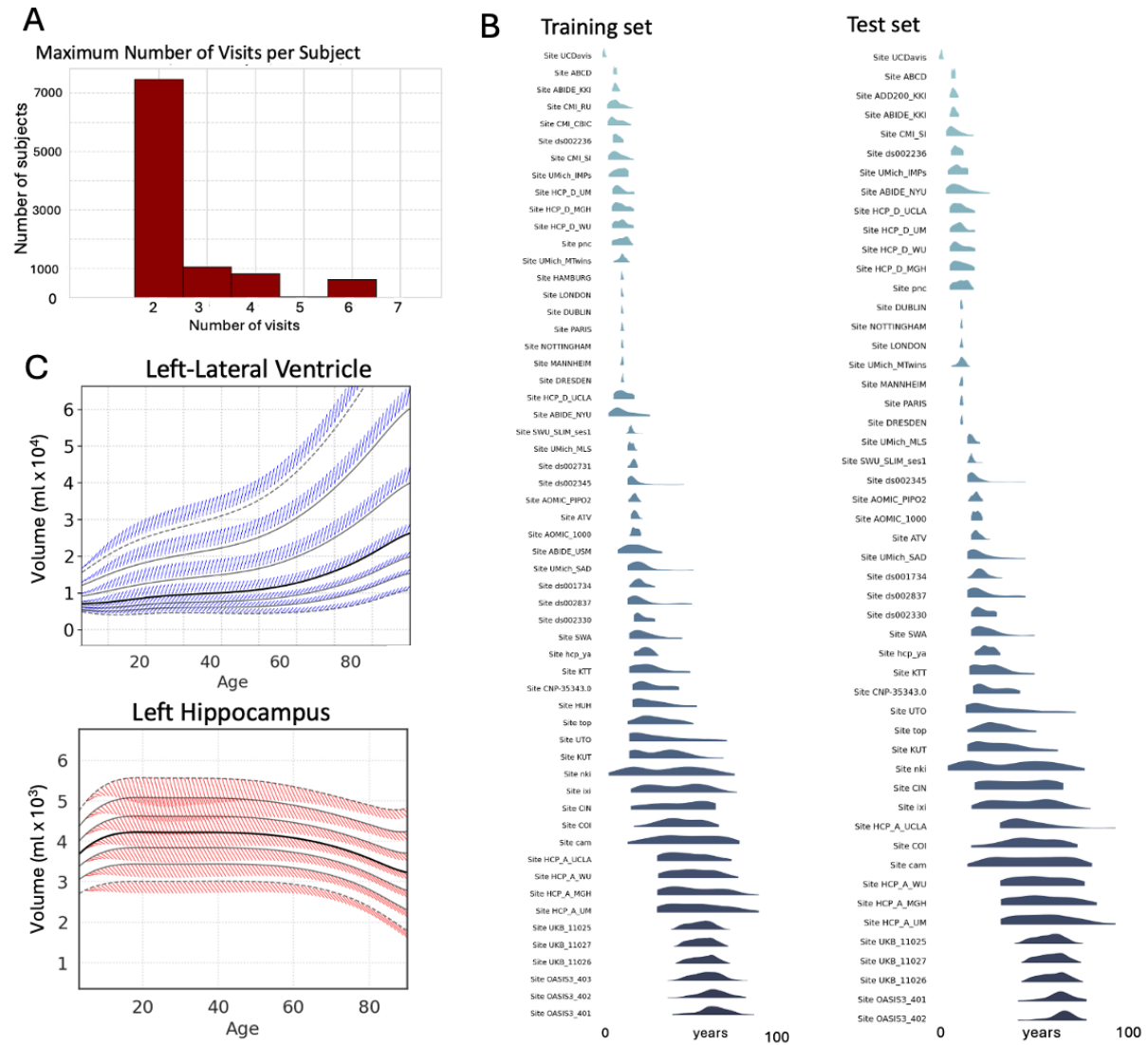

Figure 2. A: Distribution of repeat visits in the longitudinal data set. C) Distribution of age by site in the training and test set. B) Examples of thrive lines for a two year period for the Left-Lateral ventricle and the Left Hippocampus.

Next we demonstrate the utility of this approach for predicting the onset and progression of brain disorders using data from the ADNI cohort (Jack et al. 2008; Petersen et al. 2010), containing subjects with mild cognitive impairment, dementia in addition to healthy individuals. We also replicate these findings in the OASIS cohort, containing healthy individuals and those having mild cognitive impairment and (mild) dementia (LaMontagne et al. 2019; Marcus et al. 2007). In Figure 3, we show a measure that quantifies the significance of the change across two time points (i.e. quantifying statistically whether a centile crossing has occurred). We refer to this statistic as Z-gain (see methods) (van Buuren 2023; Cole 2012). The Z-gain measure accurately reflects disease status (Fig 3A) and –more importantly– provides a sensitive measure for disease transitions (Fig 3C). For example, the Z-gain measure shows that individuals who transition to dementia have faster volume loss in the hippocampi relative to both healthy individuals and individuals with a stable dementia diagnosis. The same pattern was also observed in the OASIS dataset (Fig 3 B-D).

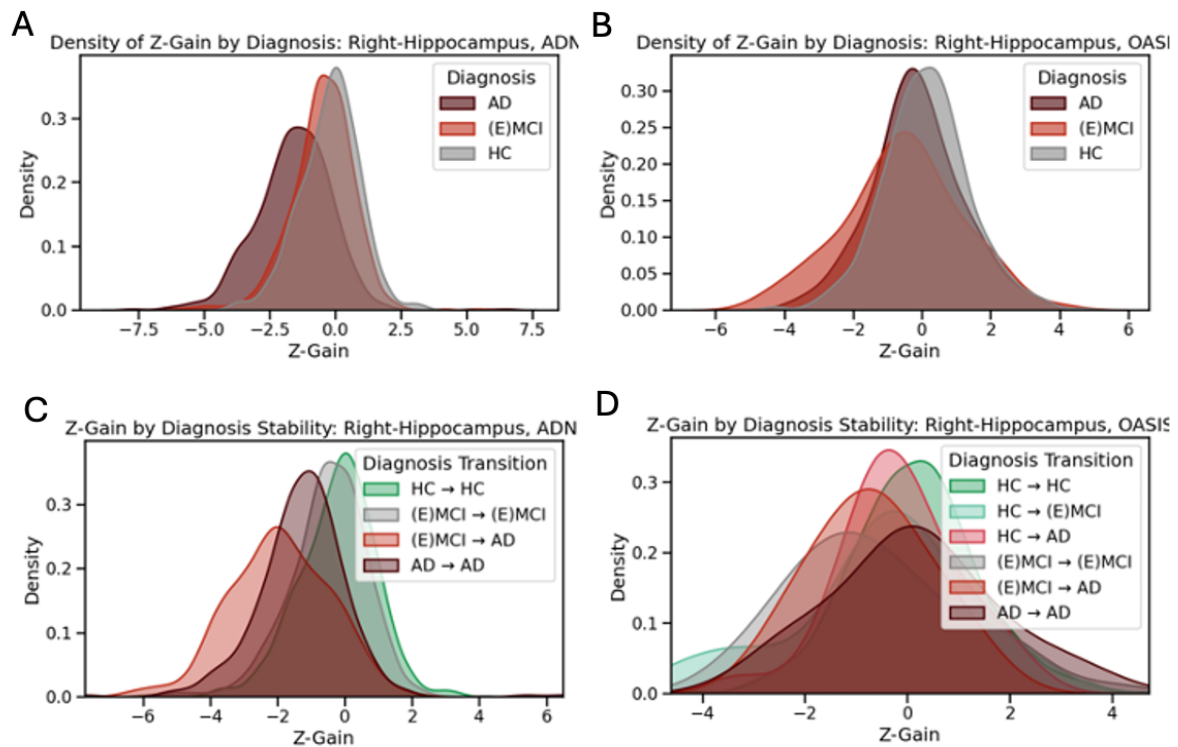

Figure 3. Z-gain distributions in the Right Hippocampus in the ADNI and the OASIS data sets. A-B: Z-gain distributions between individuals with an Alzheimer's disease diagnosis and controls in ADNI (A) and OASIS (B). C-D: Z-gain distributions by transition between time points for ADNI (A) and OASIS (B). In both cases, individuals who transition between diagnosis are losing brain volume faster than those that do not.

Next, we quantified the discriminative value of the Z-gain score across multiple brain regions by computing the area under the receiver operating characteristic curve across all pairs of timepoints and regions in the ADNI dataset. The Z-gain scores provided considerably better discrimination of disease transitions relative to the more conventional measure of the volume or cortical thickness score at each timepoint, especially lateral temporal regions (Fig 4, S3.3).

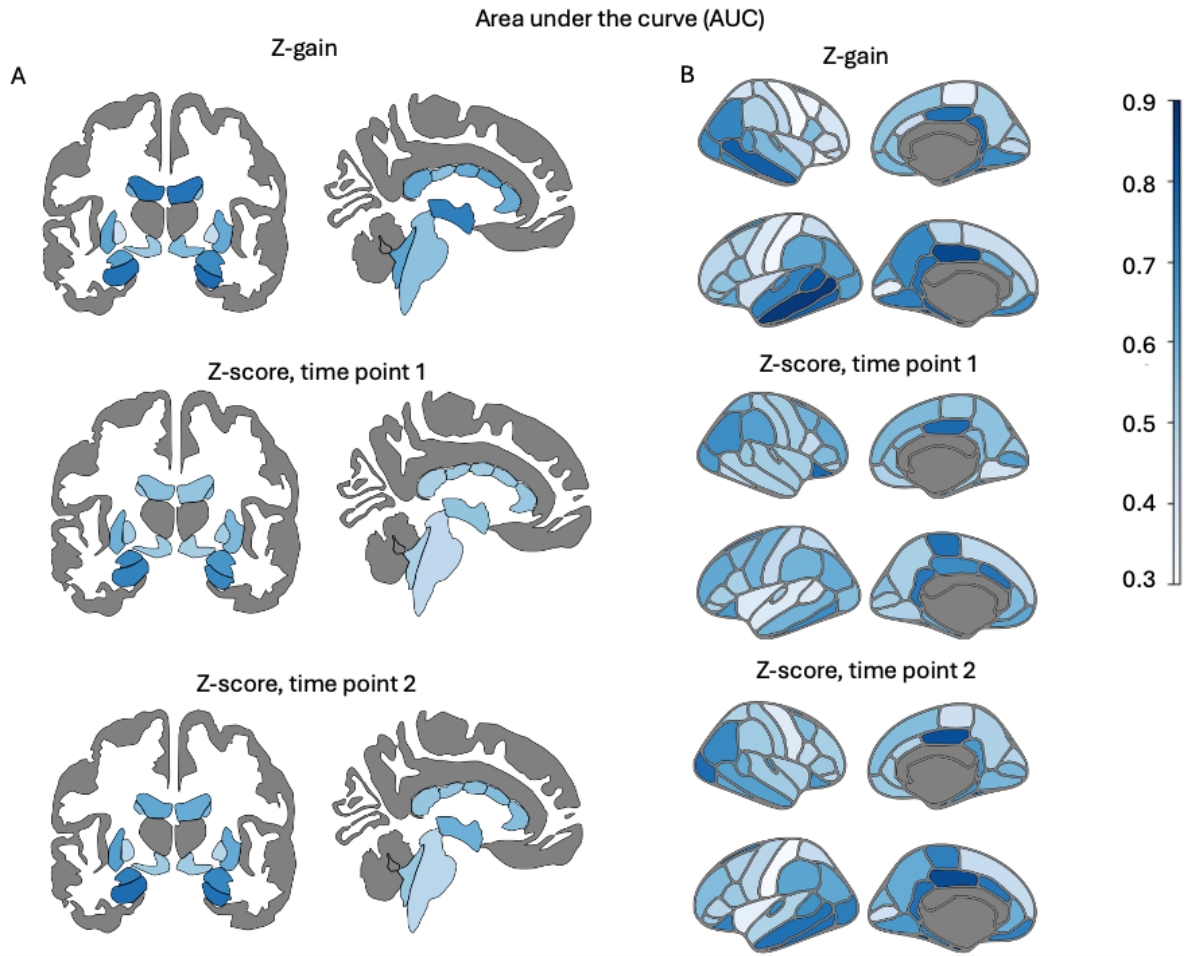

Figure 4. Area under the curve of differentiation between early mild cognitive impairment and Alzheimer's disorder, in the ADNI data set. A) subcortical regions B) Desikan cortical regions.

Next, we take this framework further towards precision diagnostics and show that our approach becomes more sensitive for detecting individual disease transitions with an increasing quantity of longitudinal data. To achieve this we use a conditional variant of the Z-gain score that takes into account multiple timepoints per individual (see methods) and apply this to trajectories derived from the lateral ventricle. In Fig 5, we show trajectories for four illustrative subjects having more than 6 timepoints within the population centiles of this region (Fig 5A, and zoomed in Fig 5B). Figures 5C and D illustrate how adding multiple time points can detect transition to Alzheimer's as an outlier event.  For example, the subject shown in Figure 5C has seven usable follow up time points, the last of which is associated with transition to Alzheimer's disease (orange). Calculating z-gain from the sixth to the seventh timepoint alone, results in a z-gain of 1.78, just within the non-significant range. Including the all six time points before the 7th to calculate z-gain for the seventh time point leads to an adjusted z-gain of 4.44, and marks this timepoint as highly significant. This shows that accounting for multiple timepoints can assist to identify transition to dementia by shrinking the cone of uncertainty for the future timepoint. Figures 5E and F demonstrate how including multiple time points of the past shrinks the windows of uncertainty, even for an event that lies already outside the range of non-extreme predicted range.

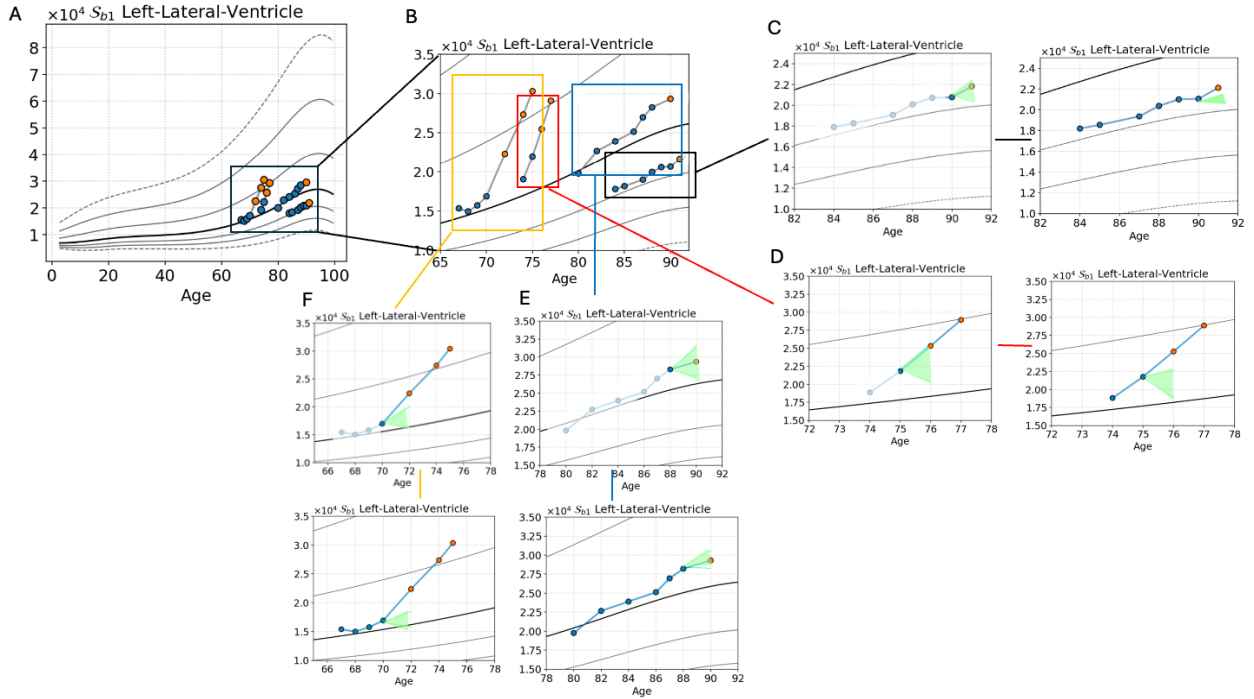

Figure 5. Examples of conditional forecasting. A. Example of four trajectories of four different patients for the size of the Left-Lateral-Ventricle. Blue: assessment points when the diagnosis was pre-Alzheimer's. Orange: assessment points when the diagnosis was considered Alzheimer's disease. B. zoom in of four trajectories. C-F green triangles: forecast window in which the next time point will lie with 95% probability, based on previous observations. Overall, the prediction window narrows when taking more time points into account to create the forecast. C-D: the time point that marks transition to dementia (time point 7 in C, and time point 3 in D) lies within the forecast when taking into account only the last time point, and outside when taking into account all previous healthy timepoints. E-F: The prediction window narrows when taking into account additional previous time points.

## 3. Discussion

In this work, we have developed a comprehensive framework for modeling change over time in biological measures, leveraging and modernising the classical notions of velocity centiles and thrive lines to model multiple time point trajectories of brain growth and aging. This approach has many applications, but may be particularly valuable in identifying early signatures of change that could signal the onset or progression of neurological and psychiatric disorders, which are nearly all grounded in change over time (Hou et al. 2019; Jack et al. 2013; Lorenzi et al. 2015; Liu et al. 2024). Further this framework allows monitoring the progression of a disease over multiple time points. We demonstrate the value of this approach in predicting the onset of dementia and showed that (i) across many brain regions, the rate of change across two timepoints was a more sensitive marker of transition to dementia than either timepoint alone and (ii) that accounting for multiple timepoints increases the sensitivity of this method for detecting individual worsening trajectories.

Brain charting techniques have emerged rapidly in the clinical neuroimaging literature, and have been invaluable in providing insights into brain development and aging across the lifespan for brain structure (Rutherford et al. 2022; Bethlehem et al. 2022; Ge et al. 2024)), brain function (Sun, Zhao, et al. 2025; Rutherford et al. 2023), and for many different clinical populations (Marquand 2019; Fraza 2024), owing to their ability to perform statistical inferences at the level of the individual. However, most of the models available are unable to provide inferences about (unexpected) change over time, even if the models are estimated using longitudinal data. This is because current brain charts do not quantify the base rate of centile crossing, which is important because even healthy individuals do not stick exactly to a fixed trajectory over cross-sectional normative models during the lifespan, but rather can fluctuate across trajectories in a manner that varies depending on the brain region and measurement technique (Fig 1B and Supplemental Figures S2.1-3). Whilst preliminary frameworks have been proposed for drawing longitudinal inferences from from normative models (Di Biase et al. 2023; Rehák Bučková et al. 2025), the approach we propose here is unique in that it provides the ability to make longitudinal inferences over multiple time points at the level of the individual person and the ability to account for the the trajectory that each individual is following. The approach proposed by (Rehák Bučková et al. 2025) can be seen as a special case of our approach for only two timepoints and for a particular estimation technique, based on difference scores between two consecutive time points relative to the uncertainty in the prediction. While this approach is useful, it is limited to two timepoints and carries the strong assumption that controls track their centiles over time. In contrast, the approach presented here is highly general in that it only requires z-scores from a given normative model for a new individual, does not require these theoretical assumptions, can be applied to any type of normative estimation method and accounts for the entire trajectory of data for each individual.

Whilst conditional reference charts and thrive lines have been proposed for classical growth charts, the sparse sampling across the lifespan, the presence of structured noise and strong site effects inherent to neuroimaging data make estimation of these models considerably more challenging than for simple measures such as height or weight that can be measured exactly. Indeed for reliable inferences, it is crucial that the Z gain scores follow a Gaussian distribution and are free from residual site effects. This highlights the benefit of our non-Gaussian normative modeling approach, which allows us to map heteroscedastic and skewed phenotype distributions into Gaussian, z-score standardized distributions in z-space.

It should also be noted that, because our velocity models and thrive lines have been estimated in z-space, they can be used as a reference for any data that has been transferred to a z-space with similar covariates. With a basis on within-subject, within-site estimates of healthy controls, velocity models can be considered free of effect of site.

Whilst the approaches we introduce here are in principle applicable to any normative model that provides the means to map raw data to Z scores, including diverse data modalities (Cirstian et al. 2025; Rutherford et al. 2023), voxel level models (Holz et al. 2023; Fraza et al. 2024) and non-imaging measures, we acknowledge some limitations: first, all large scale normative models are defined with respect to a given reference model, which have inevitable biases, due for example to sampling characteristics of the data upon which it is based. Most cohorts in brain imaging are biased toward European populations, and there is emerging evidence that more accurate inferences can be obtained using population-specific reference models (Sun, Qin, et al. 2025; Rutherford et al. 2024). In our experiments, thrive lines are defined with respect to specify a given follow up period, and whilst it is straightforward to calculate these with respect to any period, this should ideally be matched to the follow up period for the data under consideration. In addition, like most large scale normative reference models, the datasets we aggregated to estimate the models here do not have even coverage of the lifespan, and particularly pediatric cohorts have limited coverage. Therefore, in the results we present at this time the thrive lines at very young ages should be interpreted with caution. Future work should focus on enhancing the coverage of these difficult to scan populations and also focus on increasing demographic coverage.

In conclusion, we have demonstrated a comprehensive framework for modeling biological change at the level of the individual person on the basis of normative models. We have demonstrated the utility of this approach in detecting the onset of brain disorders. This provides a general purpose method towards advancing brain charting methodology towards detecting individual differences. To maximise the utility of this contribution to the field, we provide a software implementation available in the [PCNtoolkit](#) software package, along with a set of analysis scripts for reproducing the results presented in this manuscript, available via https://github.com/predictive-clinical-neuroscience/Velocity. We also release all models d to the community for use (https://surfdrive.surf.nl/s/Mb6mZyFmJeCaPcZ).

## 4. Methods

**Data**

For the purposes of normative model estimation, we obtained a total of 72,758 brain scans from 58,793 healthy subjects that were scanned on 100 different scanners (see S3.2) from publicly available sources. The age range across the total sample spanned from 2 to 100 years. A small percentage of individuals per brain phenotype were excluded as part of quality control. This entailed removing all data points that were located more than 5 standard deviations from the grand mean of that phenotype. From these subjects, 47,998 (individuals contributed one brain scan each (cross sectional cohort). A further 10,795 subjects contributed up to 8 brain scans from consecutive time points (24,062 scans, see supplementary table S3.1). A subset of the data used in the preparation of this article were obtained from the Alzheimer's Disease NeuroimagingInitiative (ADNI) database (adni.loni.usc.edu). The ADNI was launched in 2003 as a public-privatepartnership, led by Principal Investigator Michael W. Weiner, MD. The primary goal of ADNI has been to test whether serial magnetic resonance imaging (MRI), positron emission tomography (PET), other biological markers, and clinical and neuropsychological assessment can be combined to measure the progression of mild cognitive impairment (MCI) and early Alzheimer's disease (AD).

*Clinical validation cohorts*

For clinical validation, we additionally used 327 scans (169 subjects) with mild cognitive impairment or Alzheimer's diseases from the OASIS2 and OASIS3 cohorts (LaMontagne et al. 2019) and 1435 individuals with mild cognitive impairment, stages of Alzheimer's diseases from the ADNI cohorts (Jack et al. 2008; Petersen et al. 2010).

*Pre-processing*

We ran harmonised Freesurfer pipelines on all subjects, including longitudinal processing for subjects providing multiple scans. This involves standard cross-sectional processing, estimating an average template of all scans for each subject, then repeating the cortical reconstruction using longitudinal routines. Next, we extracted cortical thickness measures for regions of interest in the Desikan (34 left hemisphere, 34 right hemisphere and one bilateral median and mean thickness measure) and Destrieux atlases (74 left hemisphere 74 right hemisphere and one bilateral medial and mean thickness measure) in addition to 51 subcortical volume measures (Destrieux et al. 2010; Fischl 2012; Desikan et al. 2006). All data points that were more than 7 standard deviations from the grand mean of that region were removed.

*Datasets for modeling*

For modeling purposes, we created two different data splits including all data. The first data set included all healthy data (full data set) in one large training set. The second pipeline aimed at estimating velocity models. Here, we created a training set containing ~ 50% of all cross-sectional data. We then created a test set with the remaining cross-sectional data, and then added the longitudinal data as well as clinical data from the OASIS and the ADNI data sets. All results in this paper are generated from the test set of this pipeline (the training set was disregarded after normative model fitting). More specifically, the cross-sectional data were used to generate population thrive lines, the longitudinal data to calculate velocity centiles, and the clinical data for clinical evaluation.

**Normative modeling**

The normative modeling approach applied in this study is based on a hierarchical Bayesian regression framework with a sin-arcsinh ('SHASH') likelihood as described in (de Boer et al. 2024) and as implemented in the PCNtoolkit (version 1.12). This includes defining a generative model and inferring the posterior distribution of all model parameters as a step of inference based on the priors and the data using Markov Chain Monte Carlo Sampling (MCMC). In the version of the SHASH model used in this manuscript ($S_{b1}$), we set fixed the $\epsilon$ and $\delta$ parameters, while $\mu$ and $\sigma$ were modelled as a regressions onto an covariate matrix $\phi(X)$, where $\phi$ is a cubic B-spline basis expansion with 7 equally spaced knots and $X$ is the input matrix. In detail, our model specification included $\mu$ and $\sigma$ to be linearly modelled on age and sex as covariates, and a random, non-centered intercept offset for $\mu$ as a batch effect. Sampling in the PCNtookit was performed using the PyMC library based on the No-U-Turn Sampler (NUTS) (de Boer et al. 2024; Hoffman and Gelman 2011). We sampled 2 chains of 1000 samples, plus 500 samples discarded for warm up. This allows the estimation of individualised Z scores, which can be used as a deviation score for each individual and also to derive velocity measures as described below.

*Velocity centiles*

The question that we aim to assess in longitudinal normative modeling is the probability p that a certain change occurs between two timepoints t1 and t2. Note that the period between t1 and t2 is not restricted to the unit length 1, but can be of arbitrary length. Hence we aim to calculate p($Y_2$| $Y_1$, L), where $Y_2$ is a point in time, $Y_1$ is any previous time point and L is the variability in change in the population between these two time points. L, in turn, is defined by a distance measure D and a time measure R. While D can be obtained from cross-sectional or longitudinal data or a mix of those, R can only be obtained from longitudinal data. We write:  L = {D,R}.

The literature differentiates between marginal (unconditional) and conditional population centile or velocity models (Cole 1993). Marginal models are standardized (e.g, for example normalized for age and sex), but cross-sectional, hence one individual contributes one data point. These models do not allow any inference about longitudinal processes, as longitudinal information is absent from the data. Conditional models, in contrast, base their prediction onto additional information, such as one or several previous time points. While marginal models show biases such as a regression to the mean effect, especially in the tails of the distribution, conditional models should not (Cole 1993).

Hence, in order to obtain conditional, unbiased velocity models, we need to obtain a conditional, unbiased estimate of L. We can obtain standardized estimates for D by converting the Y axis into z-space, for example by normative modeling. However, this leaves us still with unstandardized (unconditional) estimates of R that do not take into account differences between different timepoints T on the x-axis. One possibility, and the approach employed here, is to standardize R by basing it on the correlation between different time points T on the x axis, not on the absolute amount of time passed.

Taking all this into account we can define the *gain*; that is, the conditional, standardized velocity $z_\alpha$ between two time points *t2* and *t1* and their standardized observations $Z_{t2}$ and $Z_{t1}$.

The gain $z_\alpha$ is given by (van Buuren 2023; Cole 1993) :

$$z_\alpha = (Z_{t2} - r_t * Z_{t1}) * SD^{-1} \quad (1)$$

$$SD = \sqrt{(1-r_t^2)} \quad (2)$$

where $r_t$ is the correlation between those two time points in z-space and *SD* is the standard deviation, based on rt.

From equation (1), it can be derived that the prediction for the next observation $\hat{Z_{t2}}$ from the previous observation $Z_{t-1}$ with a z-score cut-off $z_\alpha$ can be obtained via:

$$\hat{Z}_{t2} = r_t * Z_{t1} + z_\alpha * SD \quad (3)$$

We see that we obtain conditioning both by taking into account $r_t$, the amount of change between time points when calculating the next time point, and by conducting those operations in z-space.

In order to obtain velocity centiles, we applied a threshold $z_\alpha$ of ±1.96, to obtain the predicted value that signifies the threshold to a significant change (in a magnitude outside of the changes of the healthy reference population). This leads us to:

$$\hat{Z}_{t2} = r_t * Z_{t-1} \pm 1.96 * SD \quad (4)$$

**Model fits**

Model fits were evaluated using the performance measures Explained Variance (EXPV), Mean Standardised Log Loss (MSLL) (Rasmussen and Williams 2005), Pearson Correlation Coefficient $\rho$ between true and predicted values, the Standardised Mean Squared Error (SMSE) and the Root Mean Squared Error (RMSE). All measures were comparable in range to our previously published models (Rutherford et al. 2022).

*Correlation matrices*
The computation of the Z-gain statistics above requires an estimate of the correlations of each brain measure across time. These were calculated in z-space from the z-scores derived from the normative models of the 22,479 healthy individuals that provided more than one time point. The Pearson correlation coefficient $\rho$ was calculated for a specific pair of time points $t_1$ and $t_2$, if at least 5 subjects provided scans at both time points. The time period between $t_1$ and $t_2$ was set to be one year, for an age range from 0 - 100 years. Hence, all subjects that fell within that period were considered to be of that age (binning), and were consequently allocated to one of 99 age bins. Following the example of (van Buuren 2023) and based on the *Cole correlation model (Cole 1995; van Buuren 2023)*, we then completed this correlation matrix by modeling the Fisher-transposed correlations of $\rho$, $\phi = \frac{1}{2}(1+\rho)/(1-\rho)$, between two ages $t_1$ and $t_2$ as a function of their sum $(t_2 + t_1)$, their difference $(t_2 - t_1)$ and their reciprocate $(t_2 - t_1)^{-1}$, and two multiplicative terms. Defining the terms $U_1 = \log((t_2 + t_1)/2)$, $U_2 = \log((t_2 - t_1))$ and $U_3 = (t_2 - t_1)^{-1}$, we define a model with six unknown parameters that need to be estimated from the correlation matrix:

$$\phi(t_1, t_2) = \beta_0 + \beta_1 V_1 + \beta_2 V_2 + \beta_3 V_3 + \beta_4 V_1 V_2 + \beta_5 V_1 V_1 + \epsilon \quad (5)$$
$$\epsilon \sim N(0, \sigma^2) \quad (6)$$

Given that most subjects only had follow up points of less than 10 years from their first assessment, we limited the forecast to 10 years, resulting in the lower triangle of a banded matrix with a bandwidth 10 around the diagonal, with all other entries being not defined (see average correlation per pipeline in Figure S1.1).

**Multiple time points**

For obtaining forecasts including multiple time points, we extended the framework used in (1) to multiple time points. For example, we try to predict $Y_t$ from a subset of previous timepoints n < t, $p(Y_t | Y_1, Y_2, .. Y_n, L)$.

In order to solve this question, we define an extended version of (3):

$$\hat{Z}_t = \beta_1 * Z_1 + \beta_2 * Z_2 + \ldots + \beta_n * Z_n + z_\alpha * SD \quad (7)$$

With *t > n*. Of note, not all n previous timepoints need to be included in (7), nor need they to be equally spaced.

We can obtain the $\beta$ weights in (7) making use of the relationship between the Pearson correlation coefficient $\rho$ and the $\beta$ regression coefficients in ordinary least square regression (OLS) in longitudinal data sets. We implemented a python version of the sweep operator (Goodnight 1979) for this purpose, which allows to obtain the $\beta$ values related to cross correlations for specific time point pairs by applying pivots to in the correlation matrix.

## 5. Acknowledgements

Data collection and sharing for this project was funded by the Alzheimer's Disease Neuroimaging Initiative(ADNI) (National Institutes of Health Grant U01 AG024904) and DOD ADNI (Department of Defense awardnumber W81XWH-12-2-0012). ADNI is funded by the National Institute on Aging, the National Institute ofBiomedical Imaging and Bioengineering, and through generous contributions from the following: AbbVie,Alzheimer's Association; Alzheimer's Drug Discovery Foundation; Araclon Biotech; BioClinica, Inc.; Biogen;Bristol-Myers Squibb Company; CereSpir, Inc.; Cogstate; Eisai Inc.; Elan Pharmaceuticals, Inc.; Eli Lilly andCompany; EuroImmun; F. Hoffmann-La Roche Ltd and its affiliated company Genentech, Inc.; Fujirebio; GEHealthcare; IXICO Ltd.; Janssen Alzheimer Immunotherapy Research & Development, LLC.; Johnson &Johnson Pharmaceutical Research & Development LLC.; Lumosity; Lundbeck; Merck & Co., Inc.; MesoScale Diagnostics, LLC.; NeuroRx Research; Neurotrack Technologies; Novartis PharmaceuticalsCorporation; Pfizer Inc.; Piramal Imaging; Servier; Takeda Pharmaceutical Company; and TransitionTherapeutics. The Canadian Institutes of Health Research is providing funds to support ADNI clinical sitesin Canada. Private sector contributions are facilitated by the Foundation for the National Institutes of Health(www.fnih.org). The grantee organization is the Northern California Institute for Research and Education,and the study is coordinated by the Alzheimer's Therapeutic Research Institute at the University of SouthernCalifornia. ADNI data are disseminated by the Laboratory for Neuro Imaging at the University of SouthernCalifornia.

## Supplementary Information

S1. Supplementary Figures

S1.1: average cross correlations, imputed

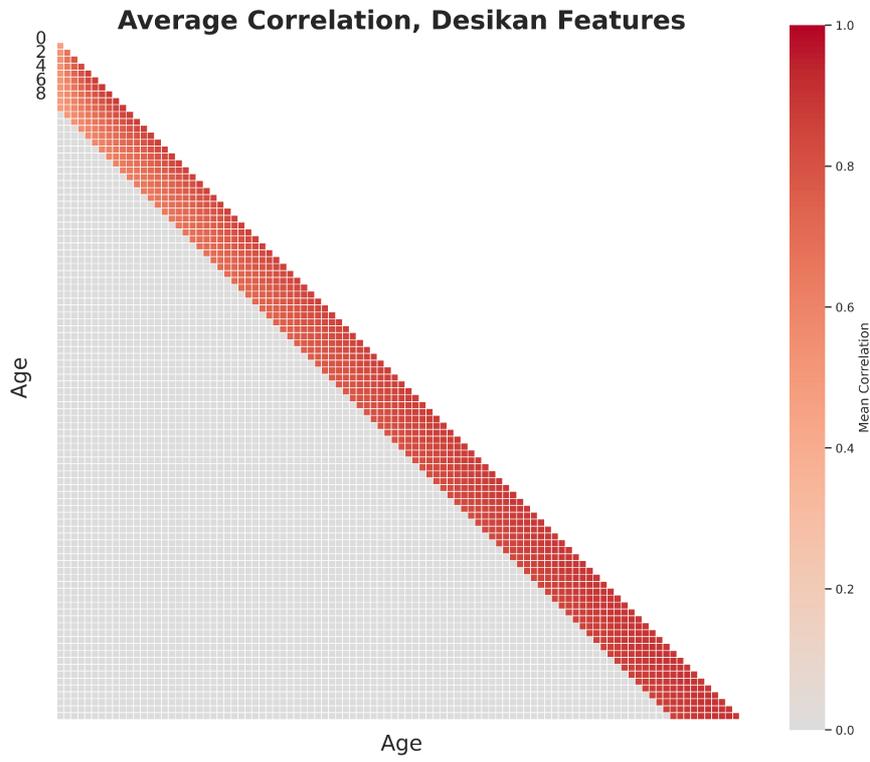

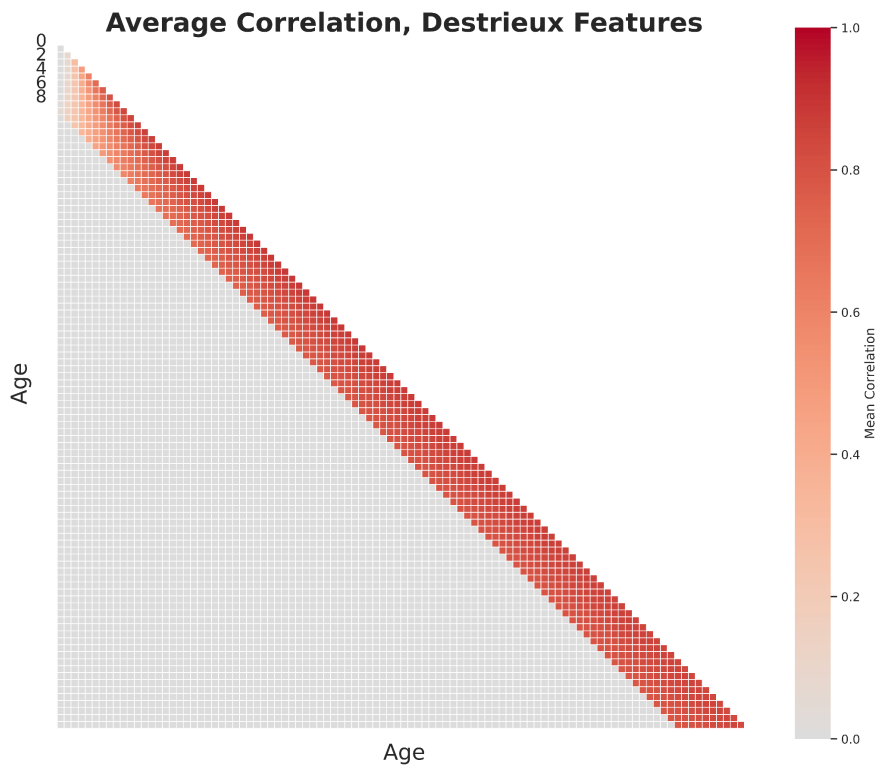

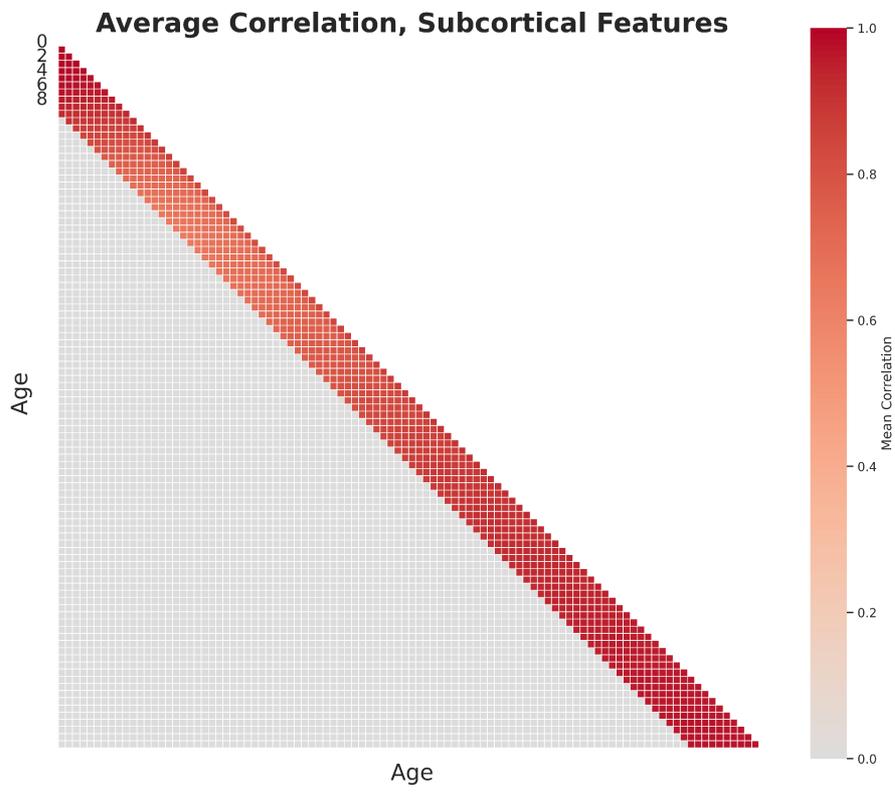

# Supplementary Figures S2: Thrive lines

## S2.1 Destrieux atlas

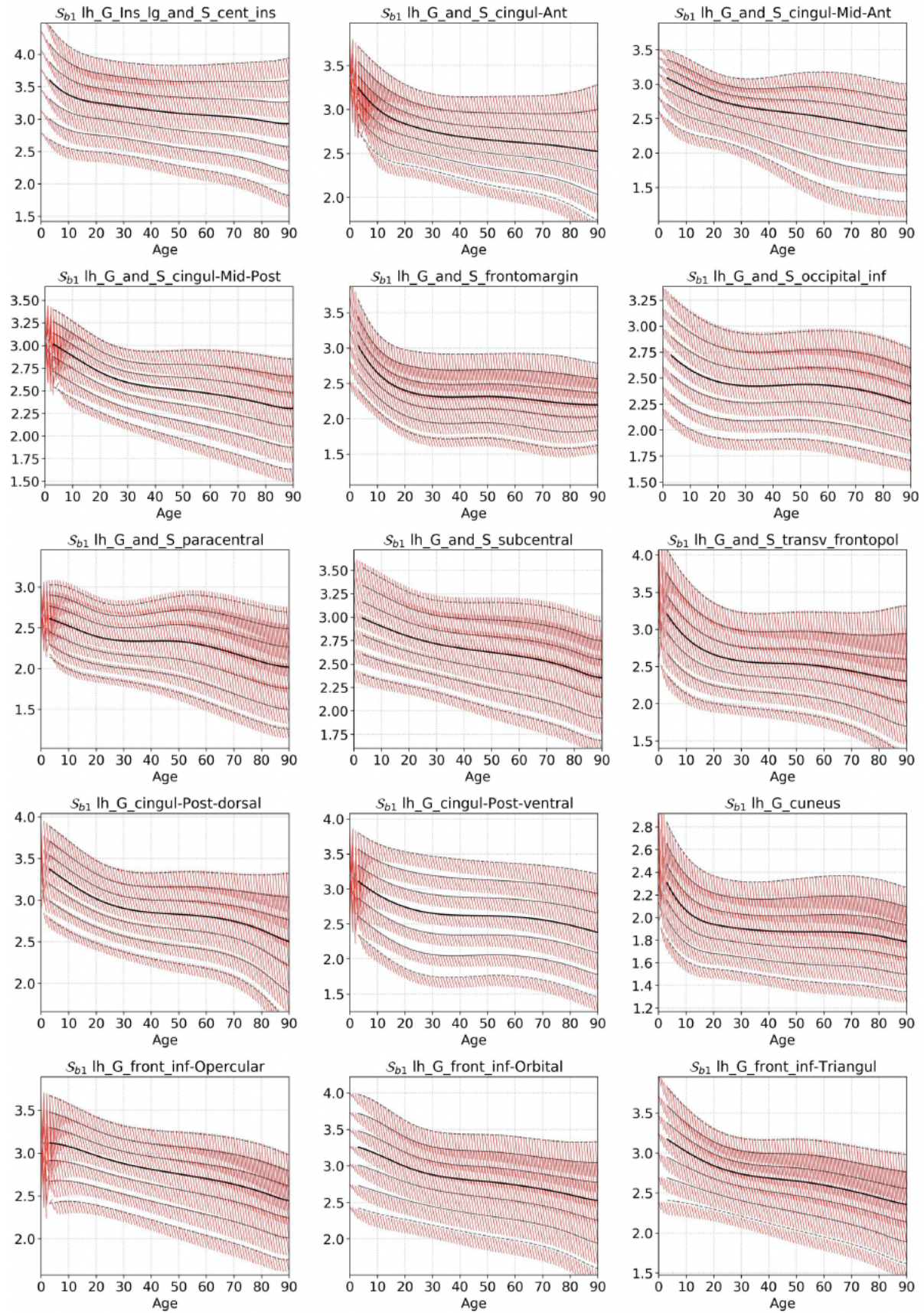

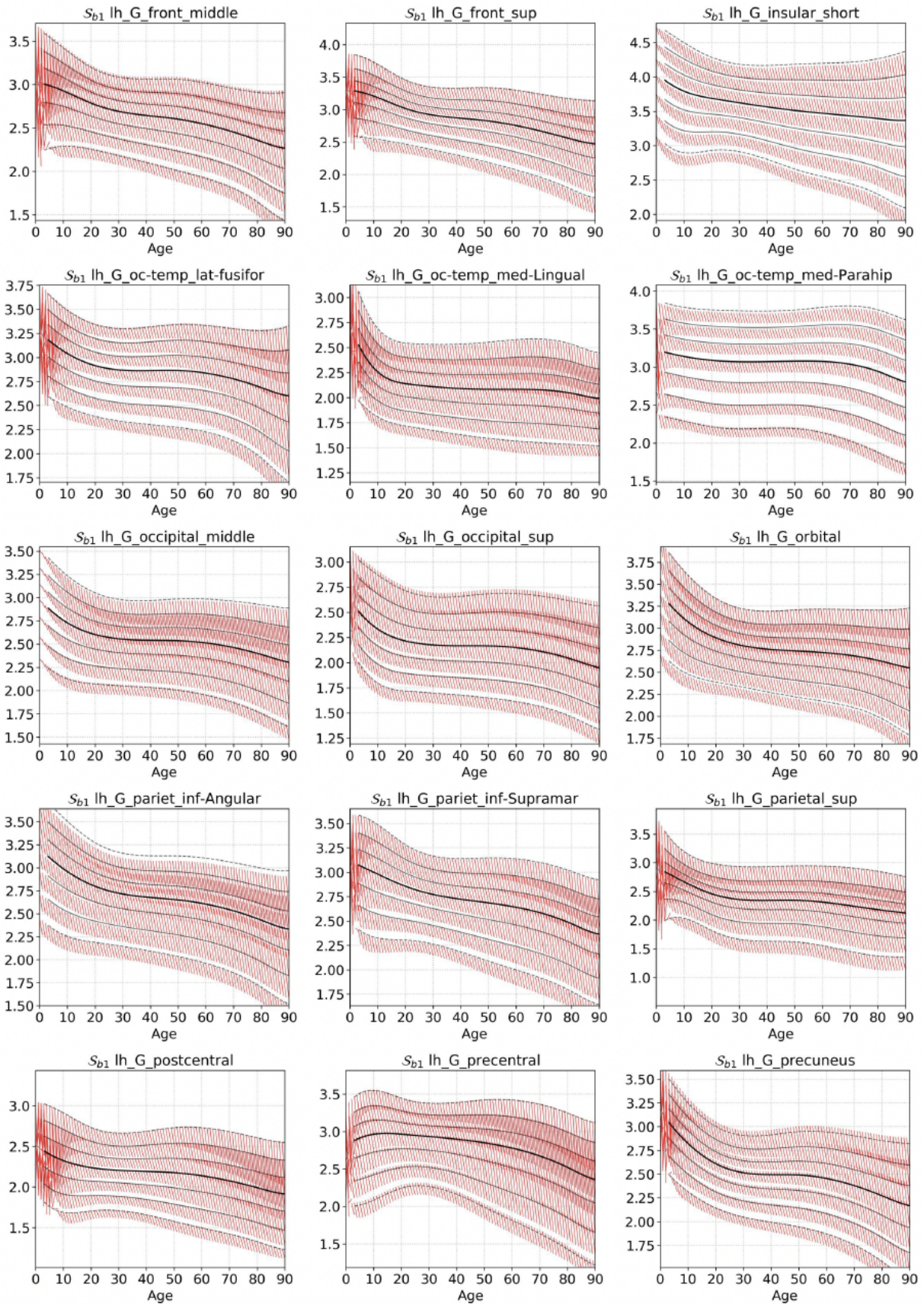

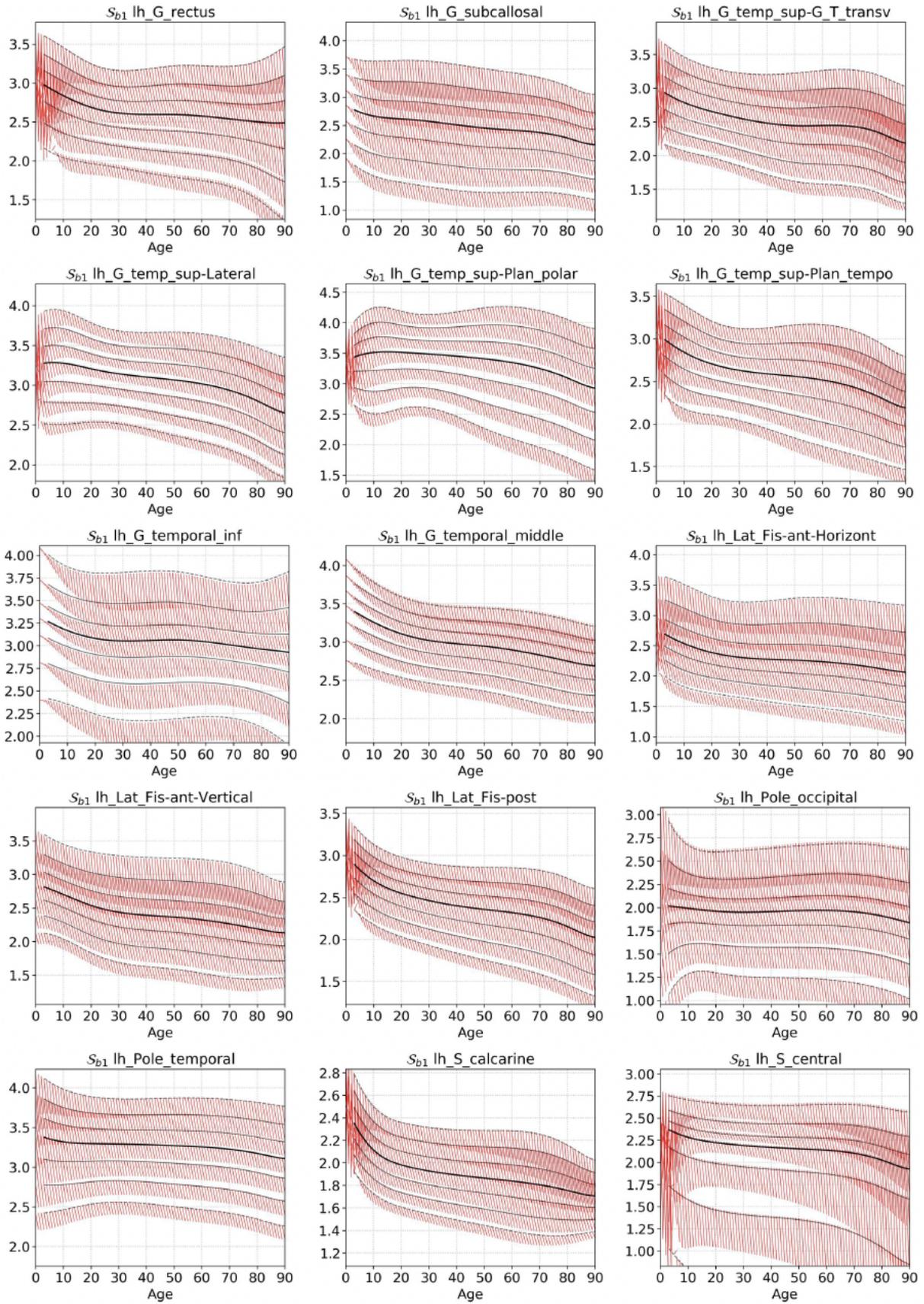

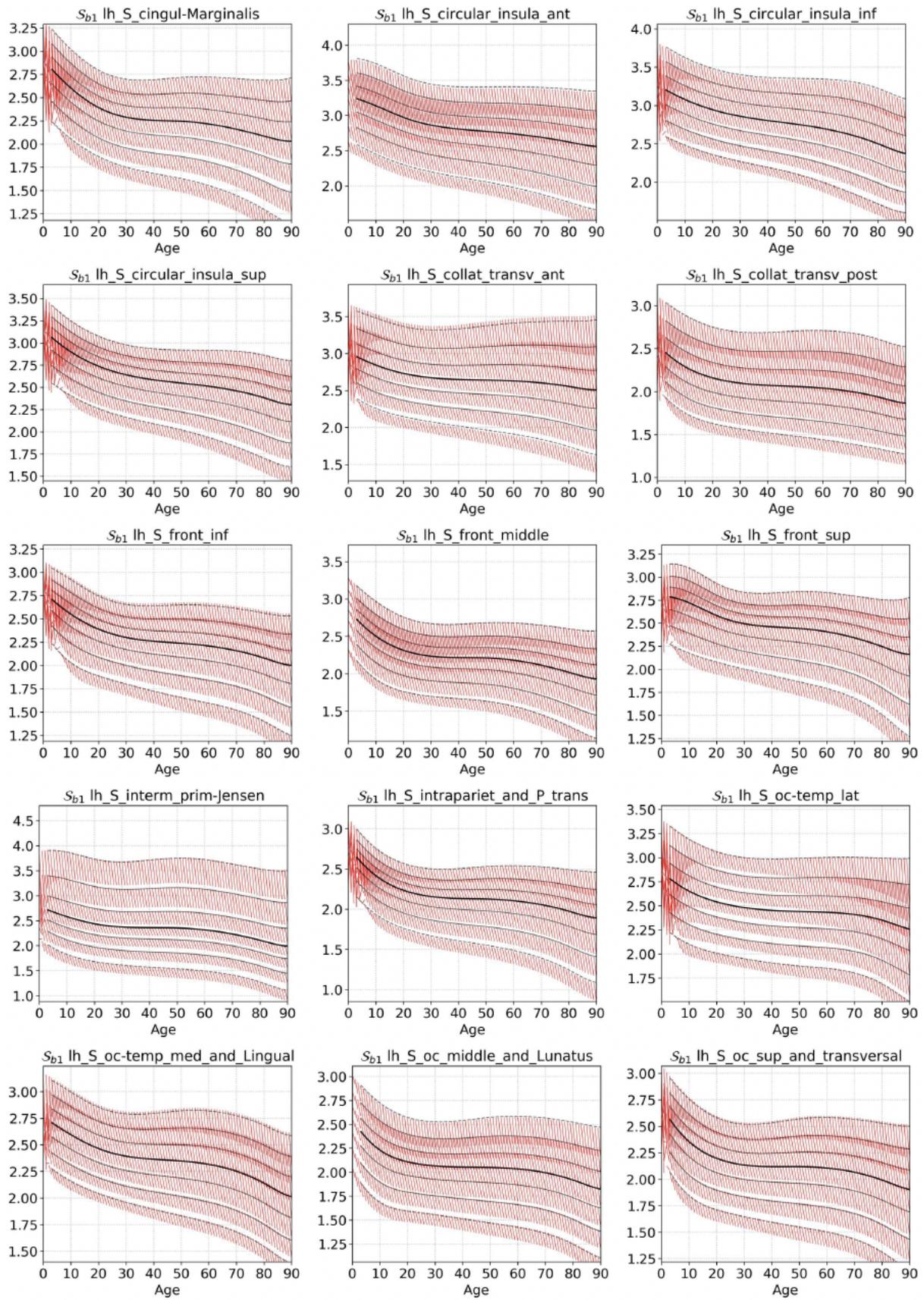

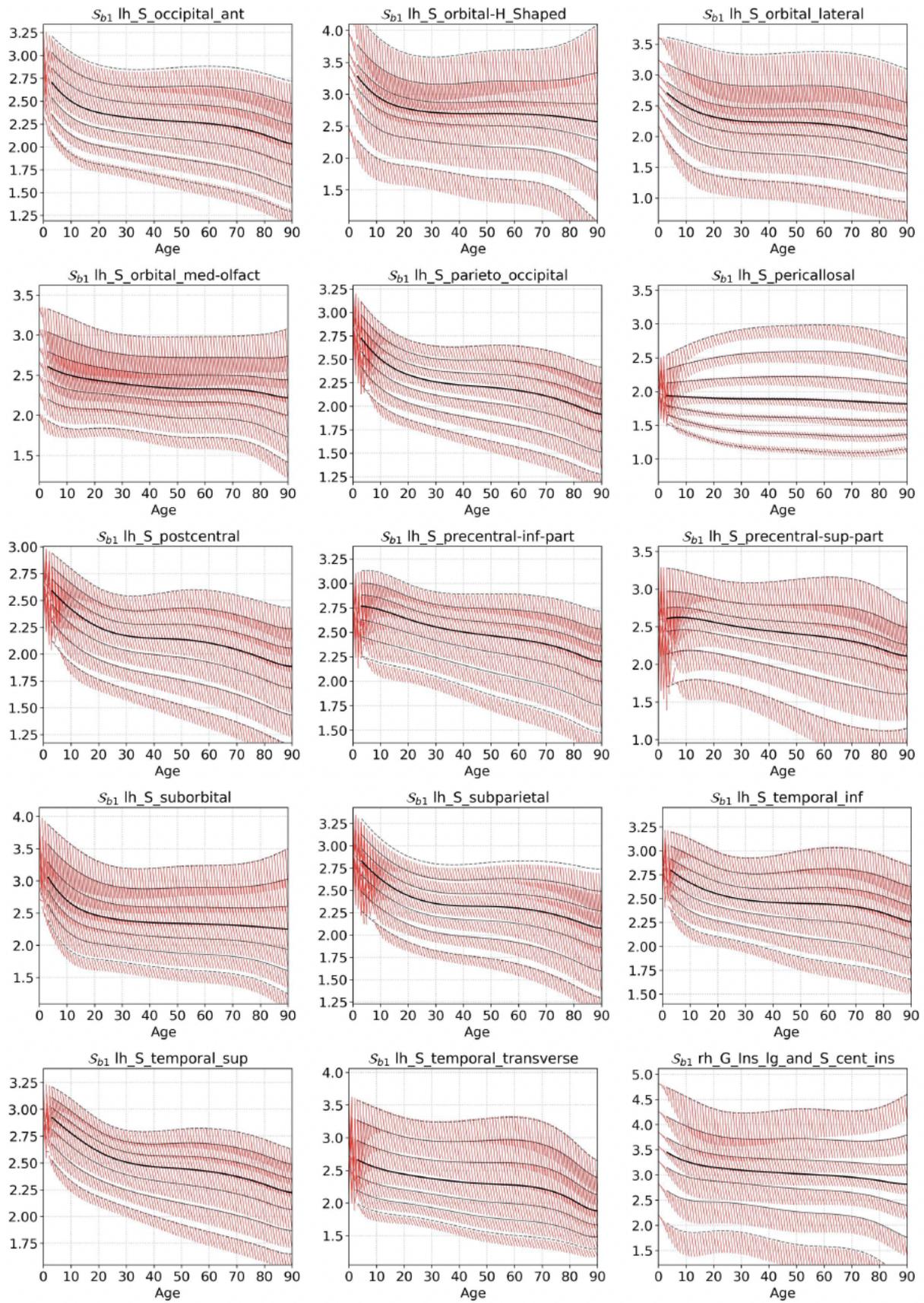

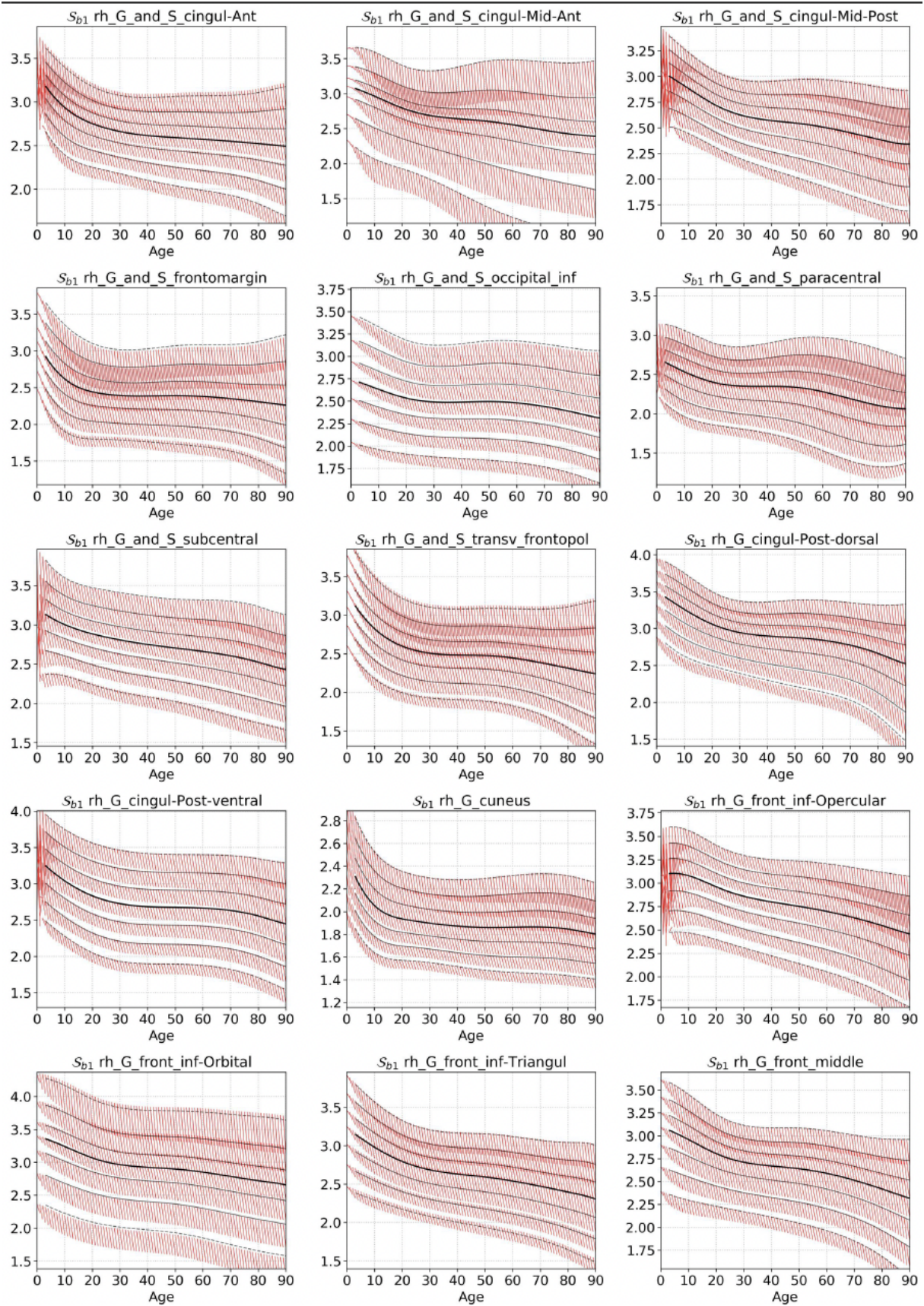

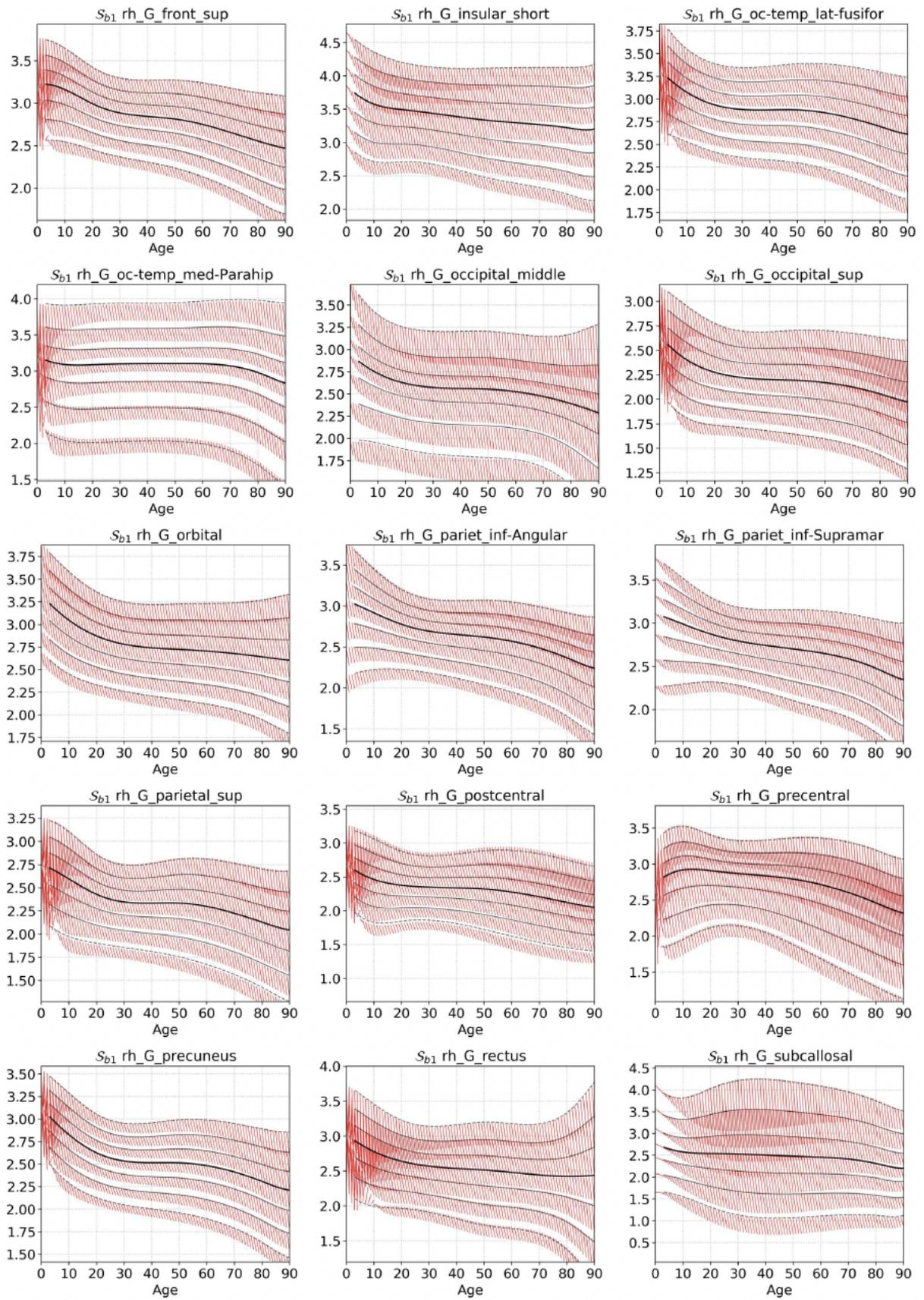

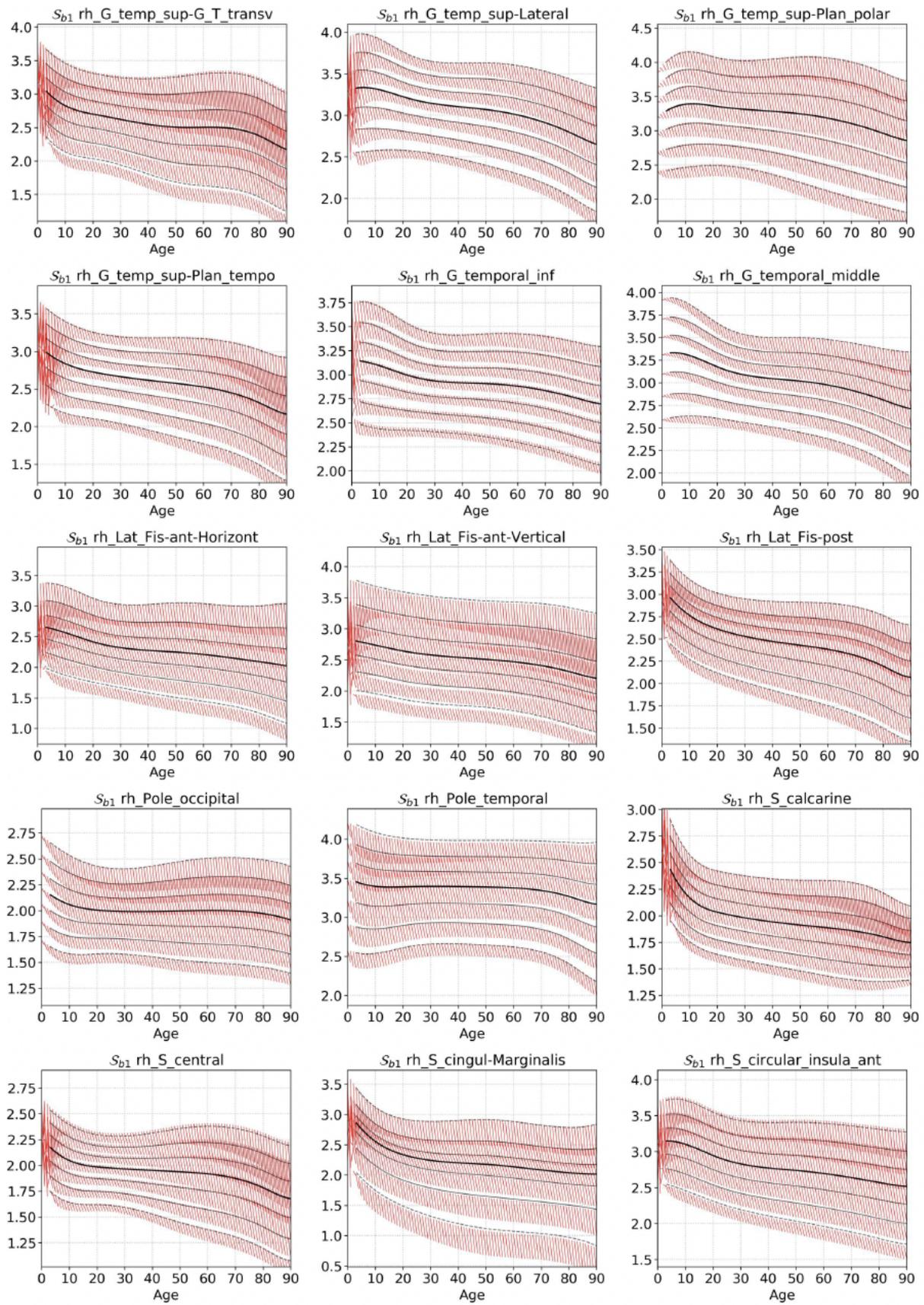

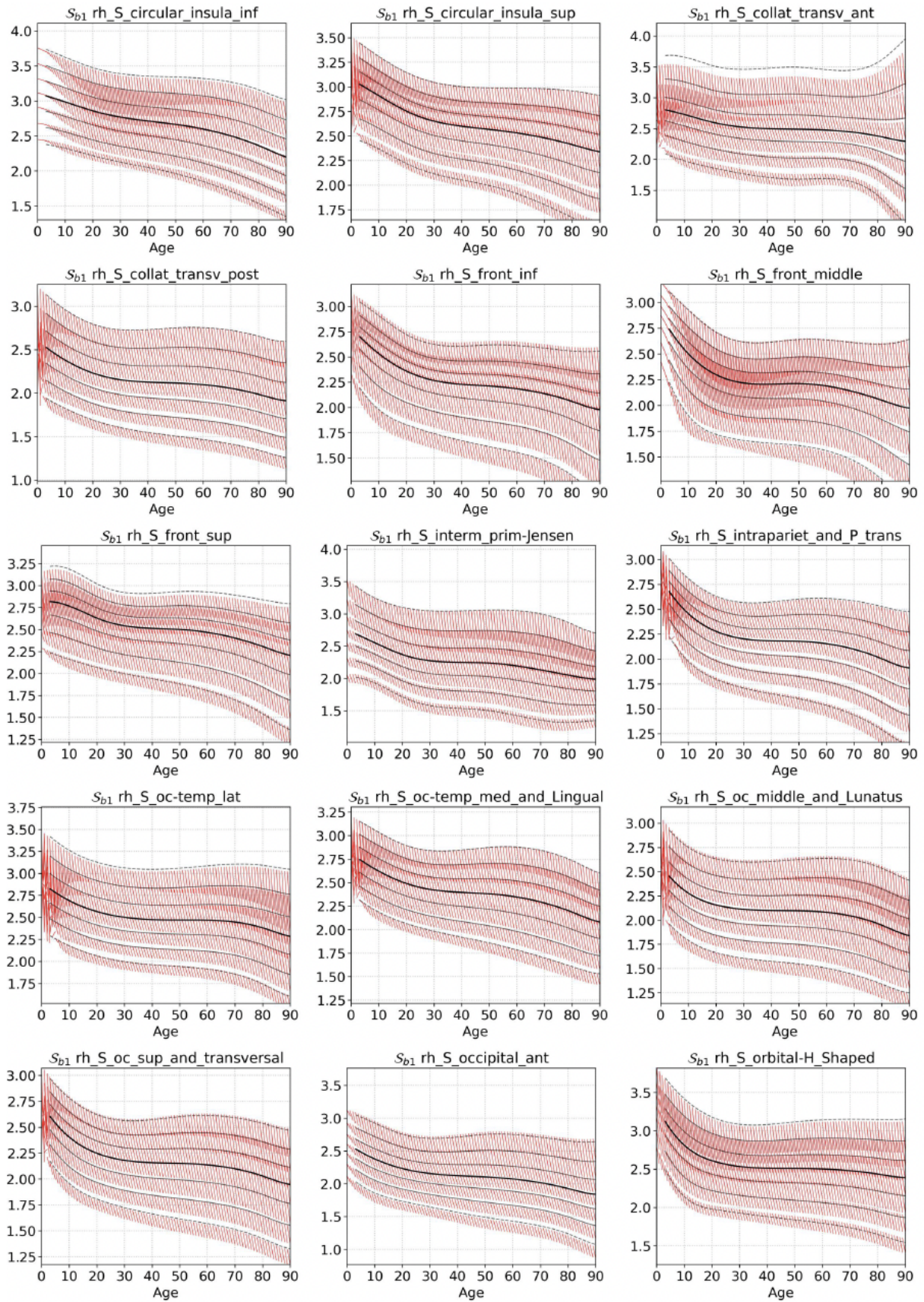

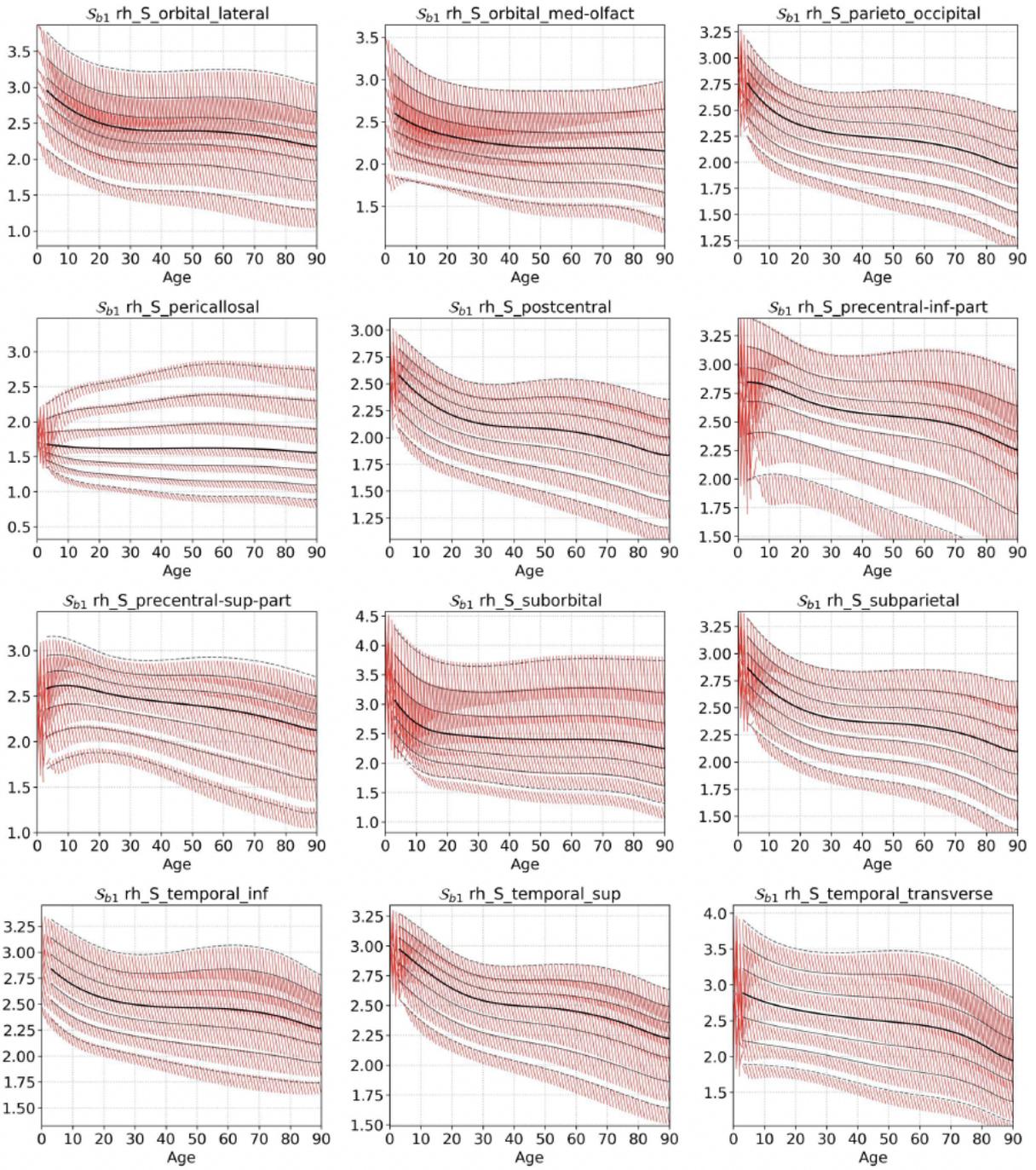

## S.2.2 Subcortical measures:

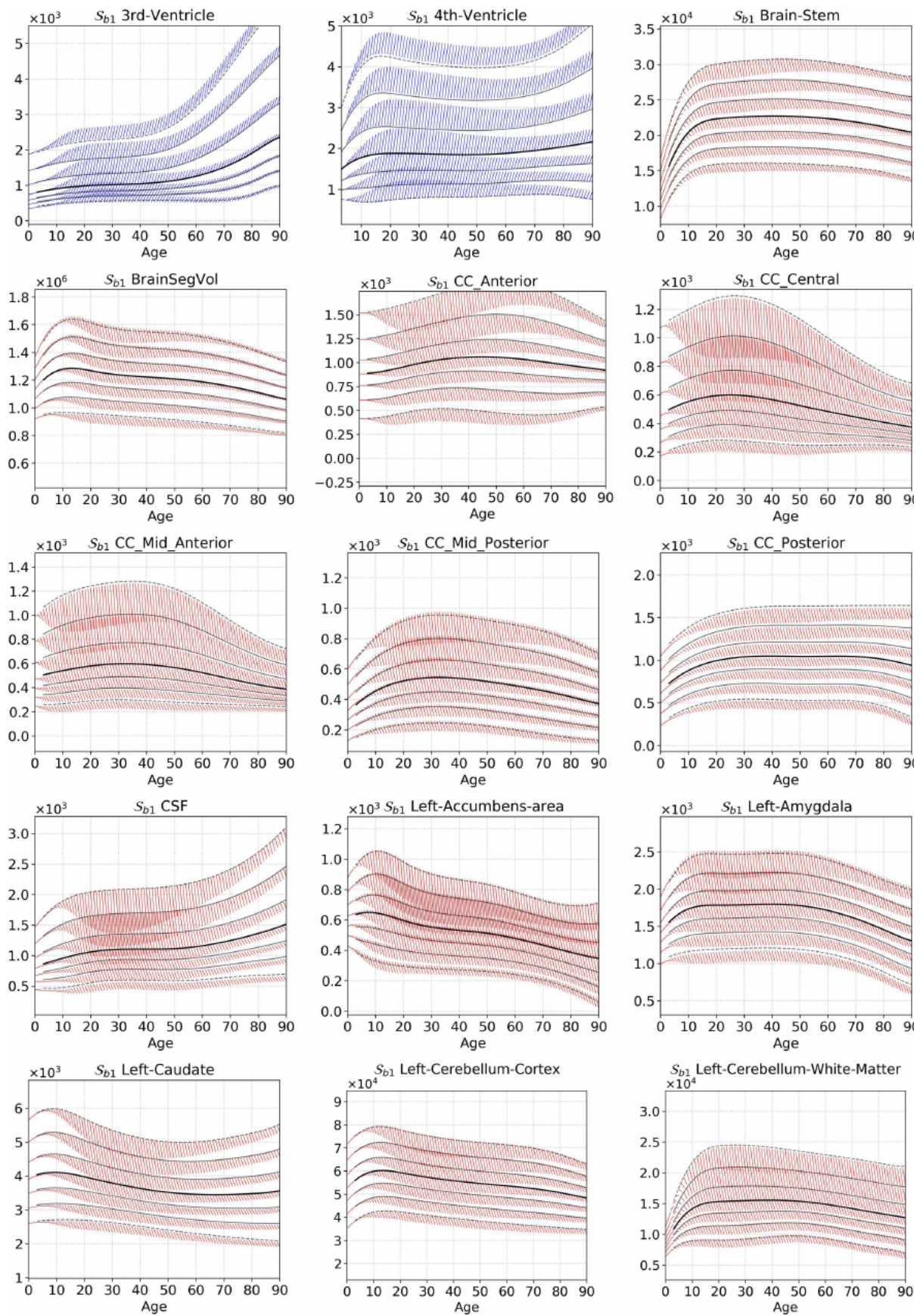

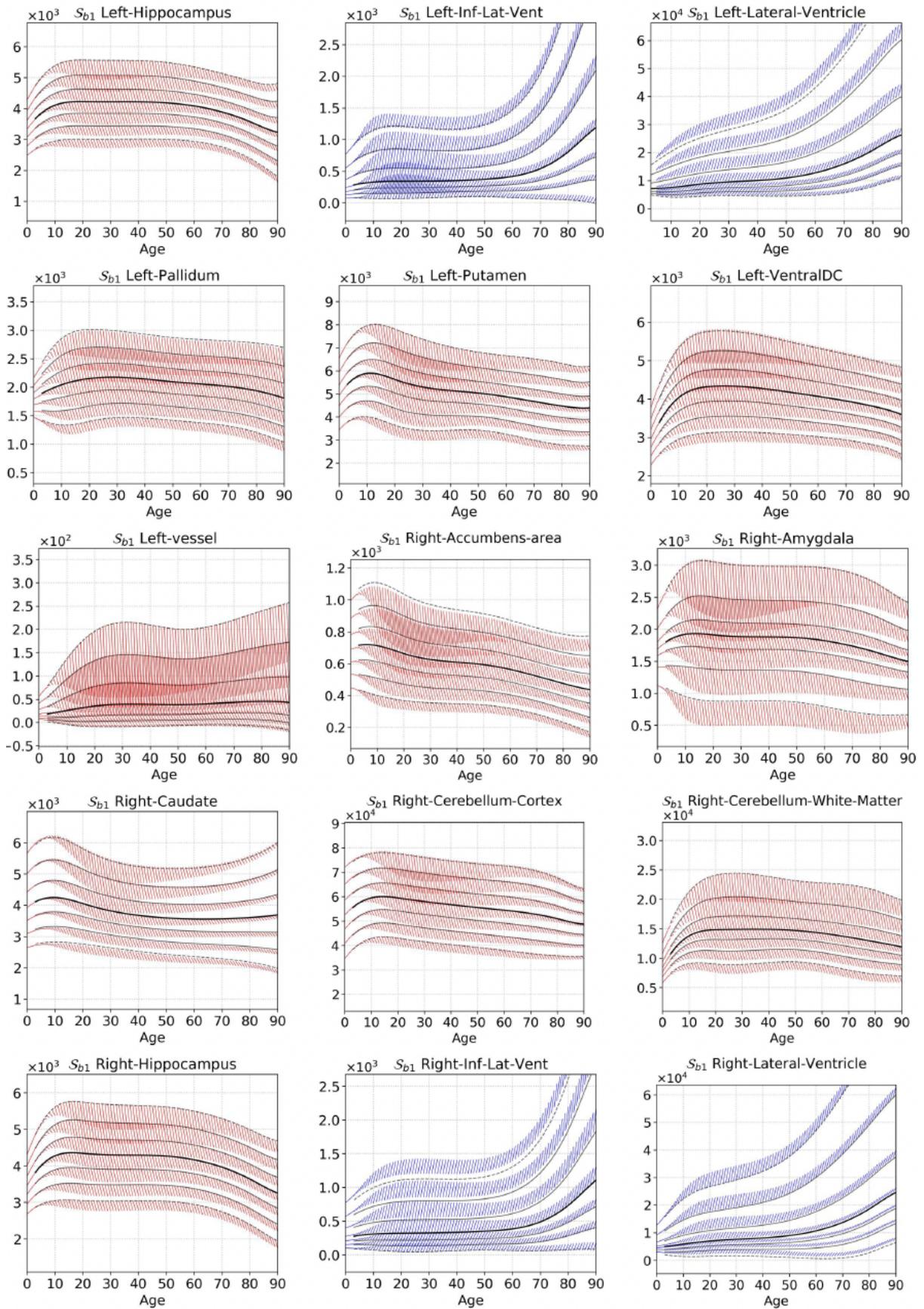

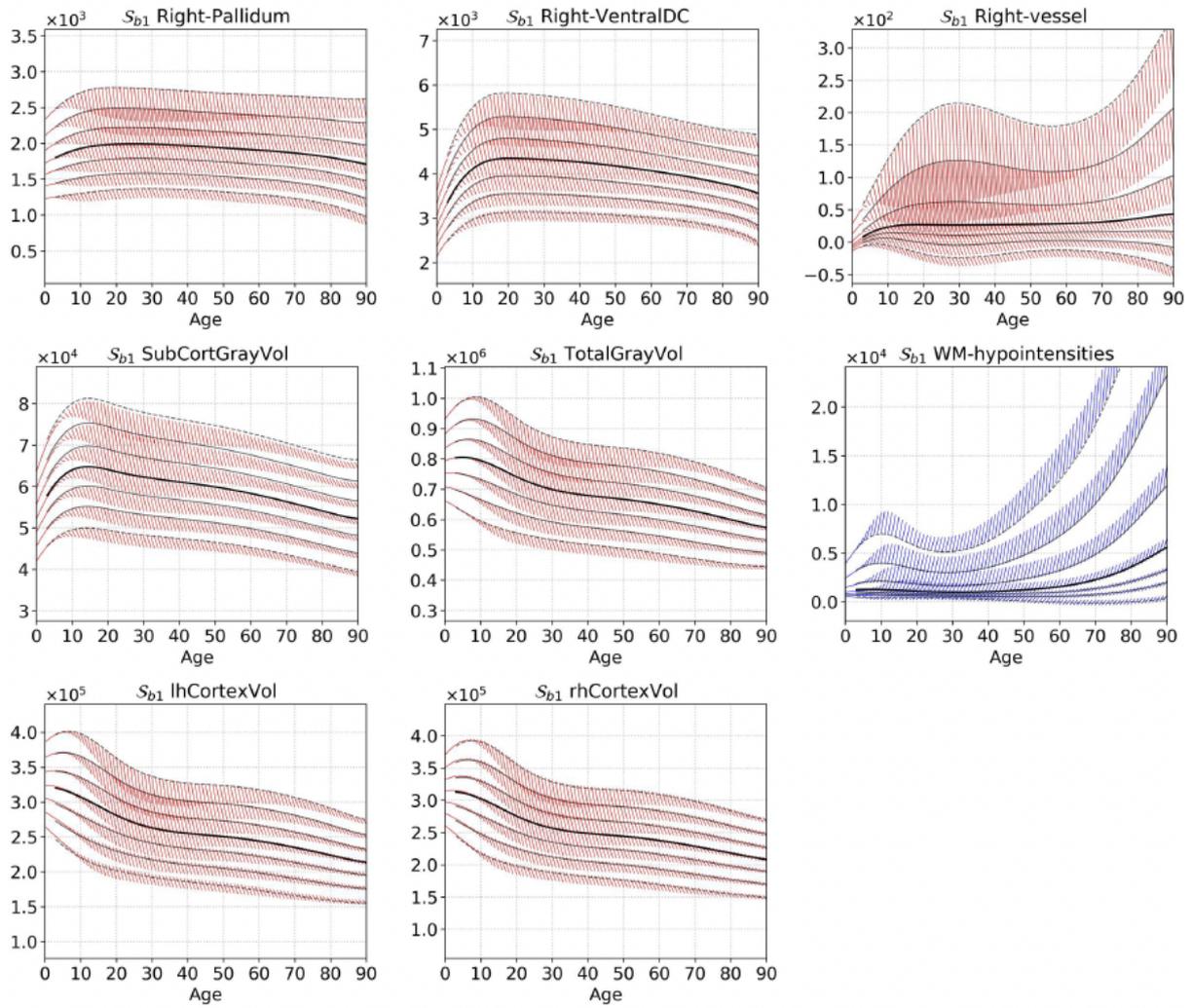

## S2.3. Desikan atlas:

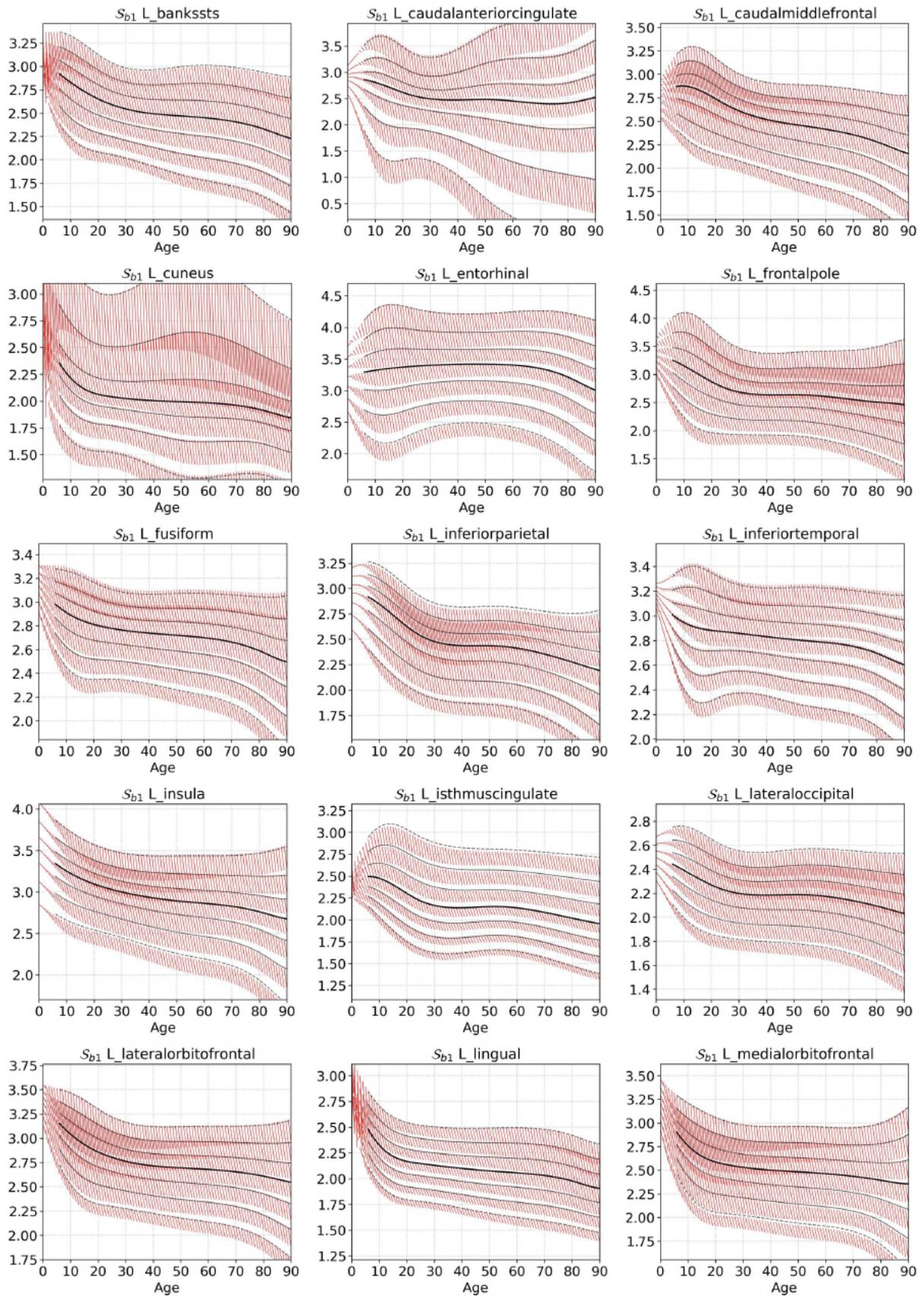

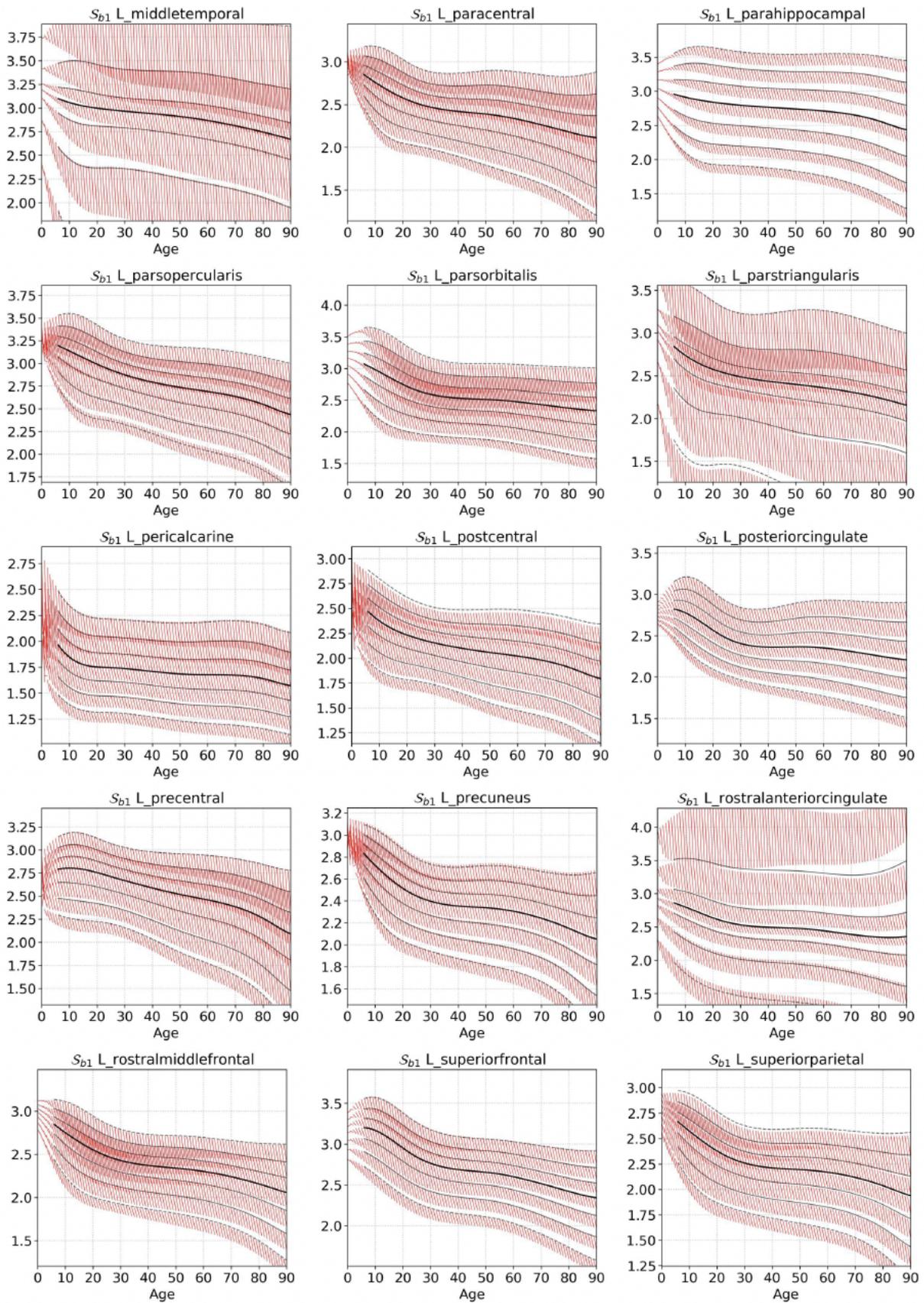

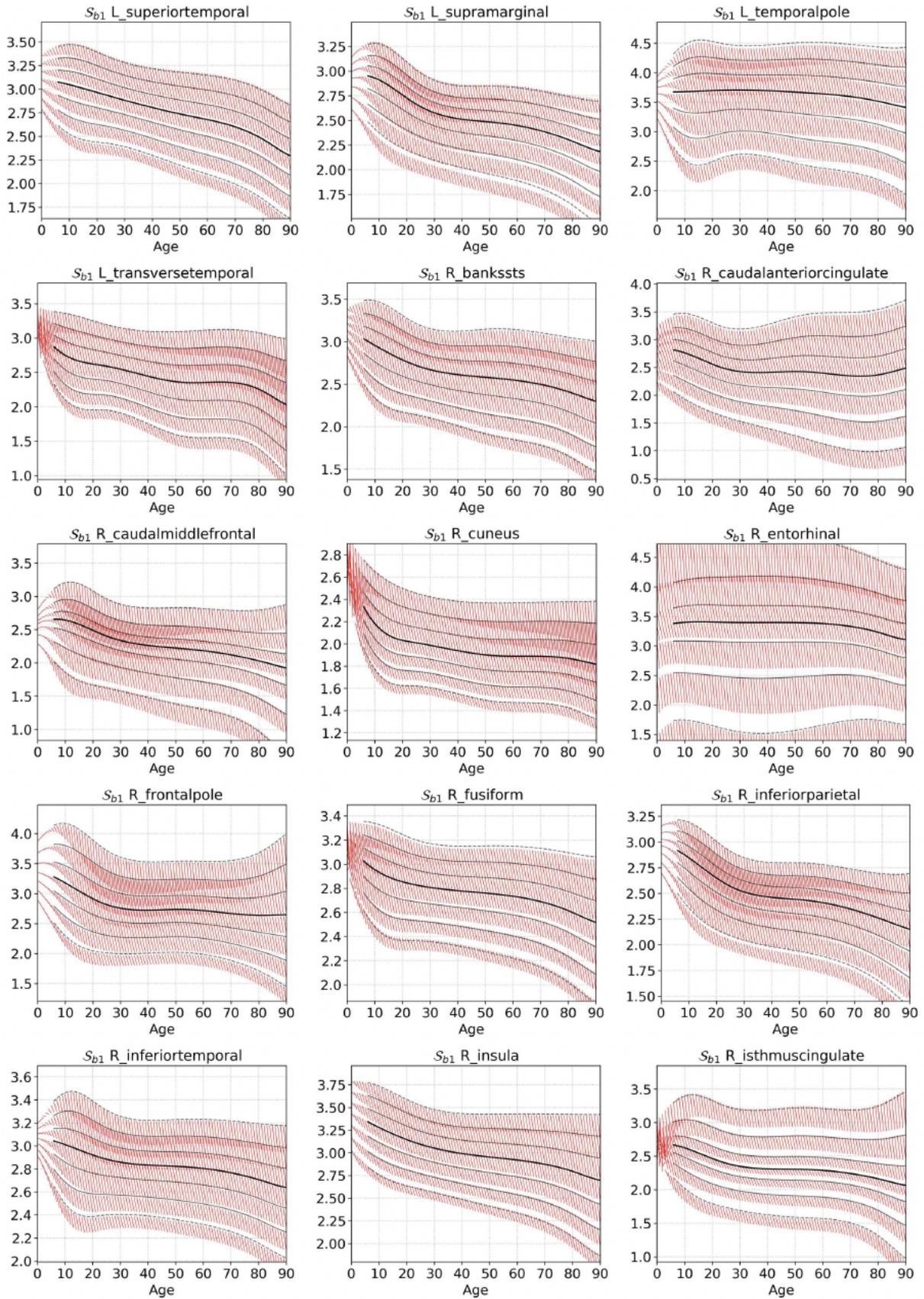

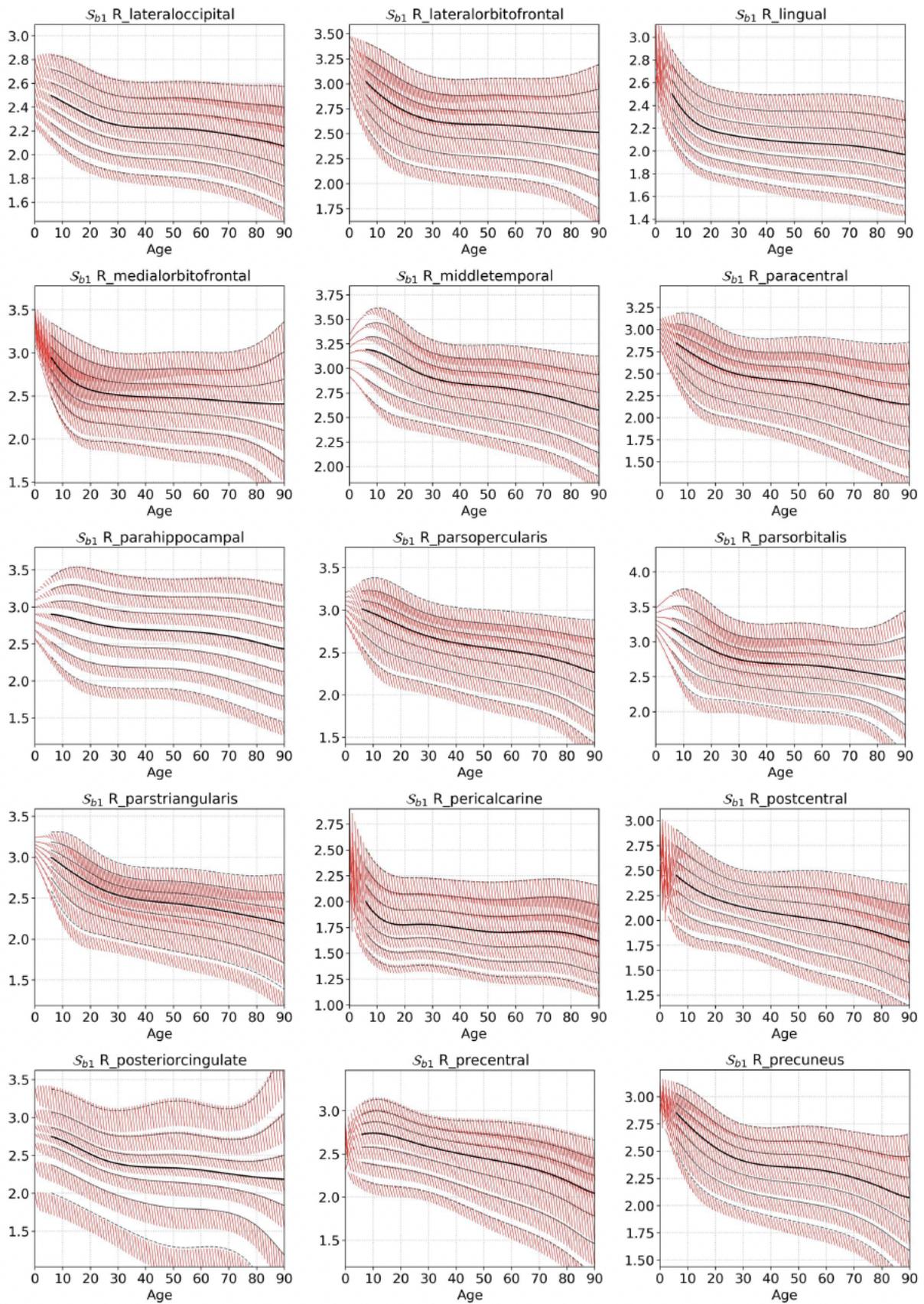
33

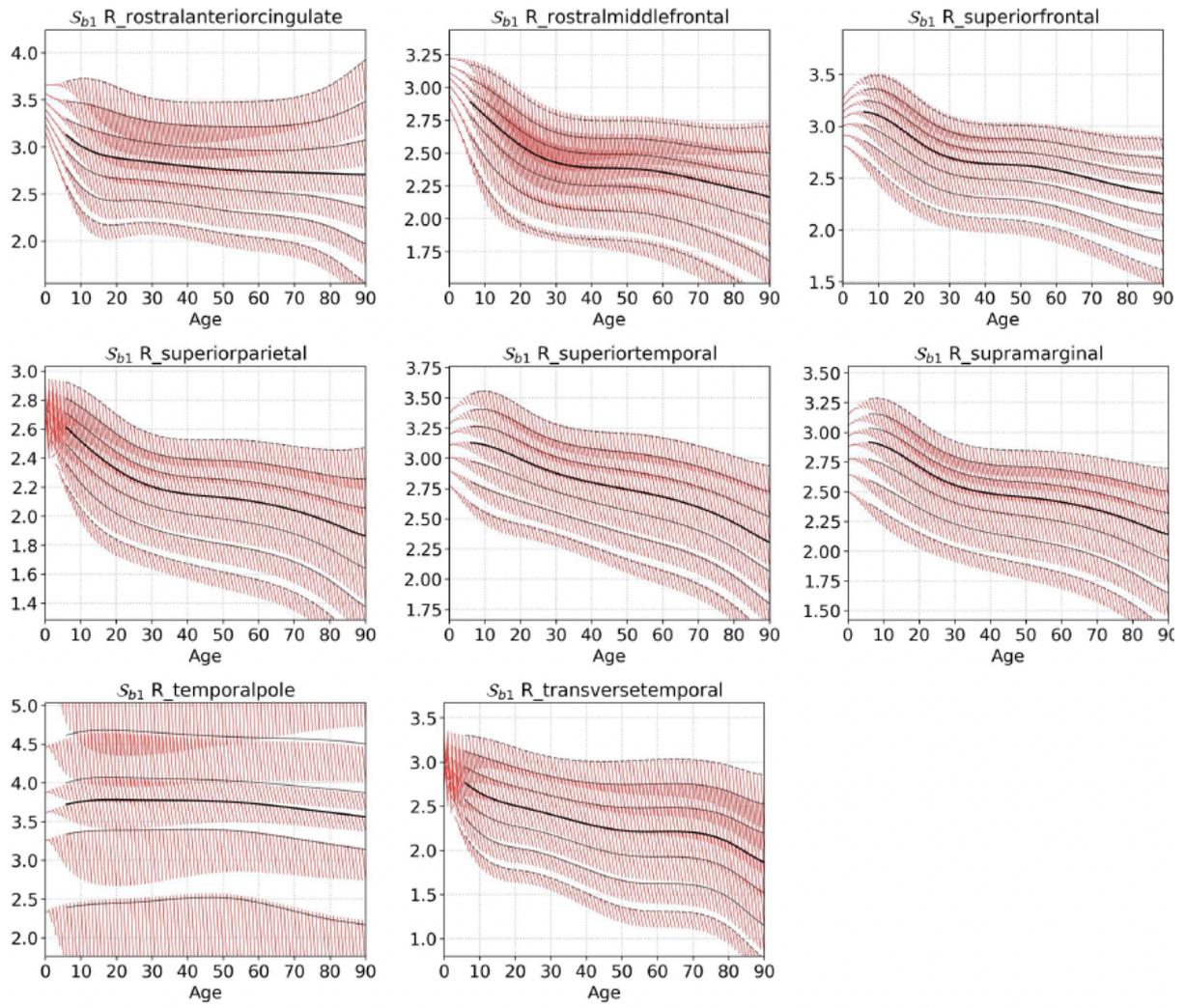

# S3. Supplementary Tables

## S3.1 Data sets and splits

| Set | Destrieux<br>n scans (n subjects) | Desikan<br>n scans (n subjects) | Subcortical<br>n scans (n subjects) |
|---|---|---|---|
| **Population reference models** (all healthy subjects) | 72758 | 70001 | 69679 |
| **Training set**<br>    cross-sectional | 28104 | 27932 | 29179 |
| **Test set** | 46382 | 41888 | 50468 |
|     cross-sectional | 18614 | 16299 | 25545 |
|     longitudinal | 24062 (10795) | 22574 (10200) | 22427 (10309) |
|     clinical (ADNI and OASIS) | 3706 | 3015 | 2506 |

## S3.2 Demographics per site*

| Dataset type | site name | n subjects | age min | age max | n site total | % female |
|---|---|---|---|---|---|---|
| cross-sectional | ABIDE_GU | 25 | 8,14 | 13,31 | 51 | 49,02 |
| cross-sectional | ABIDE_KKI | 64 | 8,07 | 12,36 | 185 | 34,59 |
| cross-sectional | ABIDE_NYU | 28 | 5,91 | 29,13 | 135 | 20,74 |
| cross-sectional | ABIDE_USM | 3 | 22,06 | 28,82 | 59 | 5,08 |
| cross-sectional | ADD200_KKI | 27 | 8,12 | 12,36 | 61 | 44,26 |
| cross-sectional | ADD200_NYU | 20 | 7,17 | 13,25 | 38 | 52,63 |
| cross-sectional | AOMIC_1000 | 483 | 20,00 | 26,00 | 925 | 52,22 |
| cross-sectional | AOMIC_PIPO2 | 120 | 18,25 | 26,25 | 209 | 57,42 |
| cross-sectional | ATV | 17 | 20,00 | 26,00 | 77 | 22,08 |
| cross-sectional | CIN | 22 | 20,00 | 69,00 | 66 | 33,33 |
| cross-sectional | CMI_SI | 38 | 5,93 | 21,72 | 106 | 35,85 |
| cross-sectional | CNP-35343.0 | 43 | 21,00 | 50,00 | 90 | 47,78 |
| cross-sectional | CNP-35426.0 | 11 | 21,00 | 45,00 | 20 | 55,00 |
| cross-sectional | COI | 78 | 20,00 | 79,00 | 124 | 62,90 |
| cross-sectional | HCP_A_MGH | 85 | 36,58 | 100,00 | 170 | 50,00 |
| cross-sectional | HCP_A_UCLA | 71 | 36,17 | 100,00 | 124 | 57,26 |
| cross-sectional | HCP_A_UM | 116 | 36,33 | 100,00 | 199 | 58,29 |
| cross-sectional | HCP_A_WU | 112 | 36,25 | 87,17 | 178 | 62,92 |
| cross-sectional | HCP_D_MGH | 109 | 8,08 | 21,67 | 216 | 50,46 |
| cross-sectional | HCP_D_UCLA | 62 | 8,00 | 21,83 | 126 | 49,21 |
| cross-sectional | HCP_D_UM | 85 | 8,50 | 21,83 | 155 | 54,84 |
| cross-sectional | HCP_D_WU | 75 | 8,17 | 21,92 | 154 | 48,70 |
| cross-sectional | HCP_EP_IU | 11 | 16,83 | 30,92 | 24 | 45,83 |
| cross-sectional | HCP_EP_MGH | 3 | 23,83 | 34,17 | 11 | 27,27 |
| cross-sectional | HCP_EP_McL | 4 | 25,33 | 26,50 | 13 | 30,77 |
| cross-sectional | HKH | 17 | 28,00 | 62,00 | 29 | 58,62 |
| cross-sectional | HRC | 36 | 26,00 | 58,00 | 49 | 73,47 |
| cross-sectional | HUH | 38 | 20,00 | 61,00 | 67 | 56,72 |
| cross-sectional | KCL | 25 | 20,00 | 64,00 | 41 | 60,98 |
| cross-sectional | KTT | 39 | 19,00 | 51,00 | 118 | 33,05 |
| cross-sectional | KUT | 65 | 18,00 | 68,00 | 158 | 41,14 |
| cross-sectional | SWA | 15 | 20,00 | 55,00 | 100 | 15,00 |
| cross-sectional | SWU_SLIM_ses1 | 317 | 17,00 | 27,00 | 547 | 57,95 |
| cross-sectional | UCDavis | 62 | 2,23 | 4,49 | 133 | 46,62 |
| cross-sectional | UMich_CWS | 17 | 3,33 | 7,47 | 29 | 58,62 |
| cross-sectional | UMich_IMPs | 114 | 6,00 | 18,00 | 205 | 55,61 |
| cross-sectional | UMich_MLS | 57 | 18,00 | 21,90 | 157 | 36,31 |
| cross-sectional | UMich_MTwins | 274 | 10,00 | 18,00 | 578 | 47,40 |
| cross-sectional | UMich_SAD | 43 | 18,00 | 49,00 | 114 | 37,72 |
| cross-sectional | UMich_SZG | 22 | 18,00 | 56,00 | 43 | 51,16 |
| cross-sectional | UTO | 106 | 19,00 | 79,00 | 202 | 52,48 |
| cross-sectional | cam | 329 | 18,00 | 88,00 | 641 | 51,33 |
| cross-sectional | delta | 17 | 45,00 | 63,00 | 48 | 35,42 |
| cross-sectional | ds001734 | 60 | 18,00 | 37,00 | 108 | 55,56 |
| cross-sectional | ds002236 | 38 | 8,67 | 15,50 | 84 | 45,24 |
| cross-sectional | ds002330 | 37 | 21,00 | 34,00 | 66 | 56,06 |
| cross-sectional | ds002345 | 131 | 18,00 | 38,00 | 207 | 63,29 |
| cross-sectional | ds002731 | 28 | 18,00 | 25,00 | 59 | 47,46 |
| cross-sectional | ds002837 | 41 | 18,00 | 50,00 | 83 | 49,40 |
| cross-sectional | hcp_ya | 589 | 22,00 | 36,00 | 1078 | 54,64 |
| cross-sectional | ixi | 312 | 19,98 | 86,32 | 556 | 56,12 |
| cross-sectional | nki | 308 | 6,00 | 85,00 | 481 | 64,03 |
| cross-sectional | pnc | 696 | 8,00 | 21,00 | 1357 | 51,29 |
| cross-sectional | top | 134 | 19,00 | 59,00 | 292 | 45,89 |
| longitudinal | ABCD | 6523 | 8,92 | 15,58 | 13726 | 47,52 |
| longitudinal | IMG_DRESDEN | 506 | 13,87 | 24,00 | 1034 | 48,94 |
| longitudinal | IMG_DUBLIN | 365 | 13,90 | 23,00 | 804 | 45,40 |
| longitudinal | IMG_HAMBURG | 535 | 13,36 | 24,00 | 1088 | 49,17 |
| longitudinal | IMG_LONDON | 587 | 13,62 | 24,00 | 1094 | 53,66 |
| longitudinal | IMG_MANNHEIM | 570 | 13,27 | 24,00 | 996 | 57,23 |
| longitudinal | IMG_NOTTINGHAM | 664 | 14,02 | 25,00 | 1263 | 52,57 |
| longitudinal | OASIS2 | 147 | 60,00 | 97,00 | 209 | 70,33 |
| longitudinal | OASIS3_401 | 321 | 46,00 | 96,00 | 520 | 61,73 |
| longitudinal | OASIS3_402 | 179 | 52,00 | 92,00 | 325 | 55,08 |
| longitudinal | OASIS3_403 | 237 | 50,00 | 88,00 | 388 | 61,08 |
| longitudinal | OASIS3_405 | 46 | 46,00 | 92,00 | 71 | 64,79 |
| longitudinal | IMG_PARIS | 1111 | 13,45 | 24,00 | 2078 | 53,46 |
| longitudinal | UKB_11025 | 1132 | 47,00 | 81,00 | 2271 | 49,85 |
| longitudinal | UKB_11026 | 51 | 49,00 | 73,00 | 82 | 62,20 |
| longitudinal | UKB_11027 | 649 | 48,00 | 80,00 | 1160 | 55,95 |
| longitudinal | ABCD | 2584 | 8,92 | 15,50 | 5279 | 48,95 |
| cross-sectional | OASIS2 | 1 | 75,00 | 75,00 | 1 | 100,00 |
| cross-sectional | OASIS3_401 | 71 | 46,00 | 95,00 | 123 | 57,72 |
| cross-sectional | OASIS3_402 | 55 | 46,00 | 84,00 | 95 | 57,89 |
| cross-sectional | OASIS3_403 | 36 | 43,00 | 78,00 | 52 | 69,23 |
| cross-sectional | OASIS3_405 | 9 | 63,00 | 97,00 | 12 | 75,00 |
| cross-sectional | UKB_11025 | 10226 | 45,00 | 82,00 | 19685 | 51,95 |
| cross-sectional | UKB_11026 | 2344 | 48,00 | 82,00 | 4321 | 54,25 |
| cross-sectional | UKB_11027 | 4242 | 48,00 | 81,00 | 7794 | 54,43 |
| clinical | OASIS2 | 84 | 65,00 | 98,00 | 181 | 46,41 |
| clinical | OASIS3_401 | 19 | 52,00 | 89,00 | 77 | 24,68 |
| clinical | OASIS3_402 | 11 | 65,00 | 81,00 | 26 | 42,31 |
| clinical | OASIS3_403 | 10 | 57,00 | 81,00 | 24 | 41,67 |
| clinical | OASIS3_405 | 6 | 50,00 | 83,00 | 19 | 31,58 |
| clinical | ADNI | 1655 | 55,00 | 96,00 | 3703 | 44,69 |

* numbers differ slightly between atlas types

## S3.3. Receiver Operating Characteristics
Area under the curve (early mild cognitive impairment vs. Alzheimer's disease in the ADNI data set)

## S3.3.1 Subcortical Regions

| Region | AUC_z_gain | AUC_Z1 | AUC_Z2 |
|---|---|---|---|
| Left-Lateral-Ventricle | 0,7400 | 0,5550 | 0,6204 |
| Left-Inf-Lat-Vent | 0,7648 | 0,5887 | 0,6845 |
| Left-Cerebellum-White-Matter | 0,4580 | 0,4465 | 0,4371 |
| Left-Cerebellum-Cortex | 0,6084 | 0,4731 | 0,5114 |
| Left-Caudate | 0,5455 | 0,5531 | 0,5781 |
| Left-Putamen | 0,6249 | 0,5950 | 0,6414 |
| Left-Pallidum | 0,4172 | 0,5212 | 0,4721 |
| 3rd-Ventricle | 0,7177 | 0,5420 | 0,5956 |
| 4th-Ventricle | 0,6324 | 0,5000 | 0,5365 |
| Brain-Stem | 0,5491 | 0,4637 | 0,4734 |

| Region | | | |
|---|---|---|---|
| Left-Hippocampus | 0,7466 | 0,7115 | 0,7600 |
| Left-Amygdala | 0,6369 | 0,7022 | 0,7219 |
| CSF | 0,2824 | 0,4218 | 0,3569 |
| Left-Accumbens-area | 0,5914 | 0,5674 | 0,6127 |
| Left-VentralDC | 0,5275 | 0,5272 | 0,5256 |
| Left-vessel | 0,4955 | 0,5163 | 0,5087 |
| Left-choroid-plexus | 0,4742 | 0,4323 | 0,4284 |
| Right-Lateral-Ventricle | 0,7411 | 0,5391 | 0,6059 |
| Right-Inf-Lat-Vent | 0,7502 | 0,6310 | 0,7008 |
| Right-Cerebellum-White-Matter | 0,4658 | 0,4477 | 0,4341 |
| Right-Cerebellum-Cortex | 0,6170 | 0,4903 | 0,5321 |
| Right-Caudate | 0,5932 | 0,5209 | 0,5602 |
| Right-Putamen | 0,6100 | 0,5741 | 0,6107 |
| Right-Pallidum | 0,4510 | 0,5169 | 0,4773 |
| Right-Hippocampus | 0,7582 | 0,7000 | 0,7484 |
| Right-Amygdala | 0,6893 | 0,6650 | 0,6991 |
| Right-Accumbens-area | 0,5963 | 0,5899 | 0,6190 |
| Right-VentralDC | 0,5666 | 0,5070 | 0,5198 |
| Right-vessel | 0,5240 | 0,4837 | 0,5022 |
| Right-choroid-plexus | 0,4719 | 0,4093 | 0,4248 |
| 5th-Ventricle | 0,5295 | 0,4149 | 0,4784 |
| WM-hypointensities | 0,2875 | 0,4305 | 0,3813 |
| non-WM-hypointensities | 0,5131 | 0,4501 | 0,4616 |
| Optic-Chiasm | 0,4327 | 0,3488 | 0,3385 |
| CC_Posterior | 0,6113 | 0,5007 | 0,5383 |
| CC_Mid_Posterior | 0,5547 | 0,5246 | 0,5397 |
| CC_Central | 0,6054 | 0,5159 | 0,5522 |
| CC_Mid_Anterior | 0,6324 | 0,5266 | 0,5629 |
| CC_Anterior | 0,6065 | 0,5338 | 0,5599 |
| BrainSegVol | 0,6442 | 0,5085 | 0,5474 |
| lhCortexVol | 0,6932 | 0,5864 | 0,6585 |
| rhCortexVol | 0,6946 | 0,5793 | 0,6507 |
| SubCortGrayVol | 0,6991 | 0,6329 | 0,6806 |
| TotalGrayVol | 0,7074 | 0,5773 | 0,6557 |
| SupraTentorialVol | 0,6409 | 0,5187 | 0,5554 |
| SupraTentorialVolNotVent | 0,6881 | 0,5273 | 0,5893 |
| BrainSegVol-to-eTIV | 0,6716 | 0,6104 | 0,6702 |
| MaskVol-to-eTIV | 0,3786 | 0,5824 | 0,5719 |
| lhSurfaceHoles | 0,5279 | 0,4715 | 0,4652 |
| rhSurfaceHoles | 0,5364 | 0,4726 | 0,4680 |
| EstimatedTotalIntraCranialVol | 0,4892 | 0,4624 | 0,4626 |

## S3.3.2 Desikan Kiliany Atlas

| Region | AUC_z_gain | AUC_Z1 | AUC_Z2 |
|---|---|---|---|
| L_bankssts | 0,813 | 0,375 | 0,674 |
| L_caudalanteriorcingulate | 0,603 | 0,744 | 0,739 |
| L_caudalmiddlefrontal | 0,491 | 0,580 | 0,493 |
| L_cuneus | 0,629 | 0,557 | 0,584 |
| L_entorhinal | 0,861 | 0,725 | 0,866 |
| L_fusiform | 0,806 | 0,640 | 0,833 |
| L_inferiorparietal | 0,636 | 0,606 | 0,620 |
| L_inferiortemporal | 0,782 | 0,686 | 0,753 |
| L_isthmuscingulate | 0,747 | 0,734 | 0,752 |
| L_lateraloccipital | 0,637 | 0,563 | 0,723 |
| L_lateralorbitofrontal | 0,621 | 0,655 | 0,711 |
| L_lingual | 0,721 | 0,496 | 0,656 |
| L_medialorbitofrontal | 0,686 | 0,622 | 0,715 |
| L_middletemporal | 0,879 | 0,587 | 0,752 |
| L_parahippocampal | 0,776 | 0,589 | 0,700 |
| L_paracentral | 0,475 | 0,752 | 0,717 |
| L_parsopercularis | 0,548 | 0,505 | 0,529 |
| L_parsorbitalis | 0,530 | 0,490 | 0,520 |
| L_parstriangularis | 0,546 | 0,509 | 0,477 |
| L_pericalcarine | 0,357 | 0,513 | 0,423 |
| L_postcentral | 0,317 | 0,459 | 0,305 |
| L_posteriorcingulate | 0,846 | 0,687 | 0,837 |
| L_precentral | 0,386 | 0,590 | 0,476 |

| Region | | | |
|---|---|---|---|
| L_precuneus | 0,700 | 0,505 | 0,609 |
| L_rostralanteriorcingulate | 0,524 | 0,574 | 0,600 |
| L_rostralmiddlefrontal | 0,474 | 0,616 | 0,567 |
| L_superiorfrontal | 0,656 | 0,641 | 0,715 |
| L_superiorparietal | 0,438 | 0,467 | 0,446 |
| L_superiortemporal | 0,714 | 0,381 | 0,492 |
| L_supramarginal | 0,622 | 0,587 | 0,624 |
| L_frontalpole | 0,280 | 0,550 | 0,378 |
| L_temporalpole | 0,820 | 0,439 | 0,640 |
| L_transversetemporal | 0,667 | 0,505 | 0,684 |
| L_insula | 0,421 | 0,379 | 0,353 |
| R_bankssts | 0,649 | 0,532 | 0,603 |
| R_caudalanteriorcingulate | 0,434 | 0,567 | 0,491 |
| R_caudalmiddlefrontal | 0,332 | 0,593 | 0,564 |
| R_cuneus | 0,520 | 0,539 | 0,482 |
| R_entorhinal | 0,775 | 0,595 | 0,675 |
| R_fusiform | 0,774 | 0,604 | 0,751 |
| R_inferiorparietal | 0,705 | 0,684 | 0,697 |
| R_inferiortemporal | 0,699 | 0,463 | 0,621 |
| R_isthmuscingulate | 0,763 | 0,565 | 0,657 |
| R_lateraloccipital | 0,691 | 0,656 | 0,761 |
| R_lateralorbitofrontal | 0,293 | 0,767 | 0,655 |
| R_lingual | 0,690 | 0,417 | 0,583 |
| R_medialorbitofrontal | 0,572 | 0,496 | 0,530 |
| R_middletemporal | 0,791 | 0,496 | 0,638 |
| R_parahippocampal | 0,710 | 0,511 | 0,659 |
| R_paracentral | 0,338 | 0,532 | 0,431 |
| R_parsopercularis | 0,564 | 0,453 | 0,466 |
| R_parsorbitalis | 0,459 | 0,503 | 0,470 |
| R_parstriangularis | 0,422 | 0,565 | 0,529 |
| R_pericalcarine | 0,463 | 0,637 | 0,580 |
| R_postcentral | 0,463 | 0,499 | 0,513 |
| R_posteriorcingulate | 0,751 | 0,767 | 0,831 |
| R_precentral | 0,216 | 0,516 | 0,376 |

| Region | | | |
|---|---|---|---|
| R_precuneus | 0,505 | 0,547 | 0,493 |
| R_rostralanteriorcingulate | 0,549 | 0,550 | 0,520 |
| R_rostralmiddlefrontal | 0,405 | 0,613 | 0,506 |
| R_superiorfrontal | 0,545 | 0,571 | 0,564 |
| R_superiorparietal | 0,407 | 0,595 | 0,490 |
| R_superiortemporal | 0,574 | 0,517 | 0,520 |
| R_supramarginal | 0,492 | 0,642 | 0,521 |
| R_frontalpole | 0,325 | 0,545 | 0,424 |
| R_temporalpole | 0,736 | 0,578 | 0,649 |
| R_transversetemporal | 0,586 | 0,537 | 0,660 |
| R_insula | 0,454 | 0,513 | 0,499 |

## S3.3.3 Destrieux Atlas

| Region | AUC_z_gain | AUC_Z1 | AUC_Z2 |
|---|---|---|---|
| lh_G_and_S_frontomargin | 0,603 | 0,518 | 0,578 |
| lh_G_and_S_occipital_inf | 0,643 | 0,568 | 0,637 |
| lh_G_and_S_paracentral | 0,569 | 0,589 | 0,628 |
| lh_G_and_S_subcentral | 0,613 | 0,561 | 0,610 |
| lh_G_and_S_transv_frontopol | 0,586 | 0,517 | 0,578 |
| lh_G_and_S_cingul-Ant | 0,648 | 0,558 | 0,612 |
| lh_G_and_S_cingul-Mid-Ant | 0,613 | 0,549 | 0,585 |
| lh_G_and_S_cingul-Mid-Post | 0,598 | 0,521 | 0,578 |
| lh_G_cingul-Post-dorsal | 0,624 | 0,564 | 0,624 |
| lh_G_cingul-Post-ventral | 0,594 | 0,478 | 0,534 |
| lh_G_cuneus | 0,578 | 0,580 | 0,604 |
| lh_G_front_inf-Opercular | 0,625 | 0,523 | 0,591 |
| lh_G_front_inf-Orbital | 0,575 | 0,425 | 0,484 |
| lh_G_front_inf-Triangul | 0,570 | 0,513 | 0,555 |
| lh_G_front_middle | 0,619 | 0,529 | 0,602 |
| lh_G_front_sup | 0,608 | 0,521 | 0,600 |
| lh_G_Ins_lg_and_S_cent_ins | 0,643 | 0,554 | 0,618 |
| lh_G_insular_short | 0,613 | 0,522 | 0,571 |
| lh_G_occipital_middle | 0,671 | 0,565 | 0,662 |
| lh_G_occipital_sup | 0,631 | 0,560 | 0,638 |
| lh_G_oc-temp_lat-fusifor | 0,683 | 0,629 | 0,698 |
| lh_G_oc-temp_med-Lingual | 0,600 | 0,608 | 0,645 |
| lh_G_oc-temp_med-Parahip | 0,746 | 0,563 | 0,649 |
| lh_G_orbital | 0,634 | 0,538 | 0,605 |
| lh_G_pariet_inf-Angular | 0,687 | 0,572 | 0,680 |
| lh_G_pariet_inf-Supramar | 0,648 | 0,600 | 0,677 |
| lh_G_parietal_sup | 0,574 | 0,543 | 0,592 |
| lh_G_postcentral | 0,617 | 0,539 | 0,616 |
| lh_G_precentral | 0,580 | 0,534 | 0,584 |
| lh_G_precuneus | 0,595 | 0,593 | 0,629 |
| lh_G_rectus | 0,630 | 0,502 | 0,576 |
| lh_G_subcallosal | 0,654 | 0,519 | 0,604 |
| lh_G_temp_sup-G_T_transv | 0,590 | 0,603 | 0,636 |
| lh_G_temp_sup-Lateral | 0,698 | 0,619 | 0,700 |
| lh_G_temp_sup-Plan_polar | 0,729 | 0,576 | 0,664 |
| lh_G_temp_sup-Plan_tempo | 0,641 | 0,536 | 0,631 |
| lh_G_temporal_inf | 0,696 | 0,569 | 0,659 |
| lh_G_temporal_middle | 0,702 | 0,575 | 0,691 |
| lh_Lat_Fis-ant-Horizont | 0,589 | 0,499 | 0,547 |
| lh_Lat_Fis-ant-Vertical | 0,567 | 0,593 | 0,607 |
| lh_Lat_Fis-post | 0,635 | 0,632 | 0,679 |
| lh_Pole_occipital | 0,583 | 0,588 | 0,625 |
| lh_Pole_temporal | 0,737 | 0,547 | 0,658 |
| lh_S_calcarine | 0,577 | 0,591 | 0,627 |
| lh_S_central | 0,572 | 0,589 | 0,619 |
| lh_S_cingul-Marginalis | 0,595 | 0,549 | 0,606 |
| lh_S_circular_insula_ant | 0,624 | 0,494 | 0,584 |
| lh_S_circular_insula_inf | 0,672 | 0,592 | 0,653 |
| lh_S_circular_insula_sup | 0,655 | 0,543 | 0,629 |
| lh_S_collat_transv_ant | 0,723 | 0,577 | 0,693 |
| lh_S_collat_transv_post | 0,655 | 0,552 | 0,638 |
| lh_S_front_inf | 0,647 | 0,563 | 0,638 |
| lh_S_front_middle | 0,661 | 0,519 | 0,627 |
| lh_S_front_sup | 0,650 | 0,487 | 0,606 |
| lh_S_interm_prim-Jensen | 0,559 | 0,495 | 0,528 |
| lh_S_intrapariet_and_P_trans | 0,632 | 0,590 | 0,642 |
| lh_S_oc_middle_and_Lunatus | 0,655 | 0,574 | 0,641 |
| lh_S_oc_sup_and_transversal | 0,648 | 0,591 | 0,650 |
| lh_S_occipital_ant | 0,619 | 0,545 | 0,612 |
| lh_S_oc-temp_lat | 0,688 | 0,533 | 0,653 |
| lh_S_oc-temp_med_and_Lingual | 0,672 | 0,611 | 0,679 |
| lh_S_orbital_lateral | 0,586 | 0,605 | 0,634 |
| lh_S_orbital_med-olfact | 0,653 | 0,556 | 0,638 |
| lh_S_orbital-H_Shaped | 0,674 | 0,547 | 0,641 |
| lh_S_parieto_occipital | 0,628 | 0,554 | 0,628 |
| lh_S_pericallosal | 0,619 | 0,513 | 0,573 |
| lh_S_postcentral | 0,613 | 0,577 | 0,638 |
| lh_S_precentral-inf-part | 0,638 | 0,606 | 0,659 |
| lh_S_precentral-sup-part | 0,592 | 0,480 | 0,548 |
| lh_S_suborbital | 0,655 | 0,489 | 0,571 |
| lh_S_subparietal | 0,644 | 0,518 | 0,608 |
| lh_S_temporal_inf | 0,710 | 0,496 | 0,633 |
| lh_S_temporal_sup | 0,691 | 0,591 | 0,676 |
| lh_S_temporal_transverse | 0,602 | 0,510 | 0,578 |
| rh_G_and_S_frontomargin | 0,605 | 0,486 | 0,550 |
| rh_G_and_S_occipital_inf | 0,632 | 0,595 | 0,641 |
| rh_G_and_S_paracentral | 0,567 | 0,597 | 0,613 |
| rh_G_and_S_subcentral | 0,568 | 0,581 | 0,611 |
| rh_G_and_S_transv_frontopol | 0,627 | 0,554 | 0,620 |
| rh_G_and_S_cingul-Ant | 0,633 | 0,476 | 0,547 |
| rh_G_and_S_cingul-Mid-Ant | 0,543 | 0,448 | 0,472 |
| rh_G_and_S_cingul-Mid-Post | 0,587 | 0,524 | 0,572 |
| rh_G_cingul-Post-dorsal | 0,552 | 0,500 | 0,552 |
| rh_G_cingul-Post-ventral | 0,657 | 0,516 | 0,586 |
| rh_G_cuneus | 0,564 | 0,628 | 0,650 |
| rh_G_front_inf-Opercular | 0,601 | 0,585 | 0,626 |
| rh_G_front_inf-Orbital | 0,543 | 0,485 | 0,522 |
| rh_G_front_inf-Triangul | 0,591 | 0,508 | 0,580 |
| rh_G_front_middle | 0,646 | 0,523 | 0,620 |
| rh_G_front_sup | 0,591 | 0,451 | 0,530 |
| rh_G_Ins_lg_and_S_cent_ins | 0,689 | 0,455 | 0,550 |
| rh_G_insular_short | 0,631 | 0,553 | 0,609 |
| rh_G_occipital_middle | 0,668 | 0,603 | 0,685 |
| rh_G_occipital_sup | 0,630 | 0,552 | 0,612 |
| rh_G_oc-temp_lat-fusifor | 0,674 | 0,589 | 0,673 |
| rh_G_oc-temp_med-Lingual | 0,633 | 0,631 | 0,674 |
| rh_G_oc-temp_med-Parahip | 0,738 | 0,587 | 0,656 |
| rh_G_orbital | 0,632 | 0,437 | 0,545 |
| rh_G_pariet_inf-Angular | 0,675 | 0,521 | 0,642 |
| rh_G_pariet_inf-Supramar | 0,612 | 0,561 | 0,636 |
| rh_G_parietal_sup | 0,594 | 0,590 | 0,634 |
| rh_G_postcentral | 0,615 | 0,569 | 0,626 |
| rh_G_precentral | 0,592 | 0,568 | 0,615 |
| rh_G_precuneus | 0,607 | 0,603 | 0,642 |
| rh_G_rectus | 0,625 | 0,547 | 0,607 |
| rh_G_subcallosal | 0,607 | 0,496 | 0,572 |
| rh_G_temp_sup-G_T_transv | 0,604 | 0,636 | 0,654 |
| rh_G_temp_sup-Lateral | 0,656 | 0,534 | 0,616 |
| rh_G_temp_sup-Plan_polar | 0,713 | 0,549 | 0,630 |
| rh_G_temp_sup-Plan_tempo | 0,621 | 0,602 | 0,640 |
| rh_G_temporal_inf | 0,686 | 0,524 | 0,613 |
| rh_G_temporal_middle | 0,709 | 0,578 | 0,669 |
| rh_Lat_Fis-ant-Horizont | 0,573 | 0,509 | 0,536 |
| rh_Lat_Fis-ant-Vertical | 0,602 | 0,512 | 0,571 |
| rh_Lat_Fis-post | 0,640 | 0,636 | 0,683 |
| rh_Pole_occipital | 0,638 | 0,625 | 0,685 |
| rh_Pole_temporal | 0,693 | 0,561 | 0,635 |
| rh_S_calcarine | 0,567 | 0,585 | 0,618 |
| rh_S_central | 0,582 | 0,597 | 0,635 |
| rh_S_cingul-Marginalis | 0,575 | 0,554 | 0,590 |
| rh_S_circular_insula_ant | 0,608 | 0,473 | 0,549 |
| rh_S_circular_insula_inf | 0,669 | 0,588 | 0,650 |
| rh_S_circular_insula_sup | 0,616 | 0,523 | 0,595 |
| rh_S_collat_transv_ant | 0,729 | 0,557 | 0,656 |
| rh_S_collat_transv_post | 0,655 | 0,611 | 0,671 |
| rh_S_front_inf | 0,613 | 0,545 | 0,611 |
| rh_S_front_middle | 0,628 | 0,483 | 0,564 |
| rh_S_front_sup | 0,634 | 0,473 | 0,587 |
| rh_S_interm_prim-Jensen | 0,632 | 0,552 | 0,608 |
| rh_S_intrapariet_and_P_trans | 0,620 | 0,578 | 0,619 |
| rh_S_oc_middle_and_Lunatus | 0,658 | 0,509 | 0,609 |
| rh_S_oc_sup_and_transversal | 0,669 | 0,590 | 0,652 |
| rh_S_occipital_ant | 0,643 | 0,633 | 0,683 |
| rh_S_oc-temp_lat | 0,681 | 0,569 | 0,656 |
| rh_S_oc-temp_med_and_Lingual | 0,739 | 0,620 | 0,710 |
| rh_S_orbital_lateral | 0,568 | 0,495 | 0,542 |
| rh_S_orbital_med-olfact | 0,632 | 0,528 | 0,595 |
| rh_S_orbital-H_Shaped | 0,633 | 0,480 | 0,577 |
| rh_S_parieto_occipital | 0,656 | 0,526 | 0,600 |
| rh_S_pericallosal | 0,592 | 0,402 | 0,453 |
| rh_S_postcentral | 0,604 | 0,574 | 0,632 |
| rh_S_precentral-inf-part | 0,655 | 0,566 | 0,649 |
| rh_S_precentral-sup-part | 0,641 | 0,520 | 0,615 |
| rh_S_suborbital | 0,588 | 0,458 | 0,506 |
| rh_S_subparietal | 0,613 | 0,554 | 0,605 |
| rh_S_temporal_inf | 0,690 | 0,500 | 0,629 |
| rh_S_temporal_sup | 0,701 | 0,603 | 0,683 |
| rh_S_temporal_transverse | 0,597 | 0,561 | 0,617 |

Right Hippocampus

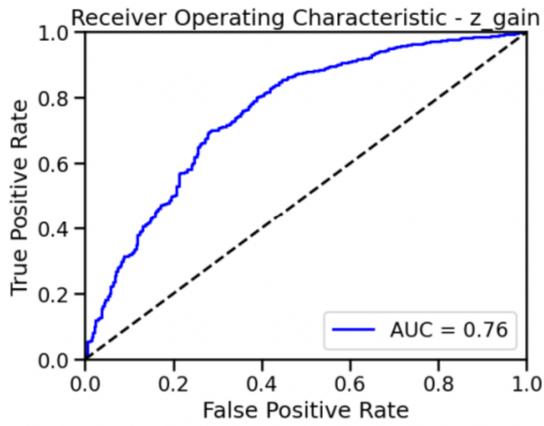
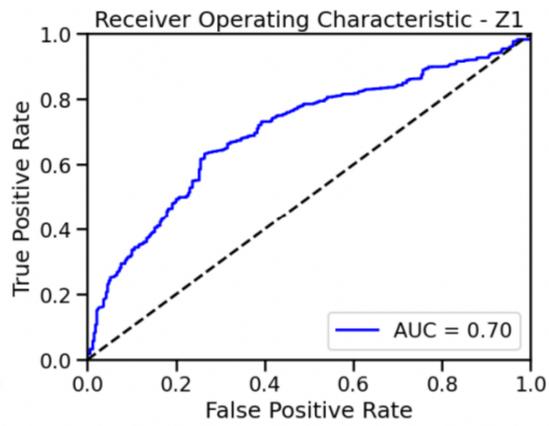
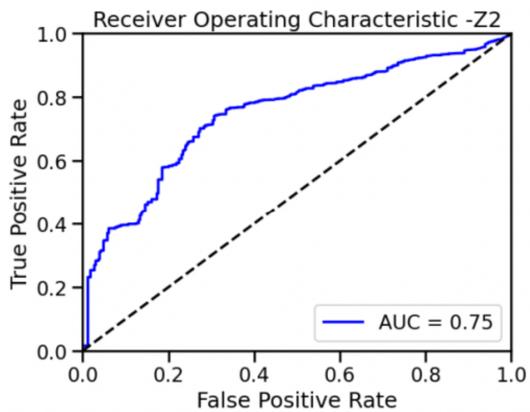

Left Lateral Ventricle

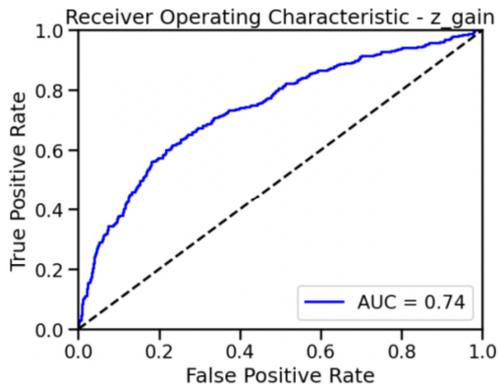
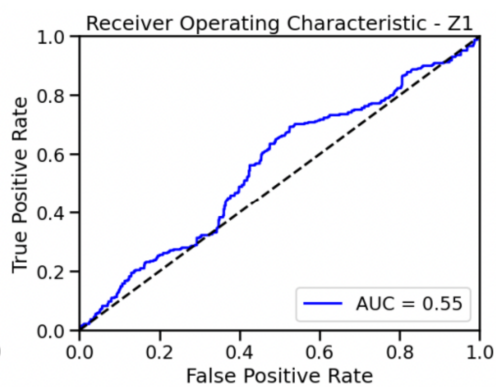
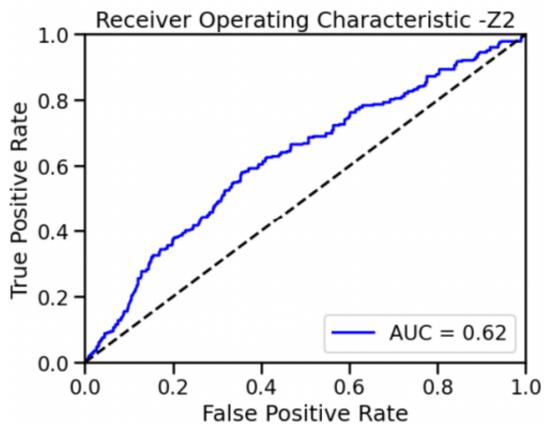

S4. Quality metrics: Full data set (reference models)

## Supplemental Table S4.1. Performance Metrics Desikan Kiliani Atlas

| Region | EXPV | MACE | MAPE | MSLL | NLL | R2 | RMSE | Rho | SMSE | ShapiroW |
|---|---|---|---|---|---|---|---|---|---|---|
| L_temporalpole | 0,0638 | 0,0024 | 0,0696 | -1,0710 | 1,3635 | 0,0638 | 0,314 | 0,2108 | 0,936 | 0,999 |
| L_frontalpole | 0,1457 | 0,0016 | 0,0690 | -1,1979 | 1,3317 | 0,1457 | 0,256 | 0,3394 | 0,854 | 1,000 |
| L_postcentral | 0,2380 | 0,0018 | 0,0512 | -1,6426 | 1,2720 | 0,2380 | 0,146 | 0,4417 | 0,762 | 0,999 |
| L_transversetemporal | 0,1810 | 0,0003 | 0,0709 | -1,2496 | 1,2977 | 0,1810 | 0,230 | 0,3698 | 0,819 | 1,000 |
| L_rostralanteriorcingulate | 0,1750 | 0,0021 | 0,0578 | -1,3325 | 1,3100 | 0,1750 | 0,215 | 0,3669 | 0,825 | 1,000 |
| L_entorhinal | 0,1056 | 0,0021 | 0,0703 | -1,1115 | 1,3537 | 0,1056 | 0,292 | 0,2989 | 0,894 | 0,999 |
| R_posteriorcingulate | 0,2303 | 0,0022 | 0,0520 | -1,4467 | 1,2480 | 0,2303 | 0,174 | 0,4500 | 0,770 | 0,999 |
| L_superiortemporal | 0,2714 | 0,0021 | 0,0450 | -1,4745 | 1,2475 | 0,2714 | 0,165 | 0,4470 | 0,729 | 0,999 |
| R_bankssts | 0,1863 | 0,0012 | 0,0545 | -1,4495 | 1,3013 | 0,1863 | 0,188 | 0,3531 | 0,814 | 1,000 |
| L_fusiform | 0,2589 | 0,0010 | 0,0378 | -1,6620 | 1,2464 | 0,2589 | 0,137 | 0,4407 | 0,741 | 1,000 |
| R_precentral | 0,2296 | 0,0043 | 0,0498 | -1,4553 | 1,2389 | 0,2295 | 0,171 | 0,4272 | 0,770 | 0,999 |
| L_caudalmiddlefrontal | 0,3130 | 0,0019 | 0,0447 | -1,4520 | 1,2075 | 0,3130 | 0,157 | 0,4791 | 0,687 | 0,999 |
| R_pericalcarine | 0,3015 | 0,0022 | 0,0657 | -1,5690 | 1,2373 | 0,3015 | 0,145 | 0,4912 | 0,699 | 0,999 |
| L_precuneus | 0,3122 | 0,0010 | 0,0429 | -1,5501 | 1,1927 | 0,3122 | 0,140 | 0,4905 | 0,688 | 1,000 |
| L_rostralmiddlefrontal | 0,3670 | 0,0013 | 0,0413 | -1,5065 | 1,1616 | 0,3670 | 0,136 | 0,5288 | 0,633 | 1,000 |
| L_bankssts | 0,1769 | 0,0014 | 0,0550 | -1,4894 | 1,3047 | 0,1769 | 0,182 | 0,3593 | 0,823 | 1,000 |
| R_parsorbitalis | 0,2463 | 0,0034 | 0,0511 | -1,3915 | 1,2702 | 0,2463 | 0,186 | 0,4241 | 0,754 | 0,999 |
| L_parsopercularis | 0,2734 | 0,0008 | 0,0431 | -1,5272 | 1,2299 | 0,2734 | 0,153 | 0,4243 | 0,727 | 1,000 |
| R_insula | 0,2319 | 0,0011 | 0,0415 | -1,5408 | 1,2701 | 0,2319 | 0,162 | 0,3976 | 0,768 | 1,000 |
| L_pericalcarine | 0,2932 | 0,0024 | 0,0661 | -1,5666 | 1,2423 | 0,2932 | 0,147 | 0,4797 | 0,707 | 0,999 |
| L_superiorparietal | 0,3246 | 0,0013 | 0,0456 | -1,5545 | 1,1943 | 0,3246 | 0,139 | 0,5227 | 0,675 | 1,000 |
| L_supramarginal | 0,3109 | 0,0019 | 0,0426 | -1,5068 | 1,2037 | 0,3109 | 0,148 | 0,4719 | 0,689 | 1,000 |
| R_caudalmiddlefrontal | 0,3201 | 0,0026 | 0,0454 | -1,4384 | 1,2010 | 0,3201 | 0,157 | 0,4853 | 0,680 | 0,999 |
| L_parsorbitalis | 0,2195 | 0,0023 | 0,0522 | -1,4018 | 1,2884 | 0,2195 | 0,191 | 0,3963 | 0,781 | 0,999 |
| Mean_Thickness | 0,3930 | 0,0006 | 0,0298 | -1,7541 | 1,1400 | 0,3930 | 0,102 | 0,5239 | 0,607 | 1,000 |
| R_caudalanteriorcingulate | 0,0843 | 0,0021 | 0,0873 | -1,1076 | 1,3082 | 0,0843 | 0,283 | 0,2928 | 0,916 | 0,999 |
| L_superiorfrontal | 0,3717 | 0,0014 | 0,0403 | -1,4108 | 1,1646 | 0,3717 | 0,150 | 0,5188 | 0,628 | 1,000 |
| L_caudalanteriorcingulate | 0,0742 | 0,0028 | 0,1015 | -0,9170 | 1,2808 | 0,0742 | 0,335 | 0,2967 | 0,926 | 1,000 |
| R_parstriangularis | 0,3337 | 0,0011 | 0,0450 | -1,4485 | 1,1900 | 0,3337 | 0,153 | 0,5023 | 0,666 | 1,000 |
| L_precentral | 0,2374 | 0,0028 | 0,0479 | -1,4712 | 1,2345 | 0,2374 | 0,167 | 0,4285 | 0,763 | 0,999 |
| L_posteriorcingulate | 0,2341 | 0,0011 | 0,0523 | -1,4540 | 1,2577 | 0,2341 | 0,174 | 0,4410 | 0,766 | 1,000 |
| L_insula | 0,2341 | 0,0015 | 0,0414 | -1,5421 | 1,2675 | 0,2341 | 0,161 | 0,4006 | 0,766 | 1,000 |
| R_lateralorbitofrontal | 0,2638 | 0,0017 | 0,0422 | -1,5710 | 1,2507 | 0,2638 | 0,151 | 0,3918 | 0,736 | 1,000 |
| R_lingual | 0,3628 | 0,0010 | 0,0472 | -1,6022 | 1,1869 | 0,3628 | 0,128 | 0,5701 | 0,637 | 1,000 |
| R_medialorbitofrontal | 0,2650 | 0,0017 | 0,0497 | -1,4506 | 1,2418 | 0,2650 | 0,168 | 0,4329 | 0,735 | 0,999 |
| R_parsopercularis | 0,2707 | 0,0013 | 0,0445 | -1,5122 | 1,2380 | 0,2707 | 0,157 | 0,4309 | 0,729 | 1,000 |

| region | EXPV | MACE | MAPE | MSLL | NLL | R2 | RMSE | Rho | SMSE | Shapiro W |
|---|---|---|---|---|---|---|---|---|---|---|
| R_middletemporal | 0,2585 | 0,0028 | 0,0401 | -1,5764 | 1,2605 | 0,2585 | 0,152 | 0,4282 | 0,741 | 0,999 |
| R_cuneus | 0,3189 | 0,0015 | 0,0547 | -1,5954 | 1,2238 | 0,3189 | 0,138 | 0,5227 | 0,681 | 1,000 |
| L_inferiortemporal | 0,2553 | 0,0019 | 0,0414 | -1,5724 | 1,2610 | 0,2553 | 0,153 | 0,4232 | 0,745 | 1,000 |
| R_rostralanteriorcingulate | 0,2084 | 0,0016 | 0,0587 | -1,2605 | 1,2886 | 0,2084 | 0,221 | 0,4092 | 0,792 | 1,000 |
| L_cuneus | 0,2306 | 0,0017 | 0,0573 | -1,6619 | 1,2821 | 0,2306 | 0,145 | 0,4189 | 0,769 | 1,000 |
| R_postcentral | 0,2358 | 0,0032 | 0,0536 | -1,6051 | 1,2744 | 0,2358 | 0,152 | 0,4447 | 0,764 | 0,999 |
| R_parahippocampal | 0,1416 | 0,0012 | 0,0697 | -1,2658 | 1,3295 | 0,1416 | 0,239 | 0,3466 | 0,858 | 1,000 |
| L_middletemporal | 0,2336 | 0,0021 | 0,0429 | -1,5625 | 1,2773 | 0,2336 | 0,159 | 0,4263 | 0,766 | 0,999 |
| R_frontalpole | 0,1624 | 0,0016 | 0,0703 | -1,1685 | 1,3209 | 0,1624 | 0,258 | 0,3497 | 0,838 | 0,999 |
| L_paracentral | 0,2738 | 0,0011 | 0,0535 | -1,3929 | 1,2300 | 0,2738 | 0,175 | 0,4861 | 0,726 | 1,000 |
| R_transversetemporal | 0,1812 | 0,0007 | 0,0731 | -1,1987 | 1,2925 | 0,1812 | 0,241 | 0,3541 | 0,819 | 1,000 |
| R_entorhinal | 0,1023 | 0,0019 | 0,0749 | -1,0366 | 1,3575 | 0,1023 | 0,316 | 0,2797 | 0,898 | 1,000 |
| L_lateraloccipital | 0,1814 | 0,0015 | 0,0432 | -1,8749 | 1,3129 | 0,1814 | 0,125 | 0,3685 | 0,819 | 1,000 |
| R_superiorfrontal | 0,3957 | 0,0020 | 0,0405 | -1,3726 | 1,1420 | 0,3957 | 0,149 | 0,5370 | 0,604 | 1,000 |
| L_parahippocampal | 0,1080 | 0,0005 | 0,0822 | -1,1374 | 1,3494 | 0,1080 | 0,282 | 0,2920 | 0,892 | 1,000 |
| R_temporalpole | 0,0839 | 0,0030 | 0,0678 | -1,0524 | 1,3541 | 0,0839 | 0,313 | 0,2568 | 0,916 | 0,999 |
| R_isthmuscingulate | 0,2356 | 0,0008 | 0,0595 | -1,3625 | 1,2616 | 0,2356 | 0,191 | 0,4549 | 0,764 | 0,999 |
| R_inferiortemporal | 0,2191 | 0,0014 | 0,0390 | -1,6715 | 1,2878 | 0,2191 | 0,146 | 0,3894 | 0,781 | 1,000 |
| R_inferiorparietal | 0,3341 | 0,0024 | 0,0403 | -1,5559 | 1,1862 | 0,3341 | 0,136 | 0,5023 | 0,666 | 1,000 |
| L_medialorbitofrontal | 0,2081 | 0,0023 | 0,0488 | -1,5612 | 1,2842 | 0,2081 | 0,163 | 0,3725 | 0,792 | 0,999 |
| R_supramarginal | 0,3070 | 0,0010 | 0,0426 | -1,5150 | 1,2083 | 0,3070 | 0,148 | 0,4672 | 0,693 | 1,000 |
| L_parstriangularis | 0,2873 | 0,0014 | 0,0469 | -1,4733 | 1,2256 | 0,2873 | 0,159 | 0,4548 | 0,713 | 1,000 |
| Median_Thickness | 0,3796 | 0,0005 | 0,0307 | -1,7529 | 1,1559 | 0,3796 | 0,105 | 0,4968 | 0,620 | 1,000 |
| R_fusiform | 0,2280 | 0,0024 | 0,0371 | -1,7273 | 1,2739 | 0,2280 | 0,135 | 0,3933 | 0,772 | 0,999 |
| R_paracentral | 0,2529 | 0,0016 | 0,0519 | -1,4422 | 1,2441 | 0,2529 | 0,172 | 0,4519 | 0,747 | 1,000 |
| R_lateraloccipital | 0,1854 | 0,0012 | 0,0431 | -1,8485 | 1,3106 | 0,1854 | 0,128 | 0,3719 | 0,815 | 1,000 |
| L_inferiorparietal | 0,3208 | 0,0022 | 0,0407 | -1,5688 | 1,1923 | 0,3208 | 0,137 | 0,4916 | 0,679 | 1,000 |
| R_precuneus | 0,3012 | 0,0018 | 0,0417 | -1,5938 | 1,2051 | 0,3012 | 0,137 | 0,4747 | 0,699 | 1,000 |
| L_lingual | 0,3233 | 0,0009 | 0,0480 | -1,6495 | 1,2168 | 0,3233 | 0,129 | 0,5242 | 0,677 | 1,000 |
| R_superiortemporal | 0,2948 | 0,0017 | 0,0437 | -1,4440 | 1,2271 | 0,2948 | 0,164 | 0,4395 | 0,705 | 0,999 |
| L_isthmuscingulate | 0,2207 | 0,0014 | 0,0580 | -1,4274 | 1,2711 | 0,2207 | 0,183 | 0,4104 | 0,779 | 1,000 |
| R_rostralmiddlefrontal | 0,4095 | 0,0022 | 0,0417 | -1,4286 | 1,1199 | 0,4095 | 0,137 | 0,5454 | 0,591 | 1,000 |
| L_lateralorbitofrontal | 0,2796 | 0,0009 | 0,0408 | -1,5693 | 1,2399 | 0,2796 | 0,148 | 0,4147 | 0,720 | 1,000 |
| R_superiorparietal | 0,3357 | 0,0011 | 0,0462 | -1,5388 | 1,1881 | 0,3357 | 0,139 | 0,5397 | 0,664 | 1,000 |

Supplemental Table S4.2. Performance Metrics Destrieux Atlas

| region | EXPV | MACE | MAPE | MSLL | NLL | R2 | RMSE | Rho | SMSE | Shapiro W |
|---|---|---|---|---|---|---|---|---|---|---|
| lh_G_temp_sup-G_T_transv | 0,240 | 0,001 | 0,075 | -1,112 | 1,267 | 0,240 | 0,246 | 0,453 | 0,760 | 1,000 |

| Region | | | | | | | | | |
|---|---|---|---|---|---|---|---|---|---|
| lh_G_precuneus | 0,347 | 0,002 | 0,049 | -1,354 | 1,185 | 0,347 | 0,165 | 0,510 | 0,653 | 1,000 |
| lh_G_temp_sup-Lateral | 0,250 | 0,002 | 0,052 | -1,281 | 1,272 | 0,250 | 0,208 | 0,429 | 0,750 | 1,000 |
| lh_G_and_S_frontomargin | 0,312 | 0,002 | 0,055 | -1,332 | 1,223 | 0,312 | 0,180 | 0,491 | 0,688 | 0,999 |
| lh_G_temp_sup-Plan_tempo | 0,246 | 0,001 | 0,061 | -1,264 | 1,258 | 0,246 | 0,209 | 0,414 | 0,754 | 1,000 |
| lh_G_temporal_inf | 0,329 | 0,001 | 0,046 | -1,325 | 1,215 | 0,329 | 0,178 | 0,474 | 0,671 | 1,000 |
| lh_G_subcallosal | 0,192 | 0,003 | 0,103 | -0,931 | 1,294 | 0,192 | 0,313 | 0,402 | 0,808 | 0,999 |
| lh_G_rectus | 0,296 | 0,002 | 0,057 | -1,235 | 1,227 | 0,296 | 0,201 | 0,473 | 0,704 | 0,999 |
| lh_G_temp_sup-Plan_polar | 0,175 | 0,002 | 0,072 | -0,961 | 1,282 | 0,175 | 0,303 | 0,370 | 0,825 | 1,000 |
| lh_G_and_S_paracentral | 0,247 | 0,001 | 0,065 | -1,284 | 1,248 | 0,247 | 0,202 | 0,473 | 0,753 | 0,999 |
| lh_G_and_S_cingul-Mid-Ant | 0,275 | 0,003 | 0,059 | -1,174 | 1,182 | 0,275 | 0,207 | 0,496 | 0,725 | 0,999 |
| lh_G_oc-temp_med-Parahip | 0,133 | 0,001 | 0,065 | -1,237 | 1,330 | 0,133 | 0,247 | 0,293 | 0,867 | 1,000 |
| lh_G_front_inf-Orbital | 0,279 | 0,002 | 0,058 | -1,197 | 1,250 | 0,279 | 0,217 | 0,487 | 0,721 | 1,000 |
| lh_G_cingul-Post-dorsal | 0,382 | 0,001 | 0,046 | -1,226 | 1,161 | 0,382 | 0,178 | 0,494 | 0,618 | 1,000 |
| lh_G_cuneus | 0,286 | 0,003 | 0,058 | -1,608 | 1,246 | 0,286 | 0,142 | 0,484 | 0,714 | 0,999 |
| lh_G_and_S_cingul-Mid-Post | 0,360 | 0,001 | 0,045 | -1,383 | 1,160 | 0,360 | 0,155 | 0,493 | 0,640 | 1,000 |
| lh_G_pariet_inf-Supramar | 0,351 | 0,002 | 0,046 | -1,345 | 1,188 | 0,351 | 0,167 | 0,495 | 0,649 | 1,000 |
| lh_G_oc-temp_med-Lingual | 0,343 | 0,001 | 0,055 | -1,510 | 1,200 | 0,343 | 0,144 | 0,548 | 0,657 | 1,000 |
| lh_G_orbital | 0,369 | 0,002 | 0,045 | -1,339 | 1,183 | 0,369 | 0,164 | 0,535 | 0,631 | 1,000 |
| lh_G_postcentral | 0,227 | 0,001 | 0,065 | -1,412 | 1,286 | 0,227 | 0,187 | 0,442 | 0,773 | 1,000 |
| lh_G_oc-temp_lat-fusifor | 0,309 | 0,002 | 0,046 | -1,356 | 1,215 | 0,309 | 0,175 | 0,507 | 0,691 | 1,000 |
| lh_G_cingul-Post-ventral | 0,230 | 0,002 | 0,077 | -1,099 | 1,284 | 0,230 | 0,256 | 0,396 | 0,770 | 1,000 |
| lh_G_and_S_cingul-Ant | 0,338 | 0,003 | 0,043 | -1,409 | 1,194 | 0,338 | 0,159 | 0,457 | 0,662 | 0,999 |
| lh_G_and_S_transv_frontopol | 0,255 | 0,002 | 0,061 | -1,253 | 1,264 | 0,255 | 0,211 | 0,464 | 0,745 | 0,999 |
| lh_G_parietal_sup | 0,350 | 0,001 | 0,051 | -1,362 | 1,186 | 0,350 | 0,164 | 0,550 | 0,650 | 1,000 |
| lh_G_Ins_lg_and_S_cent_ins | 0,149 | 0,001 | 0,065 | -1,194 | 1,330 | 0,149 | 0,256 | 0,334 | 0,851 | 1,000 |
| lh_G_and_S_occipital_inf | 0,218 | 0,001 | 0,055 | -1,496 | 1,295 | 0,218 | 0,175 | 0,393 | 0,782 | 1,000 |
| lh_G_precentral | 0,237 | 0,005 | 0,059 | -1,185 | 1,230 | 0,237 | 0,221 | 0,441 | 0,763 | 0,998 |
| lh_G_and_S_subcentral | 0,276 | 0,001 | 0,050 | -1,377 | 1,247 | 0,276 | 0,181 | 0,419 | 0,724 | 1,000 |
| lh_G_front_inf-Opercular | 0,324 | 0,002 | 0,046 | -1,348 | 1,209 | 0,324 | 0,173 | 0,471 | 0,676 | 0,999 |
| lh_G_front_middle | 0,403 | 0,003 | 0,045 | -1,301 | 1,143 | 0,403 | 0,160 | 0,566 | 0,597 | 0,999 |
| lh_G_occipital_middle | 0,251 | 0,003 | 0,046 | -1,576 | 1,264 | 0,251 | 0,153 | 0,441 | 0,749 | 0,999 |
| lh_G_insular_short | 0,140 | 0,002 | 0,056 | -1,211 | 1,326 | 0,140 | 0,252 | 0,329 | 0,860 | 1,000 |
| lh_G_front_sup | 0,429 | 0,002 | 0,043 | -1,226 | 1,125 | 0,429 | 0,165 | 0,566 | 0,571 | 0,999 |
| lh_G_pariet_inf-Angular | 0,395 | 0,003 | 0,046 | -1,289 | 1,152 | 0,395 | 0,164 | 0,564 | 0,605 | 0,999 |

| Region | | | | | | | | | |
|---|---|---|---|---|---|---|---|---|---|
| lh_Lat_Fis-ant-Vertical | 0,220 | 0,001 | 0,071 | -1,191 | 1,281 | 0,220 | 0,234 | 0,396 | 0,780 | 1,000 |
| lh_Lat_Fis-post | 0,321 | 0,001 | 0,053 | -1,355 | 1,200 | 0,321 | 0,171 | 0,450 | 0,679 | 1,000 |
| lh_Pole_occipital | 0,159 | 0,002 | 0,067 | -1,611 | 1,326 | 0,159 | 0,167 | 0,363 | 0,841 | 0,999 |
| lh_G_occipital_sup | 0,199 | 0,002 | 0,062 | -1,514 | 1,306 | 0,199 | 0,176 | 0,408 | 0,801 | 1,000 |
| lh_G_front_inf-Triangul | 0,353 | 0,001 | 0,051 | -1,261 | 1,190 | 0,353 | 0,181 | 0,533 | 0,647 | 1,000 |
| lh_S_calcarine | 0,418 | 0,001 | 0,054 | -1,460 | 1,144 | 0,418 | 0,135 | 0,593 | 0,582 | 1,000 |
| lh_S_central | 0,268 | 0,002 | 0,056 | -1,561 | 1,221 | 0,268 | 0,147 | 0,439 | 0,732 | 0,999 |
| lh_Lat_Fis-ant-Horizont | 0,154 | 0,001 | 0,081 | -1,167 | 1,307 | 0,154 | 0,256 | 0,319 | 0,846 | 1,000 |
| lh_S_cingul-Marginalis | 0,354 | 0,002 | 0,054 | -1,323 | 1,165 | 0,354 | 0,166 | 0,530 | 0,646 | 1,000 |
| lh_Pole_temporal | 0,227 | 0,003 | 0,050 | -1,278 | 1,279 | 0,227 | 0,213 | 0,402 | 0,773 | 0,999 |
| lh_S_collat_transv_ant | 0,157 | 0,001 | 0,064 | -1,275 | 1,304 | 0,157 | 0,229 | 0,319 | 0,843 | 1,000 |
| lh_G_temporal_middle | 0,312 | 0,002 | 0,046 | -1,339 | 1,225 | 0,312 | 0,179 | 0,497 | 0,688 | 0,999 |
| lh_S_circular_insula_inf | 0,317 | 0,001 | 0,055 | -1,226 | 1,214 | 0,317 | 0,198 | 0,497 | 0,683 | 1,000 |
| lh_S_circular_insula_ant | 0,276 | 0,002 | 0,056 | -1,221 | 1,240 | 0,276 | 0,210 | 0,416 | 0,724 | 1,000 |
| lh_S_collat_transv_post | 0,206 | 0,001 | 0,068 | -1,418 | 1,292 | 0,206 | 0,190 | 0,360 | 0,794 | 1,000 |
| lh_S_intrapariet_and_P_trans | 0,382 | 0,002 | 0,048 | -1,402 | 1,128 | 0,382 | 0,145 | 0,535 | 0,618 | 1,000 |
| lh_S_occipital_ant | 0,208 | 0,000 | 0,060 | -1,455 | 1,290 | 0,208 | 0,183 | 0,388 | 0,792 | 1,000 |
| lh_S_front_middle | 0,383 | 0,002 | 0,051 | -1,342 | 1,158 | 0,383 | 0,158 | 0,549 | 0,617 | 1,000 |
| lh_S_front_sup | 0,427 | 0,002 | 0,043 | -1,332 | 1,106 | 0,427 | 0,146 | 0,581 | 0,573 | 0,999 |
| lh_S_orbital_med-olfact | 0,184 | 0,004 | 0,058 | -1,494 | 1,305 | 0,184 | 0,181 | 0,359 | 0,816 | 0,999 |
| lh_S_orbital_lateral | 0,262 | 0,002 | 0,075 | -1,167 | 1,252 | 0,262 | 0,226 | 0,467 | 0,738 | 1,000 |
| lh_S_oc_middle_and_Lunatus | 0,280 | 0,001 | 0,058 | -1,474 | 1,239 | 0,280 | 0,162 | 0,487 | 0,720 | 1,000 |
| lh_S_subparietal | 0,330 | 0,001 | 0,051 | -1,381 | 1,185 | 0,330 | 0,163 | 0,492 | 0,670 | 1,000 |
| rh_G_and_S_occipital_inf | 0,173 | 0,001 | 0,060 | -1,423 | 1,323 | 0,173 | 0,199 | 0,342 | 0,827 | 1,000 |
| lh_S_temporal_transverse | 0,146 | 0,001 | 0,096 | -1,033 | 1,319 | 0,146 | 0,298 | 0,322 | 0,854 | 1,000 |
| rh_G_and_S_transv_frontopol | 0,297 | 0,002 | 0,056 | -1,284 | 1,233 | 0,297 | 0,193 | 0,501 | 0,703 | 1,000 |
| lh_S_front_inf | 0,374 | 0,002 | 0,046 | -1,422 | 1,149 | 0,374 | 0,146 | 0,515 | 0,626 | 1,000 |
| lh_S_oc_sup_and_transversal | 0,307 | 0,001 | 0,053 | -1,483 | 1,210 | 0,307 | 0,153 | 0,476 | 0,693 | 1,000 |
| rh_G_and_S_frontomargin | 0,306 | 0,002 | 0,060 | -1,251 | 1,227 | 0,306 | 0,197 | 0,470 | 0,694 | 1,000 |
| rh_G_and_S_paracentral | 0,244 | 0,001 | 0,066 | -1,294 | 1,254 | 0,244 | 0,202 | 0,473 | 0,756 | 1,000 |
| rh_G_and_S_cingul-Mid-Post | 0,341 | 0,001 | 0,043 | -1,435 | 1,170 | 0,341 | 0,151 | 0,473 | 0,659 | 1,000 |
| lh_S_temporal_sup | 0,377 | 0,000 | 0,041 | -1,477 | 1,144 | 0,377 | 0,137 | 0,502 | 0,623 | 1,000 |
| lh_S_oc-temp_med_and_Lingual | 0,305 | 0,001 | 0,051 | -1,443 | 1,213 | 0,305 | 0,160 | 0,447 | 0,695 | 1,000 |
| rh_G_and_S_cingul-Ant | 0,384 | 0,002 | 0,043 | -1,343 | 1,154 | 0,384 | 0,157 | 0,488 | 0,616 | 0,999 |
| rh_G_and_S_cingul-Mid-Ant | 0,310 | 0,003 | 0,050 | -1,303 | 1,183 | 0,310 | 0,178 | 0,482 | 0,690 | 0,999 |

| Region | | | | | | | | | | |
|---|---|---|---|---|---|---|---|---|---|---|
| lh_S_interm_prim-Jensen | 0,173 | 0,002 | 0,091 | -0,908 | 1,259 | 0,173 | 0,313 | 0,390 | 0,827 | 1,000 |
| rh_G_cingul-Post-ventral | 0,187 | 0,001 | 0,070 | -1,213 | 1,313 | 0,187 | 0,241 | 0,391 | 0,813 | 1,000 |
| rh_G_front_inf-Orbital | 0,257 | 0,002 | 0,062 | -1,182 | 1,267 | 0,257 | 0,227 | 0,476 | 0,743 | 0,999 |
| rh_G_front_inf-Opercular | 0,334 | 0,002 | 0,046 | -1,350 | 1,203 | 0,334 | 0,171 | 0,491 | 0,666 | 1,000 |
| lh_S_orbital-H_Shaped | 0,423 | 0,001 | 0,052 | -1,105 | 1,130 | 0,423 | 0,188 | 0,615 | 0,577 | 1,000 |
| lh_S_parieto_occipital | 0,339 | 0,001 | 0,052 | -1,415 | 1,184 | 0,339 | 0,156 | 0,485 | 0,661 | 1,000 |
| lh_S_oc-temp_lat | 0,239 | 0,002 | 0,054 | -1,452 | 1,273 | 0,239 | 0,176 | 0,395 | 0,761 | 1,000 |
| rh_G_and_S_subcentral | 0,270 | 0,001 | 0,051 | -1,371 | 1,252 | 0,270 | 0,184 | 0,414 | 0,730 | 1,000 |
| rh_G_insular_short | 0,126 | 0,001 | 0,065 | -1,157 | 1,344 | 0,126 | 0,273 | 0,322 | 0,874 | 1,000 |
| lh_S_postcentral | 0,363 | 0,001 | 0,052 | -1,384 | 1,158 | 0,363 | 0,154 | 0,526 | 0,637 | 1,000 |
| rh_G_cingul-Post-dorsal | 0,372 | 0,001 | 0,049 | -1,153 | 1,157 | 0,372 | 0,192 | 0,525 | 0,628 | 1,000 |
| lh_S_suborbital | 0,190 | 0,003 | 0,077 | -1,167 | 1,300 | 0,190 | 0,249 | 0,370 | 0,810 | 0,999 |
| lh_S_precentral-inf-part | 0,338 | 0,002 | 0,047 | -1,399 | 1,179 | 0,338 | 0,158 | 0,496 | 0,662 | 0,999 |
| rh_G_cuneus | 0,365 | 0,002 | 0,055 | -1,553 | 1,188 | 0,365 | 0,134 | 0,563 | 0,635 | 1,000 |
| lh_S_circular_insula_sup | 0,401 | 0,001 | 0,041 | -1,383 | 1,118 | 0,401 | 0,144 | 0,503 | 0,599 | 1,000 |
| rh_G_oc-temp_med-Lingual | 0,365 | 0,002 | 0,055 | -1,452 | 1,186 | 0,365 | 0,148 | 0,560 | 0,635 | 1,000 |
| rh_G_oc-temp_med-Parahip | 0,167 | 0,001 | 0,060 | -1,244 | 1,311 | 0,167 | 0,236 | 0,348 | 0,833 | 1,000 |
| rh_G_orbital | 0,308 | 0,002 | 0,046 | -1,403 | 1,226 | 0,308 | 0,169 | 0,479 | 0,692 | 1,000 |
| lh_S_pericallosal | 0,270 | 0,003 | 0,109 | -0,912 | 1,207 | 0,270 | 0,278 | 0,525 | 0,730 | 0,999 |
| rh_G_Ins_lg_and_S_cent_ins | 0,160 | 0,001 | 0,072 | -1,050 | 1,329 | 0,160 | 0,293 | 0,364 | 0,840 | 1,000 |
| rh_G_front_inf-Triangul | 0,344 | 0,002 | 0,052 | -1,268 | 1,201 | 0,344 | 0,183 | 0,537 | 0,656 | 0,999 |
| rh_G_front_sup | 0,421 | 0,003 | 0,043 | -1,245 | 1,132 | 0,421 | 0,164 | 0,571 | 0,579 | 0,999 |
| rh_G_front_middle | 0,390 | 0,004 | 0,045 | -1,326 | 1,151 | 0,390 | 0,159 | 0,573 | 0,610 | 0,999 |
| rh_G_oc-temp_lat-fusifor | 0,261 | 0,002 | 0,046 | -1,434 | 1,253 | 0,261 | 0,174 | 0,439 | 0,739 | 1,000 |
| rh_G_occipital_middle | 0,286 | 0,004 | 0,045 | -1,509 | 1,238 | 0,286 | 0,156 | 0,461 | 0,714 | 0,998 |
| rh_G_occipital_sup | 0,254 | 0,002 | 0,060 | -1,462 | 1,270 | 0,254 | 0,172 | 0,474 | 0,746 | 0,999 |
| rh_G_postcentral | 0,212 | 0,002 | 0,070 | -1,369 | 1,295 | 0,212 | 0,199 | 0,431 | 0,788 | 1,000 |
| rh_G_pariet_inf-Supramar | 0,345 | 0,002 | 0,046 | -1,344 | 1,190 | 0,345 | 0,168 | 0,495 | 0,655 | 1,000 |
| rh_G_temp_sup-G_T_transv | 0,226 | 0,000 | 0,073 | -1,121 | 1,271 | 0,226 | 0,247 | 0,408 | 0,774 | 1,000 |
| rh_G_temp_sup-Lateral | 0,281 | 0,002 | 0,049 | -1,275 | 1,248 | 0,281 | 0,200 | 0,446 | 0,719 | 0,999 |
| lh_S_temporal_inf | 0,241 | 0,002 | 0,052 | -1,469 | 1,260 | 0,241 | 0,171 | 0,412 | 0,759 | 0,999 |
| rh_G_temp_sup-Plan_tempo | 0,281 | 0,001 | 0,059 | -1,227 | 1,235 | 0,281 | 0,207 | 0,439 | 0,719 | 1,000 |
| rh_G_precuneus | 0,333 | 0,001 | 0,048 | -1,394 | 1,202 | 0,333 | 0,163 | 0,507 | 0,667 | 1,000 |
| rh_G_parietal_sup | 0,348 | 0,002 | 0,054 | -1,326 | 1,191 | 0,348 | 0,171 | 0,556 | 0,652 | 1,000 |
| rh_Pole_occipital | 0,159 | 0,002 | 0,060 | -1,712 | 1,327 | 0,159 | 0,151 | 0,356 | 0,841 | 1,000 |

| | | | | | | | | | |
|---|---|---|---|---|---|---|---|---|---|
| lh_S_precentral-sup-part | 0,288 | 0,004 | 0,052 | -1,377 | 1,214 | 0,288 | 0,173 | 0,479 | 0,712 | 0,998 |
| rh_G_temp_sup-Plan_polar | 0,187 | 0,002 | 0,069 | -1,044 | 1,298 | 0,187 | 0,281 | 0,358 | 0,813 | 1,000 |
| rh_G_temporal_inf | 0,292 | 0,001 | 0,046 | -1,382 | 1,244 | 0,292 | 0,177 | 0,436 | 0,708 | 1,000 |
| rh_G_rectus | 0,249 | 0,002 | 0,063 | -1,226 | 1,258 | 0,249 | 0,217 | 0,414 | 0,751 | 0,999 |
| rh_G_subcallosal | 0,110 | 0,002 | 0,118 | -0,771 | 1,302 | 0,110 | 0,388 | 0,333 | 0,890 | 1,000 |
| rh_S_calcarine | 0,419 | 0,001 | 0,055 | -1,428 | 1,145 | 0,419 | 0,139 | 0,597 | 0,581 | 1,000 |
| rh_S_circular_insula_sup | 0,392 | 0,001 | 0,041 | -1,380 | 1,133 | 0,392 | 0,147 | 0,520 | 0,608 | 1,000 |
| rh_S_collat_transv_ant | 0,185 | 0,001 | 0,058 | -1,337 | 1,290 | 0,185 | 0,208 | 0,351 | 0,815 | 1,000 |
| rh_G_precentral | 0,212 | 0,005 | 0,063 | -1,154 | 1,235 | 0,212 | 0,233 | 0,423 | 0,788 | 0,998 |
| rh_G_temporal_middle | 0,338 | 0,003 | 0,043 | -1,350 | 1,204 | 0,338 | 0,170 | 0,502 | 0,662 | 0,999 |
| rh_Pole_temporal | 0,226 | 0,003 | 0,049 | -1,299 | 1,281 | 0,226 | 0,209 | 0,393 | 0,774 | 0,999 |
| rh_S_central | 0,263 | 0,003 | 0,057 | -1,581 | 1,237 | 0,263 | 0,147 | 0,445 | 0,737 | 0,999 |
| rh_S_front_inf | 0,363 | 0,003 | 0,047 | -1,419 | 1,153 | 0,363 | 0,148 | 0,511 | 0,637 | 1,000 |
| rh_S_circular_insula_ant | 0,252 | 0,001 | 0,058 | -1,224 | 1,261 | 0,252 | 0,217 | 0,394 | 0,748 | 1,000 |
| rh_S_interm_prim-Jensen | 0,278 | 0,002 | 0,069 | -1,170 | 1,239 | 0,278 | 0,220 | 0,467 | 0,722 | 1,000 |
| rh_S_circular_insula_inf | 0,328 | 0,001 | 0,058 | -1,183 | 1,209 | 0,328 | 0,204 | 0,500 | 0,672 | 1,000 |
| rh_Lat_Fis-post | 0,356 | 0,001 | 0,050 | -1,335 | 1,168 | 0,356 | 0,164 | 0,481 | 0,644 | 1,000 |
| rh_G_pariet_inf-Angular | 0,383 | 0,003 | 0,047 | -1,283 | 1,156 | 0,383 | 0,167 | 0,567 | 0,617 | 0,999 |
| rh_Lat_Fis-ant-Horizont | 0,215 | 0,001 | 0,074 | -1,209 | 1,281 | 0,215 | 0,231 | 0,378 | 0,785 | 1,000 |
| rh_S_intrapariet_and_P_trans | 0,387 | 0,001 | 0,049 | -1,403 | 1,131 | 0,387 | 0,144 | 0,536 | 0,613 | 1,000 |
| rh_S_oc-temp_lat | 0,250 | 0,002 | 0,052 | -1,440 | 1,261 | 0,250 | 0,175 | 0,420 | 0,750 | 1,000 |
| rh_S_front_middle | 0,408 | 0,002 | 0,049 | -1,331 | 1,126 | 0,408 | 0,152 | 0,562 | 0,592 | 1,000 |
| rh_S_oc_sup_and_transversal | 0,311 | 0,001 | 0,053 | -1,482 | 1,208 | 0,311 | 0,153 | 0,462 | 0,689 | 1,000 |
| rh_S_occipital_ant | 0,218 | 0,001 | 0,060 | -1,416 | 1,278 | 0,218 | 0,186 | 0,400 | 0,782 | 1,000 |
| rh_S_oc-temp_med_and_Lingual | 0,303 | 0,002 | 0,052 | -1,440 | 1,215 | 0,303 | 0,161 | 0,489 | 0,697 | 1,000 |
| rh_S_collat_transv_post | 0,224 | 0,001 | 0,066 | -1,397 | 1,283 | 0,224 | 0,190 | 0,371 | 0,776 | 1,000 |
| rh_S_orbital-H_Shaped | 0,437 | 0,002 | 0,051 | -1,107 | 1,117 | 0,437 | 0,183 | 0,602 | 0,563 | 1,000 |
| rh_S_front_sup | 0,413 | 0,003 | 0,044 | -1,345 | 1,122 | 0,413 | 0,148 | 0,563 | 0,587 | 0,999 |
| rh_S_cingul-Marginalis | 0,342 | 0,002 | 0,052 | -1,368 | 1,171 | 0,342 | 0,161 | 0,509 | 0,658 | 1,000 |
| rh_S_oc_middle_and_Lunatus | 0,281 | 0,002 | 0,061 | -1,413 | 1,242 | 0,281 | 0,173 | 0,478 | 0,719 | 0,999 |
| rh_S_precentral-inf-part | 0,308 | 0,004 | 0,048 | -1,409 | 1,192 | 0,308 | 0,162 | 0,468 | 0,692 | 0,999 |
| rh_S_temporal_transverse | 0,182 | 0,001 | 0,094 | -0,981 | 1,314 | 0,182 | 0,305 | 0,349 | 0,818 | 1,000 |
| Median_Thickness | 0,491 | 0,001 | 0,031 | -1,550 | 1,042 | 0,491 | 0,104 | 0,554 | 0,509 | 1,000 |
| rh_S_parieto_occipital | 0,322 | 0,001 | 0,052 | -1,432 | 1,196 | 0,322 | 0,157 | 0,473 | 0,678 | 1,000 |
| rh_S_postcentral | 0,348 | 0,001 | 0,054 | -1,390 | 1,176 | 0,348 | 0,158 | 0,513 | 0,652 | 1,000 |
| rh_S_subparietal | 0,322 | 0,001 | 0,051 | -1,382 | 1,195 | 0,322 | 0,165 | 0,486 | 0,678 | 1,000 |

| Region | | | | | | | | | |
|---|---|---|---|---|---|---|---|---|---|
| rh_Lat_Fis-ant-Vertical | 0,179 | 0,001 | 0,078 | -1,132 | 1,303 | 0,179 | 0,260 | 0,359 | 0,821 | 1,000 |
| rh_S_suborbital | 0,177 | 0,001 | 0,111 | -0,774 | 1,289 | 0,177 | 0,368 | 0,393 | 0,823 | 1,000 |
| Mean_Thickness | 0,482 | 0,001 | 0,030 | -1,590 | 1,053 | 0,482 | 0,102 | 0,551 | 0,518 | 1,000 |
| rh_S_pericallosal | 0,255 | 0,003 | 0,112 | -0,880 | 1,183 | 0,255 | 0,283 | 0,514 | 0,745 | 0,999 |
| rh_S_orbital_lateral | 0,295 | 0,002 | 0,068 | -1,229 | 1,230 | 0,295 | 0,203 | 0,469 | 0,705 | 1,000 |
| rh_S_orbital_med-olfact | 0,215 | 0,002 | 0,062 | -1,407 | 1,287 | 0,215 | 0,190 | 0,370 | 0,785 | 0,999 |
| rh_S_temporal_sup | 0,403 | 0,001 | 0,041 | -1,425 | 1,123 | 0,403 | 0,138 | 0,517 | 0,597 | 1,000 |
| rh_S_temporal_inf | 0,265 | 0,002 | 0,050 | -1,461 | 1,249 | 0,265 | 0,168 | 0,435 | 0,735 | 1,000 |
| rh_S_precentral-sup-part | 0,249 | 0,006 | 0,057 | -1,331 | 1,217 | 0,249 | 0,187 | 0,467 | 0,751 | 0,996 |

Supplemental Table S4.3. Performance Metrics Subcortical Measures

| Region | EXPV | MACE | MAPE | NLL | R2 | RMSE | Rho | SMSE | Shapiro W |
|---|---|---|---|---|---|---|---|---|---|
| 3rd-Ventricle | 0.499 | 0.004 | 0.226 | 0.815 | 0.499 | 356.999 | 0.743 | 0.5010 | 0.9981 |
| 4th-Ventricle | 0.083 | 0.004 | 0.227 | 1.308 | 0.083 | 508.567 | 0.295 | 0.9167 | 0.9992 |
| 5th-Ventricle | 0.024 | 0.230 | | -0.158 | -0.011 | 2.468 | 0.205 | 1.0111 | 0.5160 |
| Brain-Stem | 0.364 | 0.001 | 0.079 | 1.184 | 0.364 | 2079.852 | 0.599 | 0.6357 | 0.9998 |
| BrainSegVol | 0.369 | 0.001 | 0.064 | 1.185 | 0.369 | 95762.217 | 0.603 | 0.6311 | 0.9998 |
| BrainSegVol-to-eTIV | 0.455 | 0.007 | 0.032 | 1.008 | 0.455 | 0.036 | 0.581 | 0.5451 | 0.9907 |
| CC_Anterior | 0.139 | 0.002 | 0.135 | 1.328 | 0.139 | 147.970 | 0.368 | 0.8612 | 0.9995 |
| CC_Central | 0.212 | 0.003 | 0.172 | 1.227 | 0.212 | 116.747 | 0.424 | 0.7879 | 0.9994 |
| CC_Mid_Anterior | 0.166 | 0.003 | 0.182 | 1.234 | 0.166 | 122.986 | 0.391 | 0.8340 | 0.9997 |
| CC_Mid_Posterior | 0.179 | 0.003 | 0.162 | 1.310 | 0.179 | 98.440 | 0.415 | 0.8209 | 0.9990 |
| CC_Posterior | 0.290 | 0.003 | 0.125 | 1.238 | 0.290 | 147.051 | 0.536 | 0.7104 | 0.9988 |
| CSF | 0.262 | 0.004 | 0.174 | 1.201 | 0.262 | 230.237 | 0.528 | 0.7383 | 0.9982 |
| EstimatedTotalIntraCranialVol | 0.357 | 0.002 | 0.067 | 1.193 | 0.357 | 127809.255 | 0.589 | 0.6429 | 0.9995 |
| Left-Accumbens-area | 0.658 | 0.002 | 0.154 | 0.841 | 0.658 | 100.935 | 0.808 | 0.3422 | 0.9992 |
| Left-Amygdala | 0.359 | 0.001 | 0.097 | 1.190 | 0.359 | 189.859 | 0.590 | 0.6405 | 0.9996 |
| Left-Caudate | 0.421 | 0.002 | 0.098 | 1.130 | 0.421 | 449.703 | 0.649 | 0.5787 | 0.9995 |
| Left-Cerebellum-Cortex | 0.320 | 0.001 | 0.073 | 1.224 | 0.320 | 5096.970 | 0.547 | 0.6804 | 0.9996 |
| Left-Cerebellum-White-Matter | 0.215 | 0.003 | 0.099 | 1.277 | 0.215 | 1869.278 | 0.463 | 0.7849 | 0.9985 |
| Left-choroid-plexus | 0.427 | 0.004 | 0.348 | 1.125 | 0.427 | 257.032 | 0.622 | 0.5735 | 0.9983 |
| Left-Hippocampus | 0.286 | 0.001 | 0.073 | 1.247 | 0.286 | 368.652 | 0.517 | 0.7139 | 0.9994 |
| Left-Inf-Lat-Vent | 0.282 | 0.006 | 0.478 | 1.009 | 0.282 | 199.135 | 0.469 | 0.7179 | 0.9976 |
| Left-Lateral-Ventricle | 0.439 | 0.008 | 0.435 | 0.777 | 0.439 | 5038.436 | 0.724 | 0.5612 | 0.9953 |
| Left-Pallidum | 0.369 | 0.002 | 0.097 | 1.178 | 0.369 | 234.557 | 0.557 | 0.6306 | 0.9994 |
| Left-Putamen | 0.592 | 0.003 | 0.088 | 0.955 | 0.592 | 567.027 | 0.767 | 0.4079 | 0.9993 |
| Left-VentralDC | 0.329 | 0.001 | 0.073 | 1.214 | 0.329 | 361.696 | 0.560 | 0.6714 | 0.9998 |
| Left-vessel | 0.197 | 0.004 | | 1.125 | 0.197 | 37.974 | 0.447 | 0.8026 | 0.9981 |
| lhCortexVol | 0.585 | 0.001 | 0.068 | 0.968 | 0.585 | 22449.217 | 0.762 | 0.4146 | 0.9996 |

| Region | | | | | | | | |
|---|---|---|---|---|---|---|---|---|
| lhSurfaceHoles | 0.386 | 0.013 | | 0.714 | 0.382 | 15.664 | 0.658 | 0.6178 | 0.9665 |
| MaskVol-to-eTIV | 0.419 | 0.008 | 0.027 | 0.952 | 0.419 | 0.043 | 0.575 | 0.5808 | 0.9898 |
| non-WM-hypointensities | 0.175 | 0.129 | | -0.007 | 0.173 | 7.775 | 0.498 | 0.8266 | 0.7873 |
| Optic-Chiasm | 0.305 | 0.001 | | 1.233 | 0.305 | 54.276 | 0.528 | 0.6948 | 0.9996 |
| rhCortexVol | 0.583 | 0.001 | 0.068 | 0.970 | 0.583 | 22470.999 | 0.761 | 0.4168 | 0.9996 |
| rhSurfaceHoles | 0.388 | 0.015 | | 0.685 | 0.385 | 14.888 | 0.648 | 0.6146 | 0.9670 |
| Right-Accumbens-area | 0.584 | 0.001 | 0.133 | 0.940 | 0.584 | 99.689 | 0.761 | 0.4164 | 0.9995 |
| Right-Amygdala | 0.322 | 0.002 | 0.093 | 1.218 | 0.322 | 204.378 | 0.560 | 0.6779 | 0.9997 |
| Right-Caudate | 0.420 | 0.002 | 0.098 | 1.131 | 0.420 | 468.168 | 0.648 | 0.5800 | 0.9995 |
| Right-Cerebellum-Cortex | 0.308 | 0.001 | 0.074 | 1.234 | 0.308 | 5275.356 | 0.539 | 0.6921 | 0.9996 |
| Right-Cerebellum-White-Matter | 0.225 | 0.005 | 0.100 | 1.260 | 0.225 | 1843.168 | 0.480 | 0.7752 | 0.9979 |
| Right-choroid-plexus | 0.529 | 0.003 | 0.309 | 1.011 | 0.529 | 259.262 | 0.667 | 0.4711 | 0.9992 |
| Right-Hippocampus | 0.271 | 0.001 | 0.074 | 1.259 | 0.271 | 385.887 | 0.502 | 0.7291 | 0.9997 |
| Right-Inf-Lat-Vent | 0.237 | 0.005 | 0.482 | 1.089 | 0.237 | 187.355 | 0.410 | 0.7628 | 0.9977 |
| Right-Lateral-Ventricle | 0.451 | 0.007 | 0.426 | 0.793 | 0.451 | 4441.496 | 0.727 | 0.5491 | 0.9950 |
| Right-Pallidum | 0.264 | 0.002 | 0.091 | 1.261 | 0.264 | 218.057 | 0.473 | 0.7363 | 0.9997 |
| Right-Putamen | 0.586 | 0.003 | 0.085 | 0.966 | 0.586 | 551.990 | 0.767 | 0.4145 | 0.9990 |
| Right-VentralDC | 0.340 | 0.001 | 0.071 | 1.203 | 0.340 | 356.495 | 0.577 | 0.6602 | 0.9997 |
| Right-vessel | 0.224 | 0.003 | | 1.013 | 0.224 | 40.446 | 0.448 | 0.7763 | 0.9978 |
| SubCortGrayVol | 0.520 | 0.001 | 0.059 | 1.045 | 0.520 | 4331.433 | 0.728 | 0.4797 | 0.9997 |
| SupraTentorialVol | 0.355 | 0.001 | 0.067 | 1.195 | 0.355 | 88780.836 | 0.593 | 0.6447 | 0.9998 |
| SupraTentorialVolNotVent | 0.372 | 0.001 | 0.067 | 1.182 | 0.372 | 86706.804 | 0.606 | 0.6282 | 0.9998 |
| TotalGrayVol | 0.575 | 0.002 | 0.060 | 0.979 | 0.575 | 52702.955 | 0.756 | 0.4248 | 0.9998 |
| WM-hypointensities | 0.312 | 0.022 | 0.548 | 0.386 | 0.311 | 1490.219 | 0.707 | 0.6890 | 0.9774 |

# S5 Quality metrics: Test set
Supplemental Table S5.1. Performance Metrics Desikan Kiliani Atlas

| Region | EXPV | MACE | NLL | R2 | RMSE | Rho | SMSE | ShapiroW |
|---|---|---|---|---|---|---|---|---|
| L_rostralanteriorcingulate | 0.11 | 0.044 | 1.318 | 0.079 | 0.22 | 0.326 | 0.921 | 0.995 |
| L_postcentral | 0.179 | 0.009 | 1.26 | 0.178 | 0.148 | 0.421 | 0.822 | 0.995 |
| L_superiortemporal | 0.288 | 0.02 | 1.242 | 0.285 | 0.162 | 0.465 | 0.715 | 0.995 |
| L_precuneus | 0.286 | 0.022 | 1.184 | 0.281 | 0.14 | 0.485 | 0.719 | 0.999 |
| L_superiorparietal | 0.274 | 0.009 | 1.172 | 0.272 | 0.14 | 0.517 | 0.728 | 0.999 |
| L_rostralmiddlefrontal | 0.353 | 0.013 | 1.161 | 0.352 | 0.136 | 0.529 | 0.648 | 0.997 |
| L_precentral | 0.195 | 0.008 | 1.166 | 0.194 | 0.164 | 0.375 | 0.806 | 0.997 |
| L_posteriorcingulate | 0.176 | 0.054 | 1.292 | 0.139 | 0.178 | 0.398 | 0.861 | 0.987 |
| L_superiorfrontal | 0.378 | 0.023 | 1.164 | 0.374 | 0.148 | 0.511 | 0.626 | 0.998 |
| R_bankssts | 0.172 | 0.029 | 1.305 | 0.16 | 0.188 | 0.339 | 0.84 | 0.998 |

| Region | | | | | | | | |
|---|---|---|---|---|---|---|---|---|
| L_supramarginal | 0.293 | 0.015 | 1.211 | 0.291 | 0.148 | 0.441 | 0.709 | 0.998 |
| L_frontalpole | 0.153 | 0.017 | 1.351 | 0.15 | 0.258 | 0.357 | 0.85 | 0.997 |
| L_bankssts | 0.175 | 0.031 | 1.32 | 0.163 | 0.184 | 0.379 | 0.837 | 0.997 |
| R_parsorbitalis | 0.253 | 0.015 | 1.284 | 0.253 | 0.187 | 0.438 | 0.747 | 0.994 |
| L_transversetemporal | 0.196 | 0.011 | 1.245 | 0.196 | 0.223 | 0.416 | 0.804 | 0.998 |
| L_temporalpole | 0.029 | 0.017 | 1.279 | 0.025 | 0.307 | 0.164 | 0.975 | 0.998 |
| L_insula | 0.221 | 0.024 | 1.225 | 0.215 | 0.158 | 0.4 | 0.785 | 0.993 |
| R_caudalanteriorcingulate | 0.033 | 0.034 | 1.16 | 0.024 | 0.268 | 0.234 | 0.976 | 0.997 |
| L_pericalcarine | 0.241 | 0.009 | 1.286 | 0.241 | 0.154 | 0.489 | 0.759 | 0.988 |
| R_caudalmiddlefrontal | 0.287 | 0.008 | 1.175 | 0.287 | 0.157 | 0.456 | 0.713 | 0.999 |
| L_paracentral | 0.263 | 0.008 | 1.184 | 0.262 | 0.172 | 0.497 | 0.738 | 0.992 |
| R_transversetemporal | 0.194 | 0.013 | 1.223 | 0.194 | 0.231 | 0.399 | 0.806 | 0.998 |
| L_fusiform | 0.213 | 0.03 | 1.295 | 0.201 | 0.14 | 0.4 | 0.799 | 0.997 |
| R_precentral | 0.188 | 0.014 | 1.177 | 0.187 | 0.169 | 0.376 | 0.813 | 0.998 |
| L_caudalmiddlefrontal | 0.287 | 0.012 | 1.177 | 0.287 | 0.156 | 0.451 | 0.713 | 0.998 |
| R_pericalcarine | 0.253 | 0.003 | 1.285 | 0.253 | 0.151 | 0.502 | 0.747 | 0.992 |
| L_cuneus | 0.212 | 0.008 | 1.276 | 0.211 | 0.146 | 0.445 | 0.789 | 0.997 |
| R_postcentral | 0.177 | 0.011 | 1.269 | 0.175 | 0.154 | 0.426 | 0.825 | 0.997 |
| L_medialorbitofrontal | 0.179 | 0.027 | 1.287 | 0.169 | 0.163 | 0.349 | 0.831 | 0.998 |
| R_supramarginal | 0.282 | 0.013 | 1.21 | 0.281 | 0.148 | 0.44 | 0.719 | 0.999 |
| L_middletemporal | 0.259 | 0.031 | 1.289 | 0.25 | 0.158 | 0.454 | 0.75 | 0.995 |
| R_frontalpole | 0.174 | 0.015 | 1.339 | 0.171 | 0.259 | 0.378 | 0.829 | 0.997 |
| L_inferiorparietal | 0.319 | 0.021 | 1.199 | 0.316 | 0.136 | 0.494 | 0.684 | 0.997 |
| R_precuneus | 0.278 | 0.023 | 1.195 | 0.271 | 0.137 | 0.47 | 0.729 | 0.998 |
| L_entorhinal | 0.042 | 0.012 | 1.369 | 0.041 | 0.302 | 0.233 | 0.959 | 0.996 |
| R_posteriorcingulate | 0.174 | 0.054 | 1.252 | 0.136 | 0.176 | 0.402 | 0.864 | 0.996 |
| L_parahippocampal | 0.086 | 0.019 | 1.293 | 0.081 | 0.279 | 0.276 | 0.919 | 0.986 |
| R_temporalpole | 0.047 | 0.018 | 1.298 | 0.043 | 0.312 | 0.214 | 0.957 | 0.998 |
| L_lingual | 0.318 | 0.025 | 1.201 | 0.312 | 0.128 | 0.554 | 0.688 | 0.993 |
| R_superiortemporal | 0.296 | 0.019 | 1.237 | 0.293 | 0.164 | 0.428 | 0.707 | 0.996 |
| L_parsopercularis | 0.253 | 0.012 | 1.221 | 0.252 | 0.153 | 0.397 | 0.748 | 0.998 |
| R_insula | 0.231 | 0.031 | 1.24 | 0.222 | 0.16 | 0.398 | 0.778 | 0.991 |
| L_inferiortemporal | 0.239 | 0.026 | 1.28 | 0.231 | 0.153 | 0.404 | 0.769 | 0.998 |
| R_rostralanteriorcingulate | 0.178 | 0.026 | 1.304 | 0.169 | 0.222 | 0.402 | 0.831 | 0.995 |
| R_cuneus | 0.307 | 0.009 | 1.258 | 0.306 | 0.14 | 0.565 | 0.694 | 0.994 |
| L_caudalanteriorcingulate | 0.036 | 0.025 | 1.098 | 0.034 | 0.308 | 0.28 | 0.966 | 0.996 |
| R_parstriangularis | 0.328 | 0.009 | 1.19 | 0.328 | 0.152 | 0.516 | 0.672 | 0.999 |
| L_lateraloccipital | 0.155 | 0.007 | 1.301 | 0.155 | 0.125 | 0.364 | 0.845 | 0.995 |
| R_superiorfrontal | 0.402 | 0.019 | 1.15 | 0.4 | 0.148 | 0.534 | 0.6 | 0.998 |
| L_lateralorbitofrontal | 0.272 | 0.023 | 1.276 | 0.266 | 0.148 | 0.405 | 0.734 | 0.997 |
| R_superiorparietal | 0.299 | 0.014 | 1.172 | 0.297 | 0.139 | 0.551 | 0.703 | 1 |
| R_entorhinal | 0 | 0.013 | 1.393 | -0.003 | 0.334 | 0.2 | 1.003 | 0.997 |
| R_fusiform | 0.199 | 0.024 | 1.288 | 0.192 | 0.136 | 0.376 | 0.808 | 0.991 |
| R_inferiortemporal | 0.2 | 0.028 | 1.32 | 0.189 | 0.147 | 0.369 | 0.811 | 0.997 |
| R_lateraloccipital | 0.172 | 0.007 | 1.293 | 0.172 | 0.127 | 0.38 | 0.828 | 0.995 |

| Region | | | | | | | | |
|---|---|---|---|---|---|---|---|---|
| L_isthmuscingulate | 0.162 | 0.052 | 1.318 | 0.121 | 0.191 | 0.389 | 0.879 | 0.995 |
| R_rostralmiddlefrontal | 0.384 | 0.006 | 1.133 | 0.384 | 0.138 | 0.553 | 0.616 | 0.999 |
| R_inferiorparietal | 0.322 | 0.019 | 1.189 | 0.319 | 0.136 | 0.502 | 0.681 | 0.997 |
| R_paracentral | 0.227 | 0.01 | 1.176 | 0.227 | 0.168 | 0.444 | 0.773 | 0.998 |
| R_isthmuscingulate | 0.184 | 0.05 | 1.308 | 0.149 | 0.197 | 0.426 | 0.851 | 0.996 |
| R_lingual | 0.348 | 0.019 | 1.167 | 0.345 | 0.126 | 0.595 | 0.655 | 0.993 |
| R_lateralorbitofrontal | 0.255 | 0.015 | 1.255 | 0.252 | 0.15 | 0.382 | 0.748 | 0.999 |
| R_middletemporal | 0.273 | 0.035 | 1.283 | 0.262 | 0.152 | 0.439 | 0.738 | 0.995 |
| R_medialorbitofrontal | 0.221 | 0.016 | 1.241 | 0.218 | 0.169 | 0.411 | 0.782 | 0.997 |
| R_parahippocampal | 0.103 | 0.03 | 1.284 | 0.09 | 0.241 | 0.329 | 0.91 | 0.987 |
| L_parstriangularis | 0.288 | 0.011 | 1.219 | 0.287 | 0.158 | 0.462 | 0.713 | 0.998 |
| Median_Thickness | 0.384 | 0.019 | 1.191 | 0.381 | 0.105 | 0.504 | 0.619 | 0.998 |
| R_parsopercularis | 0.247 | 0.007 | 1.246 | 0.247 | 0.158 | 0.4 | 0.753 | 0.999 |
| L_parsorbitalis | 0.224 | 0.016 | 1.303 | 0.223 | 0.191 | 0.407 | 0.777 | 0.997 |
| Mean_Thickness | 0.397 | 0.017 | 1.145 | 0.394 | 0.101 | 0.527 | 0.606 | 0.998 |

## Supplemental Table S5.2. Performance Metrics Subcortical Measures

| Region | EXPV | MACE | MAPE | NLL | R2 | RMSE | Rho | SMSE | ShapiroW |
|---|---|---|---|---|---|---|---|---|---|
| Left-Hippocampus | 0.370 | 0.014 | 0.078 | 1.239 | 0.368 | 5392.264 | 0.582 | 0.632 | 0.996 |
| Brain-Stem | 0.621 | 0.048 | 0.175 | 1.289 | 0.609 | 118.243 | 0.805 | 0.391 | 0.997 |
| Right-Putamen | 0.338 | 0.009 | 0.088 | 1.389 | 0.336 | 435.303 | 0.552 | 0.664 | 0.988 |
| Left-Amygdala | 0.077 | 0.032 | 0.229 | 1.383 | 0.063 | 552.738 | 0.282 | 0.937 | 0.996 |
| Left-VentralDC | 0.363 | 0.048 | 0.103 | 1.356 | 0.341 | 516.797 | 0.619 | 0.659 | 0.997 |
| Right-Caudate | 0.246 | 0.014 | 0.100 | 1.373 | 0.245 | 233.742 | 0.472 | 0.755 | 0.995 |
| CC_Posterior | 0.308 | 0.010 | 0.078 | 1.277 | 0.307 | 382.376 | 0.542 | 0.693 | 0.996 |
| Right-Cerebellum-White-Matter | 0.334 | 0.005 | 0.081 | 1.179 | 0.334 | 2099.602 | 0.576 | 0.666 | 0.996 |
| Right-VentralDC | 0.316 | 0.032 | 0.106 | 1.377 | 0.311 | 229.076 | 0.560 | 0.689 | 0.993 |
| Right-Cerebellum-Cortex | 0.354 | 0.013 | 0.086 | 1.390 | 0.351 | 417.895 | 0.570 | 0.649 | 0.987 |
| Left-choroid-plexus | 0.531 | 0.063 | 0.093 | 1.326 | 0.505 | 640.973 | 0.746 | 0.495 | 0.997 |
| CSF | 0.395 | 0.011 | 0.110 | 1.327 | 0.393 | 207.975 | 0.627 | 0.607 | 0.994 |
| Right-Inf-Lat-Vent | 0.265 | 0.175 | 0.375 | 2.658 | 0.052 | 391.661 | 0.485 | 0.948 | 0.993 |
| CC_Central | 0.549 | 0.014 | 0.231 | 0.747 | 0.549 | 375.547 | 0.736 | 0.451 | 0.993 |
| 4th-Ventricle | 0.163 | 0.044 | 0.105 | 1.302 | 0.142 | 2036.171 | 0.436 | 0.858 | 0.995 |
| WM-hypointensities | 0.282 | 0.045 | 0.184 | 1.274 | 0.261 | 270.548 | 0.549 | 0.739 | 0.989 |
| Left-vessel | 0.503 | 0.008 | | 0.703 | 0.502 | 5183.528 | 0.736 | 0.498 | 0.990 |
| 3rd-Ventricle | 0.277 | 0.014 | 0.514 | 1.146 | 0.270 | 361.854 | 0.459 | 0.730 | 0.970 |
| Right-Amygdala | 0.396 | 0.014 | 0.077 | 1.249 | 0.394 | 5219.691 | 0.598 | 0.606 | 0.996 |

| Region | | | | | | | | | |
|---|---|---|---|---|---|---|---|---|---|
| Right-vessel | 0.543 | 0.058 | 0.096 | 1.319 | 0.522 | 656.020 | 0.749 | 0.478 | 0.998 |
| Left-Putamen | 0.351 | 0.021 | 0.110 | 1.378 | 0.350 | 257.872 | 0.571 | 0.650 | 0.993 |
| Left-Cerebellum-White-Matter | 0.086 | 0.154 | | 2.007 | -0.121 | 54.947 | 0.231 | 1.121 | 0.992 |
| Left-Caudate | 0.359 | 0.049 | 0.103 | 1.368 | 0.335 | 501.175 | 0.618 | 0.665 | 0.996 |
| Left-Pallidum | 0.144 | 0.038 | 0.103 | 1.323 | 0.126 | 2051.112 | 0.406 | 0.874 | 0.995 |
| Left-Cerebellum-Cortex | 0.349 | 0.039 | | 1.056 | 0.332 | 358.352 | 0.513 | 0.668 | 0.982 |
| Right-Hippocampus | 0.500 | 0.074 | 0.152 | 1.500 | 0.461 | 123.762 | 0.755 | 0.539 | 0.998 |
| Optic-Chiasm | 0.322 | 0.018 | 0.075 | 1.282 | 0.319 | 379.100 | 0.561 | 0.681 | 0.997 |
| Right-Pallidum | 0.177 | 0.022 | 0.176 | 1.316 | 0.171 | 104.269 | 0.426 | 0.829 | 0.980 |
| CC_Mid_Posterior | 0.253 | 0.013 | 0.138 | 1.252 | 0.251 | 150.949 | 0.524 | 0.749 | 0.996 |
| Left-Accumbens-area | 0.407 | 0.003 | 0.070 | 1.223 | 0.407 | | 0.630 | 0.593 | 0.994 |
| Right-Accumbens-area | 0.609 | 0.002 | 0.063 | 1.114 | 0.609 | | 0.785 | 0.391 | 0.995 |
| CC_Anterior | 0.130 | 0.012 | 0.144 | 1.295 | 0.127 | 146.831 | 0.355 | 0.873 | 0.997 |
| Left-Lateral-Ventricle | 0.382 | 0.002 | 0.066 | 1.217 | 0.382 | | 0.612 | 0.618 | 0.995 |
| lhCortexVol | 0.608 | 0.001 | 0.072 | 1.115 | 0.608 | | 0.791 | 0.392 | 0.994 |
| SupraTentorialVolNotVent | 0.609 | 0.002 | 0.072 | 1.116 | 0.609 | | 0.794 | 0.391 | 0.993 |
| TotalGrayVol | 0.502 | 0.046 | 0.064 | 1.275 | 0.487 | 4862.451 | 0.717 | 0.513 | 0.995 |
| rhCortexVol | 0.193 | 0.087 | | 1.517 | 0.128 | 62.178 | 0.414 | 0.872 | 0.998 |
| BrainSegVol | 0.367 | 0.174 | 0.347 | 2.325 | 0.208 | 394.102 | 0.523 | 0.792 | 0.990 |
| Right-choroid-plexus | 0.521 | 0.024 | 0.036 | 1.123 | 0.518 | 0.041 | 0.664 | 0.482 | 0.984 |
| SubCortGrayVol | 0.184 | 0.029 | 0.196 | 1.207 | 0.171 | 124.555 | 0.422 | 0.829 | 0.995 |
| CC_Mid_Anterior | 0.232 | 0.024 | 0.186 | 1.232 | 0.223 | 119.263 | 0.473 | 0.777 | 0.994 |
| SupraTentorialVol | 0.247 | 0.170 | | 1.418 | 0.154 | 21.291 | 0.354 | 0.846 | 0.983 |
| BrainSegVol-to-eTIV | 0.362 | 0.002 | 0.070 | 1.234 | 0.362 | | 0.599 | 0.638 | 0.995 |
| MaskVol-to-eTIV | 0.241 | 0.168 | | 1.380 | 0.153 | 20.529 | 0.262 | 0.847 | 0.984 |
| Left-Inf-Lat-Vent | 0.324 | 0.014 | 0.069 | 1.236 | 0.322 | | 0.578 | 0.678 | 0.988 |
| non-WM-hypointensities | 0.035 | 0.181 | | 2.505 | -0.213 | 65.963 | 0.138 | 1.213 | 0.990 |
| Right-Lateral-Ventricle | 0.489 | 0.012 | 0.470 | 0.679 | 0.489 | 5907.421 | 0.733 | 0.511 | 0.991 |
| EstimatedTotalIntraCranialVol | 0.299 | 0.029 | 0.029 | 1.178 | 0.292 | 0.055 | 0.549 | 0.708 | 0.974 |
| rhSurfaceHoles | 0.255 | 0.018 | 0.540 | 0.418 | 0.245 | 3227.508 | 0.741 | 0.755 | 0.966 |
| 5th-Ventricle | 0.100 | 0.168 | | | -0.021 | 20.429 | 0.457 | 1.021 | 0.017 |
| lhSurfaceHoles | 0.004 | 0.182 | | | -0.103 | 4.789 | 0.062 | 1.103 | 0.013 |

Supplemental Table S5.3. Performance Metrics Destrieux Atlas

| Region | EXPV | MACE | MAPE | NLL | R2 | RMSE | Rho | SMSE | ShapiroW |
|---|---|---|---|---|---|---|---|---|---|
| lh_G_and_S_paracentral | 0.270 | 0.020 | 0.064 | 1.229 | 0.267 | 0.200 | 0.506 | 0.733 | 0.998 |
| lh_S_front_sup | 0.450 | 0.008 | 0.044 | 1.134 | 0.450 | 0.147 | 0.603 | 0.550 | 0.998 |
| rh_G_rectus | 0.278 | 0.014 | | 1.274 | 0.275 | 0.219 | 0.432 | 0.725 | 0.998 |
| lh_G_and_S_occipital_inf | 0.228 | 0.013 | | 1.356 | 0.226 | 0.181 | 0.426 | 0.774 | 0.995 |
| lh_S_front_middle | 0.396 | 0.003 | 0.053 | 1.200 | 0.396 | 0.162 | 0.578 | 0.604 | 0.999 |
| rh_G_precuneus | 0.395 | 0.017 | 0.049 | 1.224 | 0.393 | 0.165 | 0.571 | 0.607 | 0.997 |
| lh_G_and_S_cingul-Ant | 0.390 | 0.010 | | 1.264 | 0.389 | 0.162 | 0.509 | 0.611 | 0.996 |
| lh_S_oc_middle_and_Lunatus | 0.271 | 0.022 | | 1.286 | 0.266 | 0.167 | 0.501 | 0.734 | 0.998 |
| rh_G_temp_sup-Lateral | 0.354 | 0.022 | 0.051 | 1.307 | 0.350 | 0.204 | 0.491 | 0.650 | 0.996 |
| lh_G_and_S_transv_frontopol | 0.299 | 0.017 | | 1.283 | 0.296 | 0.212 | 0.521 | 0.704 | 0.996 |
| lh_S_intrapariet_and_P_trans | 0.403 | 0.008 | 0.049 | 1.154 | 0.402 | 0.146 | 0.560 | 0.598 | 0.998 |
| rh_G_temp_sup-G_T_transv | 0.273 | 0.007 | 0.072 | 1.239 | 0.273 | 0.243 | 0.472 | 0.727 | 0.999 |
| lh_G_and_S_subcentral | 0.335 | 0.022 | 0.051 | 1.262 | 0.330 | 0.181 | 0.446 | 0.670 | 0.998 |
| lh_S_interm_prim-Jensen | 0.174 | 0.016 | | 1.359 | 0.170 | 0.341 | 0.410 | 0.830 | 0.965 |
| rh_G_subcallosal | 0.120 | 0.028 | | 1.181 | 0.110 | 0.366 | 0.354 | 0.890 | 0.999 |
| lh_G_and_S_cingul-Mid-Post | 0.422 | 0.037 | | 1.238 | 0.411 | 0.159 | 0.532 | 0.589 | 0.991 |
| lh_S_occipital_ant | 0.244 | 0.024 | 0.061 | 1.300 | 0.237 | 0.183 | 0.429 | 0.763 | 0.999 |
| rh_G_temp_sup-Plan_tempo | 0.320 | 0.019 | 0.060 | 1.236 | 0.316 | 0.206 | 0.455 | 0.684 | 0.999 |
| lh_G_and_S_frontomargin | 0.344 | 0.011 | 0.057 | 1.260 | 0.343 | 0.182 | 0.539 | 0.657 | 0.999 |
| lh_S_front_inf | 0.400 | 0.009 | 0.048 | 1.205 | 0.400 | 0.149 | 0.540 | 0.600 | 0.999 |
| rh_G_precentral | 0.247 | 0.012 | | 1.230 | 0.247 | 0.234 | 0.437 | 0.753 | 0.996 |
| lh_G_and_S_cingul-Mid-Ant | 0.342 | 0.031 | | 1.118 | 0.338 | 0.197 | 0.553 | 0.662 | 0.992 |
| lh_S_oc_sup_and_transversal | 0.312 | 0.013 | 0.055 | 1.242 | 0.311 | 0.156 | 0.503 | 0.689 | 0.999 |
| rh_G_temp_sup-Plan_polar | 0.239 | 0.014 | 0.069 | 1.258 | 0.238 | 0.279 | 0.383 | 0.762 | 0.997 |
| lh_G_front_inf-Opercular | 0.392 | 0.020 | 0.047 | 1.224 | 0.390 | 0.173 | 0.529 | 0.610 | 0.997 |
| lh_S_orbital_med-olfact | 0.192 | 0.016 | 0.061 | 1.377 | 0.190 | 0.192 | 0.400 | 0.810 | 0.992 |
| rh_Lat_Fis-ant-Vertical | 0.177 | 0.022 | | 1.383 | 0.171 | 0.281 | 0.379 | 0.829 | 0.974 |
| lh_G_front_sup | 0.494 | 0.029 | 0.044 | 1.201 | 0.488 | 0.169 | 0.615 | 0.512 | 0.995 |
| lh_S_postcentral | 0.368 | 0.006 | 0.053 | 1.152 | 0.368 | 0.154 | 0.544 | 0.632 | 0.998 |
| rh_S_calcarine | 0.470 | 0.023 | 0.056 | 1.224 | 0.466 | 0.142 | 0.660 | 0.534 | 0.998 |
| lh_G_cingul-Post-dorsal | 0.430 | 0.047 | | 1.340 | 0.411 | 0.192 | 0.510 | 0.589 | 0.993 |
| lh_S_oc-temp_lat | 0.272 | 0.032 | | 1.369 | 0.260 | 0.185 | 0.430 | 0.740 | 0.996 |
| rh_G_temporal_inf | 0.340 | 0.022 | 0.048 | 1.298 | 0.335 | 0.184 | 0.466 | 0.665 | 0.996 |

| Region | | | | | | | | |
|---|---|---|---|---|---|---|---|---|
| lh_G_cuneus | 0.255 | 0.004 | | 1.286 | 0.255 | 0.148 | 0.507 | 0.745 | 0.994 |
| lh_S_orbital_lateral | 0.277 | 0.010 | | 1.275 | 0.276 | 0.228 | 0.504 | 0.724 | 0.998 |
| rh_Lat_Fis-ant-Horizont | 0.240 | 0.013 | | 1.279 | 0.238 | 0.230 | 0.415 | 0.762 | 0.999 |
| lh_G_cingul-Post-ventral | 0.209 | 0.020 | | 1.320 | 0.203 | 0.266 | 0.404 | 0.797 | 0.997 |
| lh_S_oc-temp_med_and_Lingual | 0.391 | 0.036 | 0.054 | 1.282 | 0.377 | 0.165 | 0.510 | 0.623 | 0.996 |
| rh_G_temporal_middle | 0.422 | 0.034 | 0.045 | 1.318 | 0.412 | 0.179 | 0.552 | 0.588 | 0.994 |
| lh_G_insular_short | 0.077 | 0.014 | 0.058 | 1.330 | 0.075 | 0.260 | 0.330 | 0.925 | 0.997 |
| lh_S_precentral-sup-part | 0.300 | 0.009 | 0.053 | 1.203 | 0.299 | 0.173 | 0.495 | 0.701 | 0.996 |
| rh_S_cingul-Marginalis | 0.382 | 0.021 | | 1.190 | 0.378 | 0.162 | 0.553 | 0.622 | 0.998 |
| lh_G_front_inf-Triangul | 0.430 | 0.023 | 0.051 | 1.227 | 0.426 | 0.181 | 0.607 | 0.574 | 0.998 |
| lh_S_parieto_occipital | 0.382 | 0.013 | | 1.226 | 0.381 | 0.159 | 0.527 | 0.619 | 0.997 |
| rh_Pole_occipital | 0.151 | 0.005 | 0.059 | 1.321 | 0.151 | 0.152 | 0.365 | 0.849 | 0.999 |
| lh_G_front_middle | 0.472 | 0.025 | 0.046 | 1.197 | 0.468 | 0.162 | 0.628 | 0.532 | 0.997 |
| lh_S_pericallosal | 0.011 | 0.047 | | 1.489 | -0.030 | 0.319 | 0.435 | 1.030 | 0.965 |
| rh_Pole_temporal | 0.216 | 0.007 | 0.052 | 1.333 | 0.216 | 0.223 | 0.406 | 0.784 | 0.993 |
| lh_G_Ins_lg_and_S_cent_ins | 0.154 | 0.022 | 0.066 | 1.316 | 0.147 | 0.258 | 0.381 | 0.853 | 0.997 |
| lh_S_precentral-inf-part | 0.363 | 0.013 | 0.048 | 1.216 | 0.361 | 0.160 | 0.505 | 0.639 | 0.998 |
| rh_S_central | 0.191 | 0.012 | 0.059 | 1.241 | 0.189 | 0.151 | 0.442 | 0.811 | 0.990 |
| lh_G_occipital_sup | 0.224 | 0.003 | 0.063 | 1.340 | 0.224 | 0.179 | 0.454 | 0.776 | 0.998 |
| lh_S_subparietal | 0.360 | 0.039 | | 1.296 | 0.345 | 0.171 | 0.509 | 0.655 | 0.997 |
| rh_S_circular_insula_inf | 0.429 | 0.016 | 0.058 | 1.230 | 0.428 | 0.202 | 0.561 | 0.572 | 0.996 |
| lh_G_occipital_middle | 0.295 | 0.020 | 0.048 | 1.310 | 0.292 | 0.158 | 0.468 | 0.708 | 0.994 |
| lh_S_suborbital | 0.194 | 0.031 | | 1.335 | 0.181 | 0.255 | 0.385 | 0.819 | 0.997 |
| rh_S_circular_insula_ant | 0.301 | 0.022 | | 1.285 | 0.295 | 0.218 | 0.425 | 0.705 | 0.997 |
| lh_G_oc-temp_lat-fusifor | 0.323 | 0.030 | 0.048 | 1.260 | 0.312 | 0.181 | 0.508 | 0.688 | 0.996 |
| lh_S_temporal_inf | 0.286 | 0.012 | | 1.303 | 0.284 | 0.173 | 0.473 | 0.716 | 0.996 |
| rh_S_circular_insula_sup | 0.441 | 0.033 | | 1.242 | 0.431 | 0.153 | 0.545 | 0.569 | 0.998 |
| lh_G_front_inf-Orbital | 0.352 | 0.014 | 0.058 | 1.279 | 0.351 | 0.216 | 0.563 | 0.649 | 0.999 |
| lh_S_orbital-H_Shaped | 0.424 | 0.015 | 0.054 | 1.206 | 0.422 | 0.192 | 0.644 | 0.578 | 0.999 |
| rh_Lat_Fis-post | 0.446 | 0.021 | 0.050 | 1.211 | 0.442 | 0.164 | 0.532 | 0.558 | 0.996 |
| lh_G_pariet_inf-Supramar | 0.417 | 0.023 | 0.047 | 1.255 | 0.413 | 0.170 | 0.529 | 0.587 | 0.998 |
| rh_G_and_S_paracentral | 0.259 | 0.027 | | 1.253 | 0.252 | 0.203 | 0.503 | 0.748 | 0.996 |
| rh_S_front_sup | 0.436 | 0.005 | | 1.176 | 0.436 | 0.151 | 0.591 | 0.564 | 0.998 |
| lh_G_oc-temp_med-Parahip | 0.151 | 0.017 | 0.068 | 1.332 | 0.147 | 0.254 | 0.318 | 0.853 | 0.998 |

| Region | | | | | | | | |
|---|---|---|---|---|---|---|---|---|
| lh_S_temporal_transverse | 0.182 | 0.020 | | 1.293 | 0.177 | 0.293 | 0.360 | 0.823 0.993 |
| rh_S_collat_transv_post | 0.271 | 0.019 | | 1.295 | 0.267 | 0.191 | 0.423 | 0.733 0.997 |
| lh_G_postcentral | 0.227 | 0.015 | 0.066 | 1.304 | 0.223 | 0.191 | 0.458 | 0.777 0.998 |
| rh_G_and_S_transv_frontopol | 0.328 | 0.016 | | 1.275 | 0.325 | 0.197 | 0.562 | 0.675 0.996 |
| rh_S_intrapariet_and_P_trans | 0.402 | 0.009 | 0.050 | 1.161 | 0.402 | 0.147 | 0.568 | 0.598 0.998 |
| lh_G_precentral | 0.274 | 0.009 | 0.060 | 1.227 | 0.274 | 0.223 | 0.462 | 0.726 0.996 |
| rh_G_and_S_cingul-Ant | 0.409 | 0.010 | 0.045 | 1.243 | 0.408 | 0.162 | 0.537 | 0.592 0.999 |
| rh_S_oc_middle_and_Lunatus | 0.298 | 0.021 | 0.063 | 1.281 | 0.294 | 0.176 | 0.507 | 0.706 0.998 |
| lh_G_rectus | 0.317 | 0.020 | | 1.247 | 0.314 | 0.203 | 0.495 | 0.686 0.996 |
| rh_G_and_S_cingul-Mid-Post | 0.400 | 0.032 | 0.044 | 1.211 | 0.391 | 0.152 | 0.499 | 0.609 0.995 |
| rh_S_occipital_ant | 0.254 | 0.029 | 0.062 | 1.282 | 0.244 | 0.187 | 0.437 | 0.756 0.999 |
| lh_G_temporal_inf | 0.358 | 0.019 | 0.048 | 1.305 | 0.354 | 0.185 | 0.498 | 0.646 0.999 |
| rh_G_front_inf-Triangul | 0.406 | 0.019 | | 1.245 | 0.404 | 0.185 | 0.609 | 0.596 0.997 |
| rh_S_parieto_occipital | 0.386 | 0.017 | 0.053 | 1.215 | 0.383 | 0.158 | 0.520 | 0.617 0.999 |
| lh_G_oc-temp_med-Lingual | 0.358 | 0.014 | | 1.193 | 0.357 | 0.143 | 0.600 | 0.643 0.996 |
| lh_S_temporal_sup | 0.461 | 0.027 | 0.043 | 1.233 | 0.454 | 0.140 | 0.575 | 0.546 0.999 |
| rh_S_collat_transv_ant | 0.194 | 0.020 | | 1.380 | 0.190 | 0.221 | 0.375 | 0.810 0.998 |
| lh_G_pariet_inf-Angular | 0.475 | 0.026 | 0.048 | 1.244 | 0.470 | 0.168 | 0.627 | 0.530 0.997 |
| rh_G_and_S_occipital_inf | 0.195 | 0.014 | 0.062 | 1.368 | 0.192 | 0.204 | 0.377 | 0.808 0.999 |
| rh_S_front_middle | 0.377 | 0.010 | | 1.228 | 0.376 | 0.161 | 0.591 | 0.624 0.997 |
| lh_G_orbital | 0.434 | 0.023 | 0.046 | 1.273 | 0.430 | 0.168 | 0.607 | 0.570 0.996 |
| rh_G_and_S_frontomargin | 0.313 | 0.022 | | 1.294 | 0.307 | 0.203 | 0.519 | 0.693 0.998 |
| rh_S_front_inf | 0.376 | 0.009 | | 1.225 | 0.375 | 0.153 | 0.545 | 0.625 0.998 |
| lh_G_parietal_sup | 0.392 | 0.006 | 0.052 | 1.212 | 0.392 | 0.165 | 0.603 | 0.608 0.998 |
| rh_G_and_S_subcentral | 0.328 | 0.025 | 0.052 | 1.267 | 0.322 | 0.183 | 0.445 | 0.678 0.999 |
| rh_S_interm_prim-Jensen | 0.300 | 0.010 | | 1.252 | 0.299 | 0.222 | 0.495 | 0.701 0.995 |
| lh_Pole_occipital | 0.136 | 0.003 | 0.067 | 1.329 | 0.136 | 0.170 | 0.362 | 0.864 0.997 |
| rh_G_occipital_middle | 0.353 | 0.020 | 0.047 | 1.282 | 0.350 | 0.160 | 0.510 | 0.650 0.994 |
| rh_S_suborbital | 0.174 | 0.016 | | 1.341 | 0.170 | 0.374 | 0.433 | 0.830 0.985 |
| lh_S_circular_insula_sup | 0.459 | 0.034 | 0.043 | 1.235 | 0.450 | 0.149 | 0.529 | 0.550 0.999 |
| rh_G_pariet_inf-Supramar | 0.405 | 0.016 | 0.047 | 1.250 | 0.404 | 0.171 | 0.531 | 0.596 0.999 |
| lh_S_central | 0.205 | 0.013 | 0.057 | 1.220 | 0.203 | 0.150 | 0.431 | 0.797 0.993 |
| rh_G_oc-temp_med-Lingual | 0.358 | 0.015 | 0.055 | 1.195 | 0.356 | 0.148 | 0.598 | 0.644 0.998 |
| rh_S_temporal_sup | 0.482 | 0.026 | 0.042 | 1.216 | 0.476 | 0.142 | 0.565 | 0.524 0.999 |

| Region | | | | | | | | |
|---|---|---|---|---|---|---|---|---|
| lh_S_collat_transv_post | 0.241 | 0.022 | | 1.309 | 0.235 | 0.191 | 0.401 | 0.765 | 0.997 |
| rh_G_postcentral | 0.205 | 0.020 | 0.071 | 1.308 | 0.199 | 0.203 | 0.443 | 0.801 | 0.999 |
| lh_S_collat_transv_ant | 0.191 | 0.022 | | 1.302 | 0.184 | 0.230 | 0.360 | 0.816 | 0.998 |
| rh_G_parietal_sup | 0.390 | 0.007 | 0.055 | 1.227 | 0.389 | 0.173 | 0.609 | 0.611 | 0.999 |
| lh_G_temp_sup-G_T_transv | 0.284 | 0.008 | 0.074 | 1.252 | 0.284 | 0.242 | 0.511 | 0.716 | 0.999 |
| rh_G_cingul-Post-ventral | 0.201 | 0.029 | | 1.349 | 0.190 | 0.249 | 0.441 | 0.810 | 0.994 |
| rh_S_oc-temp_med_and_Lingual | 0.381 | 0.036 | 0.054 | 1.266 | 0.367 | 0.166 | 0.556 | 0.633 | 0.995 |
| lh_G_subcallosal | 0.135 | 0.051 | | 1.287 | 0.098 | 0.323 | 0.430 | 0.902 | 0.998 |
| rh_G_cingul-Post-dorsal | 0.400 | 0.036 | | 1.294 | 0.387 | 0.204 | 0.532 | 0.613 | 0.995 |
| rh_S_oc-temp_lat | 0.279 | 0.035 | 0.056 | 1.349 | 0.264 | 0.183 | 0.449 | 0.736 | 0.998 |
| lh_G_temp_sup-Lateral | 0.330 | 0.019 | 0.053 | 1.327 | 0.326 | 0.211 | 0.491 | 0.674 | 0.998 |
| rh_G_cuneus | 0.317 | 0.015 | 0.056 | 1.266 | 0.315 | 0.140 | 0.580 | 0.685 | 0.998 |
| rh_S_orbital_lateral | 0.306 | 0.007 | | 1.293 | 0.306 | 0.208 | 0.508 | 0.694 | 0.998 |
| lh_G_temp_sup-Plan_polar | 0.229 | 0.014 | 0.070 | 1.211 | 0.228 | 0.296 | 0.399 | 0.772 | 0.997 |
| rh_G_front_inf-Opercular | 0.400 | 0.026 | | 1.237 | 0.395 | 0.172 | 0.549 | 0.605 | 0.996 |
| rh_S_orbital_med-olfact | 0.200 | 0.016 | | 1.379 | 0.197 | 0.202 | 0.401 | 0.803 | 0.995 |
| lh_G_precuneus | 0.411 | 0.025 | 0.050 | 1.211 | 0.405 | 0.167 | 0.571 | 0.595 | 0.998 |
| rh_G_and_S_cingul-Mid-Ant | 0.377 | 0.033 | 0.049 | 1.167 | 0.370 | 0.175 | 0.535 | 0.630 | 0.994 |
| rh_S_oc_sup_and_transversal | 0.347 | 0.012 | 0.054 | 1.223 | 0.346 | 0.154 | 0.507 | 0.654 | 0.998 |
| lh_G_temp_sup-Plan_tempo | 0.288 | 0.016 | 0.061 | 1.248 | 0.285 | 0.208 | 0.446 | 0.715 | 0.999 |
| rh_G_front_inf-Orbital | 0.297 | 0.017 | | 1.303 | 0.295 | 0.230 | 0.532 | 0.705 | 0.997 |
| rh_S_orbital-H_Shaped | 0.438 | 0.017 | 0.054 | 1.208 | 0.436 | 0.188 | 0.641 | 0.564 | 0.999 |
| lh_Lat_Fis-ant-Horizont | 0.171 | 0.013 | 0.083 | 1.301 | 0.169 | 0.256 | 0.349 | 0.831 | 1.000 |
| rh_G_front_sup | 0.488 | 0.023 | 0.044 | 1.210 | 0.484 | 0.168 | 0.623 | 0.516 | 0.995 |
| rh_S_postcentral | 0.369 | 0.007 | 0.055 | 1.175 | 0.368 | 0.158 | 0.546 | 0.632 | 0.998 |
| lh_G_temporal_middle | 0.401 | 0.018 | 0.047 | 1.312 | 0.399 | 0.184 | 0.562 | 0.601 | 0.996 |
| rh_G_front_middle | 0.451 | 0.018 | | 1.216 | 0.450 | 0.162 | 0.643 | 0.550 | 0.996 |
| rh_S_pericallosal | 0.067 | 0.050 | 0.128 | 1.403 | 0.027 | 0.308 | 0.404 | 0.973 | 0.968 |
| lh_Lat_Fis-post | 0.405 | 0.023 | 0.054 | 1.242 | 0.400 | 0.171 | 0.508 | 0.600 | 0.999 |
| rh_G_insular_short | 0.055 | 0.021 | 0.068 | 1.359 | 0.048 | 0.285 | 0.304 | 0.952 | 0.994 |
| rh_S_precentral-sup-part | 0.266 | 0.010 | | 1.228 | 0.266 | 0.187 | 0.485 | 0.734 | 0.994 |
| lh_Lat_Fis-ant-Vertical | 0.225 | 0.012 | | 1.302 | 0.223 | 0.239 | 0.409 | 0.777 | 0.994 |
| rh_G_Ins_lg_and_S_cent_ins | 0.184 | 0.018 | 0.073 | 1.335 | 0.180 | 0.296 | 0.422 | 0.820 | 0.996 |
| rh_S_precentral-inf-part | 0.337 | 0.010 | | 1.235 | 0.336 | 0.164 | 0.481 | 0.664 | 0.998 |

| Region | | | | | | | | | |
|---|---|---|---|---|---|---|---|---|---|
| lh_Pole_temporal | 0.224 | 0.010 | | 1.304 | 0.223 | 0.223 | 0.416 | 0.777 | 0.996 |
| rh_G_occipital_sup | 0.299 | 0.008 | 0.061 | 1.315 | 0.299 | 0.175 | 0.536 | 0.701 | 0.998 |
| rh_S_subparietal | 0.359 | 0.044 | | 1.273 | 0.340 | 0.171 | 0.497 | 0.660 | 0.999 |
| lh_S_circular_insula_ant | 0.329 | 0.024 | 0.057 | 1.292 | 0.322 | 0.213 | 0.445 | 0.678 | 0.998 |
| rh_G_orbital | 0.356 | 0.017 | 0.047 | 1.278 | 0.353 | 0.172 | 0.545 | 0.647 | 0.997 |
| Mean_Thickness | 0.570 | 0.023 | 0.031 | 1.144 | 0.566 | 0.103 | 0.595 | 0.434 | 0.998 |
| lh_S_calcarine | 0.469 | 0.030 | | 1.217 | 0.463 | 0.138 | 0.654 | 0.537 | 0.995 |
| rh_G_oc-temp_lat-fusifor | 0.313 | 0.026 | 0.048 | 1.265 | 0.305 | 0.178 | 0.467 | 0.695 | 0.994 |
| rh_S_temporal_inf | 0.323 | 0.024 | 0.052 | 1.314 | 0.317 | 0.170 | 0.487 | 0.683 | 0.998 |
| lh_S_circular_insula_inf | 0.412 | 0.008 | 0.054 | 1.234 | 0.412 | 0.196 | 0.554 | 0.588 | 0.997 |
| rh_G_pariet_inf-Angular | 0.466 | 0.025 | 0.048 | 1.235 | 0.462 | 0.170 | 0.628 | 0.538 | 0.997 |
| Median_Thickness | 0.570 | 0.025 | 0.032 | 1.153 | 0.565 | 0.107 | 0.589 | 0.435 | 0.999 |
| lh_S_cingul-Marginalis | 0.414 | 0.022 | | 1.205 | 0.410 | 0.167 | 0.598 | 0.590 | 0.998 |
| rh_G_oc-temp_med-Parahip | 0.192 | 0.010 | | 1.339 | 0.190 | 0.248 | 0.379 | 0.810 | 0.996 |
| rh_S_temporal_transverse | 0.218 | 0.016 | | 1.326 | 0.215 | 0.305 | 0.376 | 0.785 | 0.995 |